\documentclass[%
 reprint,
superscriptaddress,
 amsmath,amssymb,
 aps,
]{revtex4-2}

\usepackage{graphicx}
\usepackage{dcolumn}
\usepackage{bm}
\usepackage{booktabs}
\usepackage[version=4]{mhchem}

\usepackage[table,xcdraw]{xcolor}

\usepackage{tablefootnote}

\begin{document}

\preprint{APS/123-QED}

\title{Revealing the bonding nature and electronic structure of early transition metal dihydrides}

\date{\today}

\author{Curran~Kalha}%
\affiliation{%
Department of Chemistry, University College London, 20 Gordon Street, London WC1H 0AJ, United Kingdom.
}%

\author{Laura~E.~Ratcliff}
\affiliation{ 
Centre for Computational Chemistry, School of Chemistry, University of Bristol, Bristol BS8 1TS, United Kingdom.
}

\author{Giorgio~Colombi}%
\affiliation{%
Materials for Energy Conversion and Storage, Department of Chemical Engineering, Delft University of Technology, NL-2629HZ Delft, The Netherlands.
}%
\author{Christoph~Schlueter}
\affiliation{%
Deutsches Elektronen-Synchrotron DESY, Notkestra{\ss}e 85, 22607 Hamburg, Germany.
}%

\author{Bernard~Dam}%
\affiliation{%
Materials for Energy Conversion and Storage, Department of Chemical Engineering, Delft University of Technology, NL-2629HZ Delft, The Netherlands.
}%

\author{Andrei~Gloskovskii}
\affiliation{%
Deutsches Elektronen-Synchrotron DESY, Notkestra{\ss}e 85, 22607 Hamburg, Germany.
}%

\author{Tien-Lin~Lee}%
\affiliation{%
Diamond Light Source Ltd., Diamond House, Harwell Science and Innovation Campus, Didcot, OX11 0DE, United Kingdom.
}%

\author{Pardeep~K.~Thakur}%
\affiliation{%
Diamond Light Source Ltd., Diamond House, Harwell Science and Innovation Campus, Didcot, OX11 0DE, United Kingdom.
}%

\author{Prajna~Bhatt}%
\affiliation{%
Department of Chemistry, University College London, 20 Gordon Street, London WC1H 0AJ, United Kingdom.
}%

\author{Yujiang~Zhu}%
\affiliation{%
Department of Chemistry, University College London, 20 Gordon Street, London WC1H 0AJ, United Kingdom.
}%

\author{J\"urg~Osterwalder}%
\affiliation{Physik-Institut, Universit\"{a}t Z{\"u}rich, CH-8057 Z{\"u}rich, Switzerland.
}%

\author{Francesco~Offi}%
\affiliation{%
Dipartimento di Scienze, Università di Roma Tre, 00146 Rome, Italy.
}%

\author{Giancarlo~Panaccione}%
\email{panaccione@iom.cnr.it}
\affiliation{%
Istituto Officina dei Materiali (IOM)-CNR, Laboratorio TASC, in Area Science Park, S.S.14, Km 163.5, I-34149 Trieste, Italy.
}%

\author{Anna~Regoutz}%
 \email{a.regoutz@ucl.ac.uk}
\affiliation{%
Department of Chemistry, University College London, 20 Gordon Street, London WC1H 0AJ, United Kingdom.
}%

\date{\today}

\begin{abstract}

Hydrogen as a fuel plays a crucial role in driving the transition to net zero greenhouse gas emissions. To realise its potential, obtaining a means of efficient storage is paramount. One solution is using metal hydrides, owing to their good thermodynamical absorption properties and effective hydrogen storage. Although metal hydrides appear simple compared to many other energy materials, understanding the electronic structure and chemical environment of hydrogen within them remains a key challenge. This work presents a new analytical pathway to explore these aspects in technologically relevant systems using Hard X-ray Photoelectron Spectroscopy (HAXPES) on thin films of two prototypical metal dihydrides: YH\textsubscript{2-$\delta$} and TiH\textsubscript{2-$\delta$}. By taking advantage of the tunability of synchrotron radiation, a non-destructive depth profile of the chemical states is obtained using core level spectra. Combining experimental valence band spectra collected at varying photon energies with theoretical insights from density functional theory (DFT) calculations, a description of the bonding nature and the role of \textit{d} versus \textit{sp} contributions to states near the Fermi energy are provided. Moreover, a reliable determination of the enthalpy of formation is proposed by using experimental values of the energy position of metal \textit{s} band features close to the Fermi energy in the HAXPES valence band spectra.

\end{abstract}

\maketitle


\section{Introduction}

Metal hydrides present a viable route to efficient hydrogen storage,~\cite{Schlappbach2001, Joubert2002, Sakintuna2007, BELLOSTAVONCOLBE20197780} having demonstrated many essential advantages, such as reversibility of hydrogen absorption,~\cite{Wu_2021} good and tunable thermodynamical absorption properties,~\cite{Ngene2017} and a higher volumetric hydrogen density than compressed or liquefied hydrogen,~\cite{modi2021room} while their switchable optical properties allow for the development of optical hydrogen sensors.~\cite{HUIBERTS1996158, Ngene2017, Palm_2018} These properties are heavily dictated by their thermodynamic behaviour and by extension, their electronic structure. Therefore, a fundamental understanding of the electron interaction between hydrogen and the metal atoms at a local atomic level is of utmost importance for the optimisation of any process.\par

Identifying the electronic character and the chemical environment of hydrogen in metal hydrides remains a key challenge in research focused on energy materials, catalysis, and gas-storage technology. Longstanding questions remain over (i) whether hydrides should be considered as predominantly ionic or covalent, and (ii) the location of hydrogen-derived states in the \textit{d}-band of the metal or in the energy gap of the corresponding oxides.~\cite{Fujimori_1984} Moreover, the strength and stability of the metal-hydrogen bond, as influenced by the possible formation and/or coexistence of other phases (e.g.\ oxides and hydroxides), defects, and by long-range diffusion of hydrogen atoms over the hydride thickness, is an essential parameter to be characterised and controlled.\par

Hydrogen presents a formidable challenge for many characterisation techniques, with very few having sensitivity to its chemical states and bonding or being able to detect hydrogen-induced modifications on the electronic states directly.~\cite{MALINOWSKI19831, Liu_2019} One such technique is photoelectron spectroscopy (PES), which also has the added value of being non-destructive.~\cite{king_2021} PES is frequently regarded as being ``blind'' to hydrogen, given it has one electron and no available core line needed for chemical state analysis, however, this is misleading as the effect of hydrogen can be observed in the core levels of elements that are bonded with hydrogen. This has enabled PES-based studies to capture many relevant characteristics of metal hydrides to date.~\cite{Weaver1979, Weaver1981, SCHLAPBACH1982271, Butera1983, Fujimori_1984, Osterwalder1985OnHydrides, Riesterer1987, Hayoz_2000, Hayoz2001SwitchableStudy, Hayoz2002ANGLE-SCANNED:, Hayoz_2003, UNO2004101, kato2012co, Mongstad2014TheOxide, billeter2021surface} Observations include (i) their high surface reactivity and formation of detrimental stable intrinsic oxides (influencing their catalytic properties),~\cite{Butera1983} and (ii) the changes in the electronic character of the extended valence states upon formation of the hydrides (influencing their (semi)-metallic or semiconducting behaviour).~\cite{Weaver1979,Riesterer1987} However, these results are predominantly obtained with highly surface-sensitive methods, including ultra-violet or soft X-ray electron emission excitation. As metal hydrides are extremely reactive, with their surfaces prone to oxidation under ambient conditions, it is challenging to disentangle the contribution from surface overlayers from those intrinsic to the hydride itself. A precise, quantitative determination of both the thickness-dependent composition of surface oxides, hydroxides, and/or hydrides, and of the concentration/gradient of hydrogen-related features remains challenging, leaving significant uncertainties in the understanding of hydrides both experimentally and theoretically.\par

Synchrotron-based PES, particularly hard X-ray photoelectron spectroscopy (HAXPES), provides tunability of the photon energy, enabling access to a range of photon energies within the hard X-ray regime. This allows for both the probing depth and the photoionisation cross sections to be manipulated, with the latter enhancing the sensitivity to specific orbital states.~\cite{Panaccione2005,Sacchi2005,Fadley2005, PAYNE200926, Claessen_2009, Mudd2014, woicik, HAXPES_Big_Boy} Regarding the application of HAXPES to metal dihydrides, the increase in probing depth with HAXPES over traditional soft X-ray PES enables the study of truly bulk-like hydrides as the contribution from the surface, which is expected to be metal oxide-rich, is minimised. The probing depth advantage also removes constraints on the samples, as past studies using soft X-ray PES often needed samples to be cleaved or grown in-situ to avoid surface oxidation. HAXPES is well suited for measuring realistic and technologically relevant samples with no surface preparation, as in the present case, where thin films of prototypical metal hydrides have been grown ex-situ on specific substrates.\par

This work exploits the bulk sensitivity of HAXPES to probe the electronic structure of metal-hydrogen states of two technologically relevant metal dihydrides: titanium dihydride (TiH\textsubscript{2-$\delta$}) and yttrium dihydride (YH\textsubscript{2-$\delta$}). By exploiting photon energy-dependent core level analysis, pure hydride states can be disentangled from oxide and hydroxide species, and a non-destructive depth profile of the chemical states is obtained. Analysis of the valence band spectra and comparison to theoretical models from density functional theory (DFT) allows the identification of metal-hydrogen states, as well as the specific contribution of metal \textit{d} versus \textit{sp} states near the Fermi energy ($E_F$). Moreover, this work discusses the empirical model proposed by Griessen and Driessen,~\cite{Griessen_1984} which correlated the enthalpy of formation ($\Delta{H_f}$) of metal dihydrides with a characteristic energy of the electronic structure of the host metal. Here, their model is extended and it is shown that the $\Delta{H_f}$ of metal dihydrides can be directly extracted from their HAXPES valence band spectra.

\section{Results and Discussion}
\subsection{Analysis of Chemical States and Probing Depth}

\begin{figure*}[ht!]
\centering
    \includegraphics[keepaspectratio, width=1\linewidth]{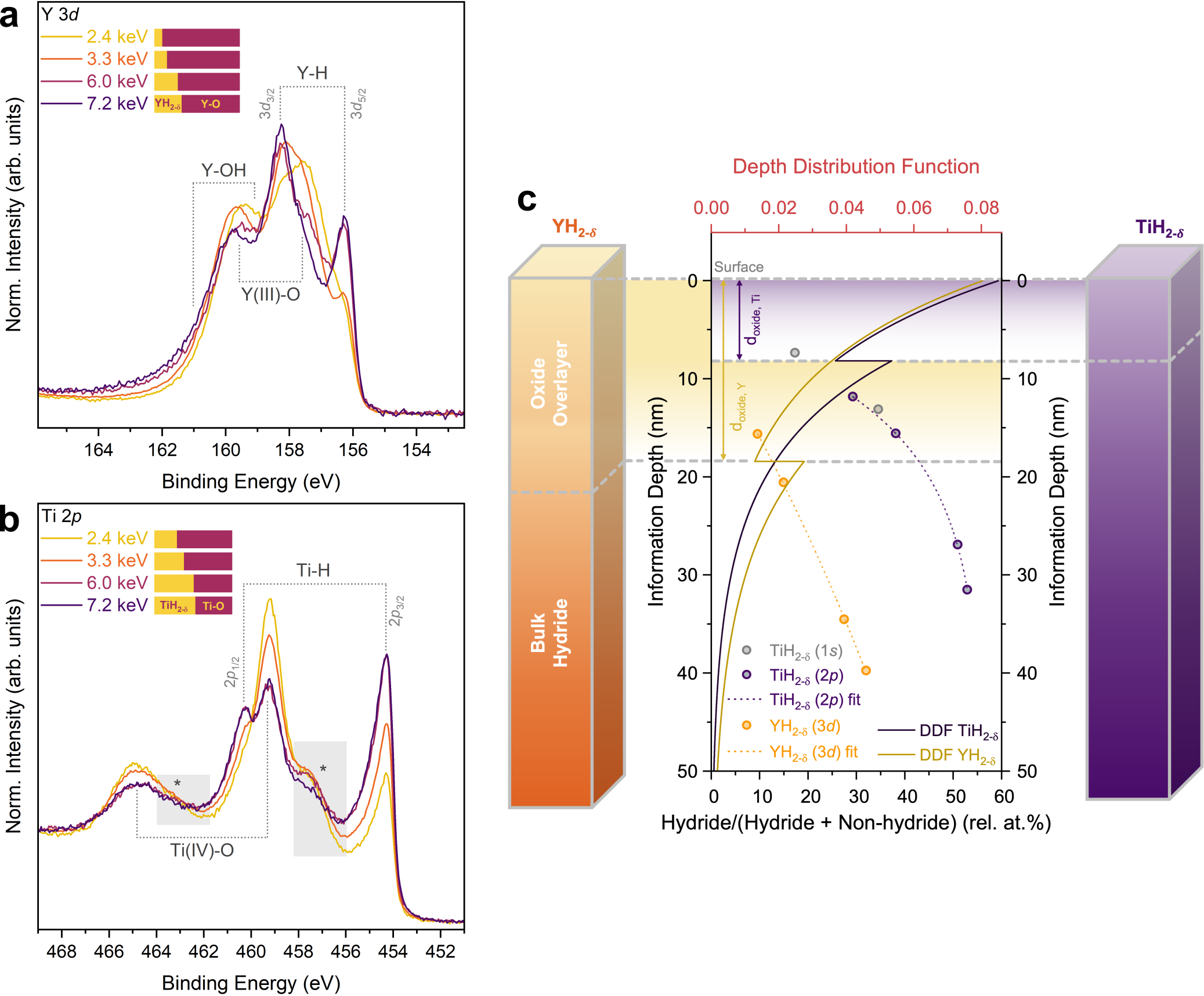}
    \caption{\textbf{Chemical states and probing depth.} \textbf{a} and \textbf{b} Annotated core level spectra of YH\textsubscript{2-$\delta$} (Y~3\textit{d}) and TiH\textsubscript{2-$\delta$} (Ti~2\textit{p}), respectively, as a function of photon energy. Spectra are normalised to their respective areas (after removing a Shirley-type background) and plotted on a calibrated BE scale. An indication of the percentage split between hydride (denoted as TiH\textsubscript{2-$\delta$} or YH\textsubscript{2-$\delta$}) and non-hydride (denoted as Ti-O or Y-O for simplicity) contributions as determined from peak fit analysis is included adjacent to the legends. Shaded regions marked with an asterisk in \textbf{b} correspond to areas where additional lower valence state metal oxide environments are present. \textbf{c} Probing depth and oxide layer thickness estimation. The evolution of the hydride intensity as a function of information depth of the respective core level kinetic energies (integrating over 95\% of the depth distribution function (DDF)) for TiH\textsubscript{2-$\delta$} and YH\textsubscript{2-$\delta$}. The solid lines show the DDF curves, and the data points represent the percentage contribution of the hydride components to the total spectral area determined from the Y~3\textit{d}, Ti~2\textit{p}, and Ti~1\textit{s} core level spectra. These points are fitted with the function $y=y_0+Ae^{-x/t}$ (dashed line). The horizontal shaded regions separated by dashed grey lines represent the estimated maximum oxide overlayer thickness, d\textsubscript{oxide}. This value was determined using the DDF derived from the core level spectra collected at $h\nu$~=~7.2~keV. The oxide thicknesses were estimated to be 8.2$\pm$2 and 18.4$\pm$2~nm for TiH\textsubscript{2-$\delta$} and YH\textsubscript{2-$\delta$}, respectively.}
    \label{fig:DDF}
\end{figure*}

The Y~3\textit{d} and Ti~2\textit{p} core levels collected at photon energies of 2.4, 3.3, 6.0, and 7.2~keV provide a non-destructive depth profile of the chemical states present in the metal hydride films. Section IV details information on the experimental methodology and HAXPES setup. The collected survey spectra, as well as additional core levels (O~1\textit{s}, Y~3\textit{p} and C~1\textit{s}) and the deep Ti~1\textit{s} core level, can be found in Supplemental Material I, II and III at Ref.~\cite{SI}, respectively. The survey spectra of both samples show all expected elements, with dominant signals coming from Ti/Y. A significant oxygen signal is also detected, owing to the unavoidable surface oxidation in realistic, technologically relevant samples. Both YH\textsubscript{2-$\delta$} and TiH\textsubscript{2-$\delta$} films display a minor signal from carbon, which decreases in intensity with increasing photon energy, suggesting it is constrained to the surface of the sample. Fluorine is also detected for YH\textsubscript{2-$\delta$} only, and this stems from the synthesis route. \par

Figs.~\ref{fig:DDF}(a) and (b) display the Y~3\textit{d} and Ti~2\textit{p} core level spectra collected as a function of photon energy for the YH\textsubscript{2-$\delta$} and  TiH\textsubscript{2-$\delta$} films, respectively. The spectra are normalised to their respective spectral areas. In both cases, a clear evolution of the relative peak intensities is observed, with the lowest binding energy (BE) peak gaining more intensity as the photon energy increases, i.e.\ for larger probing depths.~\cite{Tanuma_2011} This peak arises from the pure metal hydride states M-H, where M = Ti or Y. In good agreement with values previously reported in the literature for \ce{TiH2} and \ce{YH2},~\cite{LAMARTINE1980537, Fujimori_1984, Hayoz_2000, MA20092250, Ren_2014} the metal hydride features appear at BEs of 156.4~eV for YH\textsubscript{2-$\delta$} (Y~3\textit{d}\textsubscript{5/2}) and 454.3~eV for TiH\textsubscript{2-$\delta$} (Ti~2\textit{p}\textsubscript{3/2}), both determined from peak fit analysis of the spectra collected at 7.2~keV. These BE positions are higher than the expected BE positions of the pure metal chemical states, which for Y and Ti metal are reported on average at 155.7~\cite{Fujimori_1984, Hayoz_2000} and 453.9~eV,~\cite{BIWER1986207, SLEIGH199641} respectively. In a comparative study by Hayoz~\textit{et al.}, a +0.4~eV BE shift is observed for yttrium dihydride relative to hydrogen-free yttrium metal due to the electronic charge transfer from yttrium to hydrogen.~\cite{Hayoz_2000} Additionally, they report that a shift of +1.5~eV from the pure metal represents a yttrium trihydride chemical state. Whereas, Riesterer reports a +0.5~eV shift for TiH\textsubscript{1.9} relative to the pure Ti metal.~\cite{Riesterer1987} Given the existing literature and the fact that the present samples were fabricated under a constant Ar/H\textsubscript{2} gas ratio, it is clear that only metal dihydride chemical states are present, with both the pure metal and higher hydride species being absent from the spectra. A compilation of selected Ti~2\textit{p}\textsubscript{3/2} and Y~3\textit{d}\textsubscript{5/2} BE values reported in the literature for both Ti and Y in their metal, hydride, oxide, and hydroxide states can be found in Supplemental Material IV at Ref.~\cite{SI}. Additionally, all spectra consistently show an asymmetric line shape for the dihydride states, owing to the core hole to conduction electron coupling occurring due to the metallic character and large population of states near the $E_F$ of the hydrides. The asymmetric character was taken into account during the peak fit analysis of the spectra as detailed in Supplemental Material V at Ref.~\cite{SI}.\par

Given that no in-situ surface preparation was performed on the films, metal-oxide chemical states at higher BEs are expected as oxygen strongly interacts with hydrides, replacing hydrogen on the surface and forming an oxide overlayer. The hydrides studied here were deposited in the same manner as the samples detailed in Ref.~\cite{Cornelius_2019} but without a palladium capping layer. In this reference, Rutherford backscattering spectroscopy (RBS) measurements were performed and showed no traces of oxygen (i.e.\ below the detection limit of the technique) within the bulk or surface of the metal dihydride films. Therefore, the oxide found on the surface of these samples results from post-deposition oxidation rather than due to any intrinsic oxygen incorporated during deposition. Due to the high reactivity of the hydrides, it can be assumed that the presence of residual oxygen present in either the load lock of the deposition chamber or the argon-filled transport setup is sufficient to lead to surface oxidation. To confirm this hypothesis, air exposure tests were performed.\par 

After the initial measurements, which will be referred to in the remaining discussion as ``argon-transferred'', the samples were exposed to air for a short initial period (2~h) followed by long-term exposure (1~month). After both instances of air exposure, the samples were re-measured using the same photon energy (7.2~keV) and optics setup. A comparison of both the Ti~2\textit{p} and Y~3\textit{d} spectra collected without exposure to air and after both air exposure timescales can be viewed in Supplemental Material VI at Ref.~\cite{SI}. The spectra appear almost identical across all measurements despite the significant air exposure. Assuming the samples are homogeneous across the entire surface, the hydride to oxide peak intensities remain consistent over time with only a slight decrease in the M-H signal intensity observed after extensive air exposure (less than 5~rel. at.\% decrease in the hydride/(hydride + non-hydride) signal intensity between the argon-transferred and 1~month air exposed spectra). The lack of significant change indicates that a stable passive oxide layer was already present in the samples after transporting them under argon before the initial HAXPES measurements, covering the sample surface and protecting the metal hydride bulk. \par

The Y~3\textit{d} spectra displayed in Fig.~\ref{fig:DDF}(a) show two chemical environments in addition to the metal hydride state. Firstly, an intense yttrium oxide feature (i.e.\ Y\textsubscript{2}O\textsubscript{3}-like), labelled as Y(III)-O, with the Y~3\textit{d}\textsubscript{5/2} peak at 157.5~eV (determined from the peak fit analysis of the spectrum collected at 2.4~keV), agreeing with past reported values of the same chemical environment.~\cite{REICHL1986196, Barreca_2001} Secondly, a minor metal hydroxide environment at even higher BEs, labelled as Y-OH. Such high BE Y states are often attributed to metal hydroxide, other hydroxylated species, and metal carbonates.~\cite{GOUGOUSI20086197, Mitrovic_Veal_2014} As the C~1\textit{s} core level spectra show no detectable evidence of carbonate states and the O~1\textit{s} spectra show a clear hydroxide feature on the higher BE side of the main metal-oxide signal (see Supplemental Material II at Ref.~\cite{SI} for the C~1\textit{s} and O~1\textit{s} spectra), it is clear that in the present case, the features in the Y core level arise from hydroxide species.\par

From peak fit analysis of the Y~3\textit{d} spectra, the metal hydride contribution to the total spectral area was determined, with bar charts representing this ratio displayed adjacent to the legend in Fig.~\ref{fig:DDF}(a). The ratio was determined by comparing the raw spectral areas of the Y~3\textit{d}\textsubscript{5/2} peaks of each environment (i.e.\ without any escape depth correction). Details regarding the methods used to peak fit the Y~3\textit{d} core level can be found in Supplemental Material V at Ref.~\cite{SI}. Due to the chemical shift between the hydride and all other chemical states, as well as the depth sensitivity of HAXPES, it is possible to disentangle the contributions from the bulk hydride and oxide overlayer, with the hydride contribution increasing while the oxide contribution reduces with increasing photon energy. Although a greater probing depth is provided with increasing photon energy, oxygen-containing environments always dominate, with the hydride contribution increasing from 9.5 to 32.0 rel. at.\% when going from 2.4 to 7.2~keV photon energy. This suggests that a thick oxide overlayer is present, and this observation will be discussed further toward the end of this section.\par

The Ti~2\textit{p} spectra displayed in Fig.~\ref{fig:DDF}(b) are similarly complex, owing to the mixture of both hydride and oxide environments. The main peak close to the centre of the spectra is attributed to titanium oxide in the +4 oxidation state (i.e.\ TiO\textsubscript{2}-like),~\cite{SLEIGH199641, POUILLEAU1997235, DIEBOLD200353} labelled as Ti(IV)-O with the doublet assigned to this environment appearing at BEs of 459.3 (2\textit{p}\textsubscript{3/2}) and 465.0~eV (2\textit{p}\textsubscript{1/2}), determined from the peak fit analysis of the spectrum collected at 7.2~keV. Peak fit analysis reveals the presence of additional lower valence state metal oxide environments, which are labelled in Fig.~\ref{fig:DDF}(b) with an asterisk, appearing on the lower BE side of the main Ti(IV)-O core lines.~\cite{POUILLEAU1997235, HUANG20092825} As used for the Y~3\textit{d} spectra, bar charts representing the hydride to non-hydride composition from peak fits of the Ti~2\textit{p}\textsubscript{3/2} lines are placed adjacent to the legend. Comparing the two systems, a significantly greater metal hydride contribution to the total signal was detected for the TiH\textsubscript{2-$\delta$} sample. The Ti dihydride contribution at 2.4~keV (29.2~rel. at.\%) almost matches the value obtained for Y dihydride at 7.2~keV. This rises to 52.9~rel. at.\% when measured at a photon energy of 7.2~keV. Details regarding the methods used to peak fit the Ti~2\textit{p} core level can be found in Supplemental Material V at Ref.~\cite{SI}. The high BE Ti~1\textit{s} core level collected at 6.0 and 7.2~keV corroborates these findings (see Supplemental Material III and V at Ref.~\cite{SI} for the Ti~1\textit{s} spectra, and the peak fitting and analysis, respectively).\par

From the qualitative analysis of the core level spectra collected as a function of photon energy, the layer sequence of the hydride films and their overlayers becomes clear, as indicated in the bar charts in Figs.~\ref{fig:DDF}(a) and (b). Taking the depth information provided by HAXPES a step further, Fig.~\ref{fig:DDF}(c) offers a quantitative estimate of the oxide thickness as extracted from the Ti~2\textit{p} and Y~3\textit{d} core level spectra using a depth distribution function (DDF). The DDF estimates the probability of a photoelectron leaving the surface of a material having originated from a given depth measured normally from the surface. To calculate the DDF, the approach taken by Berens~\textit{et al.}\ was followed.~\cite{Berens_2020} This approach is based on a bilayer, whereby the metal hydride is the bottom layer which is covered by a homogeneous oxide overlayer. The required input parameters to the DDF are (a) the inelastic mean free path (IMFP) of the photoelectrons originating from their respective layers and travelling through their respective materials, (b) the number density of both layers, and (c) an estimate of the percentage of the signal originating from each layer. Two assumptions had to be made for calculating the DDF, namely (i) that the hydrides are stoichiometric TiH\textsubscript{2} and YH\textsubscript{2}, and (ii) that the metal oxide overlayer is composed of the highest valence oxidation state only (i.e.\ TiO\textsubscript{2} and Y\textsubscript{2}O\textsubscript{3}), and all non-hydride contributions to the spectral area are grouped together to represent the metal oxide. The relativistic IMFP was calculated using the TPP-2M predictive formula embedded in the QUASES-IMFP-TPP2M Ver.3.0 software package.~\cite{Shinotsuka_2015} The software already had TiO\textsubscript{2} and Y\textsubscript{2}O\textsubscript{3} within its database but not the metal dihydrides, and so the hydrides were entered into the software as new materials. A full description of the methodology, input values, approximations, and assumptions used for calculating the DDF can be found in Supplemental Material VII at Ref.~\cite{SI}. The DDF was calculated at each photon energy, with the DDF for 7.2~keV displayed in Fig.~\ref{fig:DDF}(c), and used to estimate the oxide overlayer thickness. 7.2~keV was selected on the basis that the hydride and non-hydride environments within the core level spectra measured at this photon energy were better resolved than the spectra collected at lower photon energies, thereby reducing the error associated with the calculation. Fig.~\ref{fig:DDF}(c) shows an increase in metal hydride contribution as the probing depth increases. The discontinuity in the DDF is due to the hydride and oxide having different bulk densities and atomic masses (i.e.\ number densities). At each photon energy, the information depth of the Y~3\textit{d} and Ti 2\textit{p} electrons was determined by integrating the area under the bilayer DDF curve and finding the depth that equated to 95\% of the total area. This value is reported on the $y$-axis of the graph. The $x$-axis is the hydride/(hydride + non-hydride) composition.\par

From the 7.2~keV derived DDF, the maximum oxide layer thickness can be estimated as 8.2$\pm$2 and 18.4$\pm$2~nm for TiH\textsubscript{2-$\delta$} and YH\textsubscript{2-$\delta$}, respectively, as schematically shown from the three-dimensional blocks on the left and right sides that describe the sample structure. This thickness was determined by finding the information depth of the discontinuity of the DDFs, which signifies the interface of the oxide/hydride bilayer. The difference in oxide thickness is related to differences in the oxygen solubility and oxygen diffusivity of the two bulk hydrides, which are directly influenced by the metal.~\cite{bosseboeuf2019effect, Bessouet_2021} Previous reports show that Ti has a low oxygen diffusivity and a high oxygen solubility, whereas the opposite is true for Y.~\cite{Bessouet_2021} This translates to a greater diffusion length of oxygen in Y compared to Ti, leading to a significantly thicker oxide layer to form on YH\textsubscript{2-$\delta$}. This will considerably influence the behaviour of these two metal hydrides in applications, and native oxide overlayer thicknesses should be considered when comparing the performance of different metal hydride samples.

\begin{figure*}[ht]
\centering
    \includegraphics[keepaspectratio, width=0.67\linewidth]{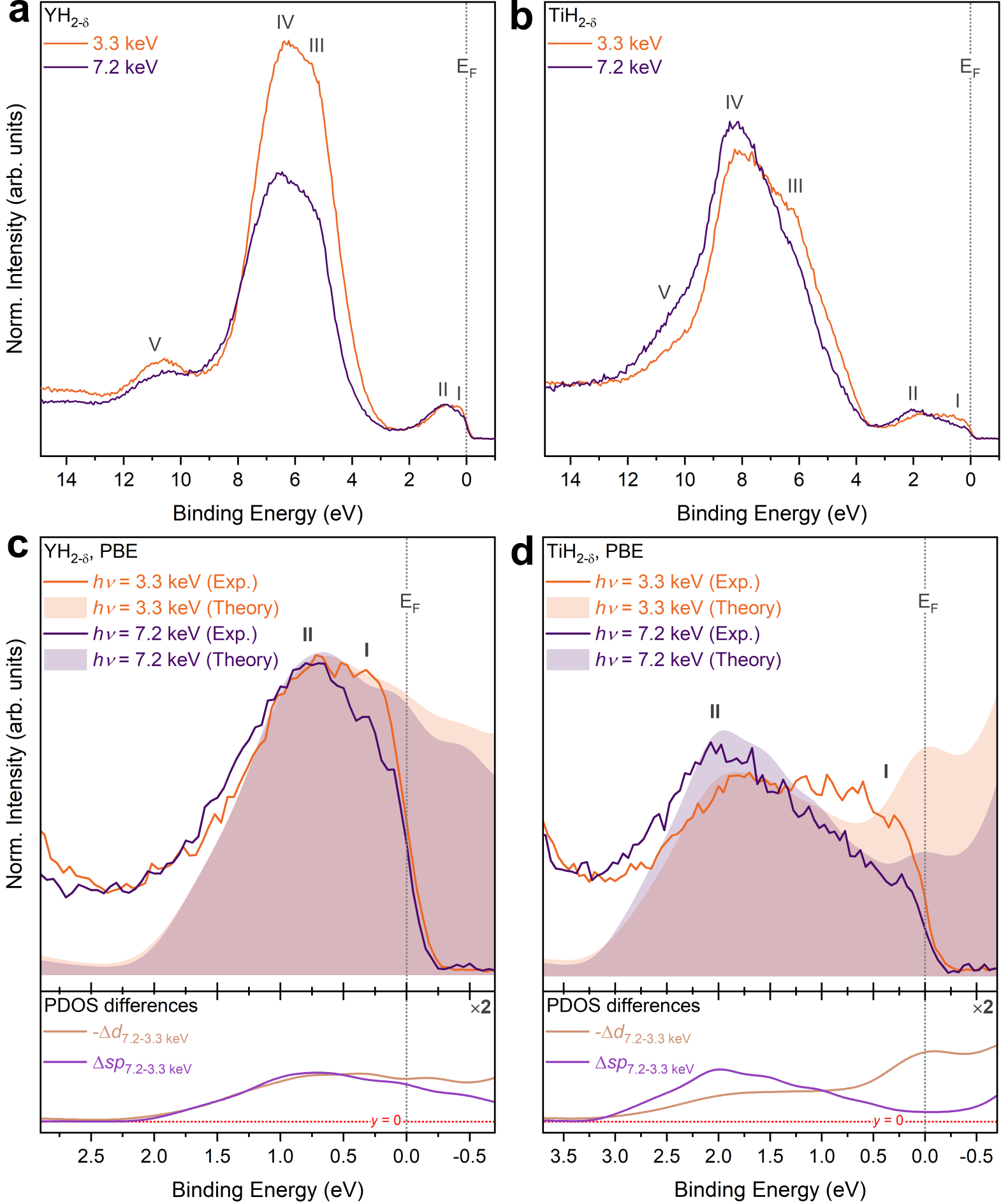}
    \caption{\textbf{Electronic structure of YH\textsubscript{2-$\delta$} and TiH\textsubscript{2-$\delta$}}. Valence band spectra collected at 3.3 and 7.2~keV are presented in \textbf{a} for YH\textsubscript{2-$\delta$} and \textbf{b} for TiH\textsubscript{2-$\delta$}. These spectra are normalised to the total area of features I and II adjacent to the $E_F$. Panels \textbf{c} and \textbf{d} displays a magnified view of the density of states adjacent to the $E_F$, along with a direct comparison with the total photoionisation cross section weighted PDOS for TiH\textsubscript{2} and YH\textsubscript{2}, calculated with PBE (i.e.\ sum of all PDOS). Underneath the main panels of \textbf{c} and \textbf{d} are the PDOS differences between the \textit{d} and \textit{sp} states when cross-section weighted at photon energies of 3.3 and 7.2~keV. Both \textbf{a} and \textbf{b}, and \textbf{c} and \textbf{d}, are plotted on the same $y$-axis scale but note the different $x$-axis scales. The PDOS differences at the bottom of panels \textbf{c} and \textbf{d} are plotted on a $\times$2 magnified $y$-axis scale compared to the main panel $y$-axis scale.}
    \label{fig:VB}
\end{figure*}

\subsection{Electronic Structure and Bond Nature}

Analysis of the core states and observation of the depth distribution of hydride versus non-hydride chemical states promises the possibility of also observing metal hydride states in the valence band (VB) and in particular close to the Fermi energy ($E_F$). It is important to emphasise that the analysis of VB spectra in solids is complex. While core level BEs are well separated for different chemical species and oxidation states, valence states are superimposed and weighted with both depth sensitivity and photoionisation cross-sections.~\cite{Offi2021} Figs.~\ref{fig:VB}(a) and (b) display the VB spectra of the YH\textsubscript{2-$\delta$} and TiH\textsubscript{2-$\delta$} films, respectively, collected at photon energies of 3.3 and 7.2~keV (VB spectra and shallow core level spectra collected at all four-photon energies are included in Supplemental Material VIII and IX at Ref.~\cite{SI}, respectively). The spectra are normalised to the spectral area of features I and II. They show several clearly identifiable features, including (I) 0.4, (II) 0.8, (III) 5.5, (IV) 6.8 and (V) 10.5~eV for YH\textsubscript{2-$\delta$} and (I) 0.8, (II) 2.0, (III) 6.4, (IV) 8.3 and (V) 10.2~eV for TiH\textsubscript{2-$\delta$}. By combining the photon energy-dependent experimental spectra with cross-section weighted projected density of states (PDOS) calculations (see Supplemental Information X and XI at Ref.~\cite{SI} for the 3.3 and 7.2~keV weighted PDOS spectra, respectively), it is possible to disentangle the individual orbital contributions to these features.\par

Starting with YH\textsubscript{2-$\delta$} shown in Fig.~\ref{fig:VB}(a), the spectra are in good agreement with past experimental VB spectra collected by Fujimori and Schlapbach for YH\textsubscript{2.1} ($h\nu$ = 1253.6~eV (Mg~K$\alpha$),~\cite{Fujimori_1984} and by Weaver~\textit{et al.} for YH\textsubscript{2} ($h\nu$~=~18-35~eV).~\cite{Weaver1979} Notably, we observe density of states (DOS) adjacent to $E_F$ (in the BE region up to 2.0~eV) similar to the aforementioned studies. Fujimori and Schlapbach identified two main features at 5.5~eV and within 2.0~eV of the $E_F$ attributing them to a hydrogen-induced band and the Y~4\textit{d}\textsuperscript{1} conduction band, respectively.~\cite{Fujimori_1984} In the present case, the core level spectra show that metal-oxide states will also contribute to the VB spectra, however, features I and II can unequivocally be assigned to yttrium hydride states, as (i) yttrium oxide has a band gap of approximately 5.6~eV and therefore does not have any states in this BE region,~\cite{Zhang_1998, Tanuma_2003, Wang_2009, MUDAVAKKAT2012893, Mongstad2014TheOxide} and (ii) the core level spectra exclude the presence of yttrium metal. Weaver~\textit{et al.}\ highlight that if such features are present in the valence band spectrum, then this is further evidence of a metal dihydride specifically rather than a mono or trihydride system.~\cite{WEAVER1980207, Weaver1981}\par

The higher BE features III and IV in the YH\textsubscript{2-$\delta$} spectra have contributions from both the metal hydride,~\cite{Weaver1979, Fujimori_1984} and metal oxide states.~\cite{Osterwalder1985OnHydrides} Given that features I and II are metal hydride related and the VB spectra are normalised to these features, the clear decrease of both features III and IV  with higher photon energies agrees with the expected decrease of the surface metal oxide contribution upon increasing information depth. Feature V is not described by the PDOS for either the metal hydride or the corresponding oxide but has previously been attributed to surface states,~\cite{Barrett1987} yttrium-hydrogen,~\cite{Baptist1988}or yttrium-carbon~\cite{Budke_2008} contributions. The cross-section weighted PDOS of the metal oxides can also be found in Supplemental Material X and XI at Ref.~\cite{SI}.\par

In parallel to the analysis of YH\textsubscript{2-$\delta$}, TiH\textsubscript{2-$\delta$} has a similar VB structure as shown in Fig.~\ref{fig:VB}(b). Features I and II can again be attributed solely to titanium hydride-derived bands, owing to the $\approx$3.0~eV band gap of TiO\textsubscript{2}.~\cite{Scanlon2013} The presence of such features as highlighted by Weaver~\textit{et al.}\ also confirms that a titanium dihydride system is present. It is noted that the BE region width encompassing features I and II is approximately 1~eV larger for TiH\textsubscript{2-$\delta$} compared to YH\textsubscript{2-$\delta$}. This widening is thought to stem from the difference in the valence state of the two hydrides, with TiH\textsubscript{2-$\delta$} being tetravalent and YH\textsubscript{2-$\delta$} being trivalent.~\cite{WEAVER1980207} Weaver~\textit{et al.} observed the same trend when comparing the width of states that fall near $E_F$ for YH\textsubscript{1.98} and ThH\textsubscript{1.91}, with the latter being tetravalent similar to TiH\textsubscript{2-$\delta$} in our case.~\cite{WEAVER1980207} Features III, IV and V in the TiH\textsubscript{2-$\delta$} spectra are analogous to the same numbered features in the YH\textsubscript{2-$\delta$} spectra. \par

The unweighted theoretical spectra calculated with DFT are displayed in Supplemental Material XII at Ref.~\cite{SI}. The PDOS for TiH\textsubscript{2} has the centre of the H~1\textit{s} state at an energy of 5.6~eV below $E_F$, whereas it is at 4.3~eV for YH\textsubscript{2}. This is commensurate with the energy band diagram and density of states of TiH\textsubscript{2} reported by Gupta,~\cite{Gupta_1979, Weaver1981} and Smithson~\textit{et al.},~\cite{Smithson_2002} respectively, where a hydrogen-induced band is centred around 6~eV from $E_F$, and with the density of states calculated for YH\textsubscript{2} by Peterman~\textit{et al.}\ wherein they report the equivalent band at $\approx$4~eV below $E_F$.~\cite{Peterman_1979} This further confirms that a hydrogen-induced state will contribute to feature III in the experimental VB spectra for both dihydrides. Furthermore, the PDOS show that compared to YH\textsubscript{2-$\delta$}, the VB states in TiH\textsubscript{2-$\delta$} span a wider BE range. This leads to the bottom of the VB overlapping with feature V in the experimental valence band spectra, which is assumed to arise from surface states and adsorbed species. Other notable theoretical works on these two metal dihydride systems can be found in Refs.~\cite{Wolf2000First-principlesBonding, Mehta2021AHydride, billeter2021surface, LUIGGIA2021102639} and are in good agreement with the results obtained here.\par

Returning to a more detailed discussion of features I and II, Figs.~\ref{fig:VB}(c) and (d) display magnified views of the experimental and theoretical VB spectra along with the corresponding cross-section weighted PDOS calculations in the bottom panel for YH\textsubscript{2-$\delta$} and TiH\textsubscript{2-$\delta$}, respectively. One noteworthy aspect of the comparison between theory and experiment throughout this work is that both are aligned to their respective intrinsic $E_F$ values. No additional alignment was needed, as is often the case due to challenges in obtaining absolute energy scales from theory or fully trustworthy energy alignments in experiments. The exceptional quality of the agreement between experiment and theory for the hydrides is further compounded by changes in experimental intensity near the $E_F$ with photon energy being clearly reflected in the photoionisation cross section corrected PDOS.\par

The drop in intensity at higher photon energy is not constant across features I and II, but a stronger drop at the top of the VB (feature I) is noticeable. This is due to differences in the photoionisation cross sections, where \textit{d} states decrease at a higher rate relative to both \textit{s} and \textit{p} states at higher photon energies. From the complete cross-section weighted PDOS at 3.3 and 7.2~keV (see Supplemental Material X and XI at Ref.~\cite{SI}, respectively), Y~5\textit{s} (Ti~4\textit{s}) states contribute to feature II. In contrast, feature I has contributions from both Y~4\textit{d} (Ti~3\textit{d}) and Y~5\textit{p} (Ti~4\textit{p}) states. The net decrease of feature I at higher photon energy with respect to feature II suggests a larger \textit{d} state contribution compared to \textit{p} states, pointing to a $sd$ hybridisation scheme, with only marginal participation of the $p$ orbitals. This reflects a more complex bonding situation in these hydrides than previously thought, in that not only does the metal $d$-derived band contribute to the energy region within 3.0~eV from the $E_F$, as suggested by Weaver~\textit{et al.} for both group III B~\cite{Weaver1979} and group IV B~\cite{Weaver1981} transition metal dihydrides, but that extended $s$ and $p$ states are also important.~\cite{Fujimori_1984}\par

\begin{figure*}
\centering
    \includegraphics[keepaspectratio, width=\linewidth]{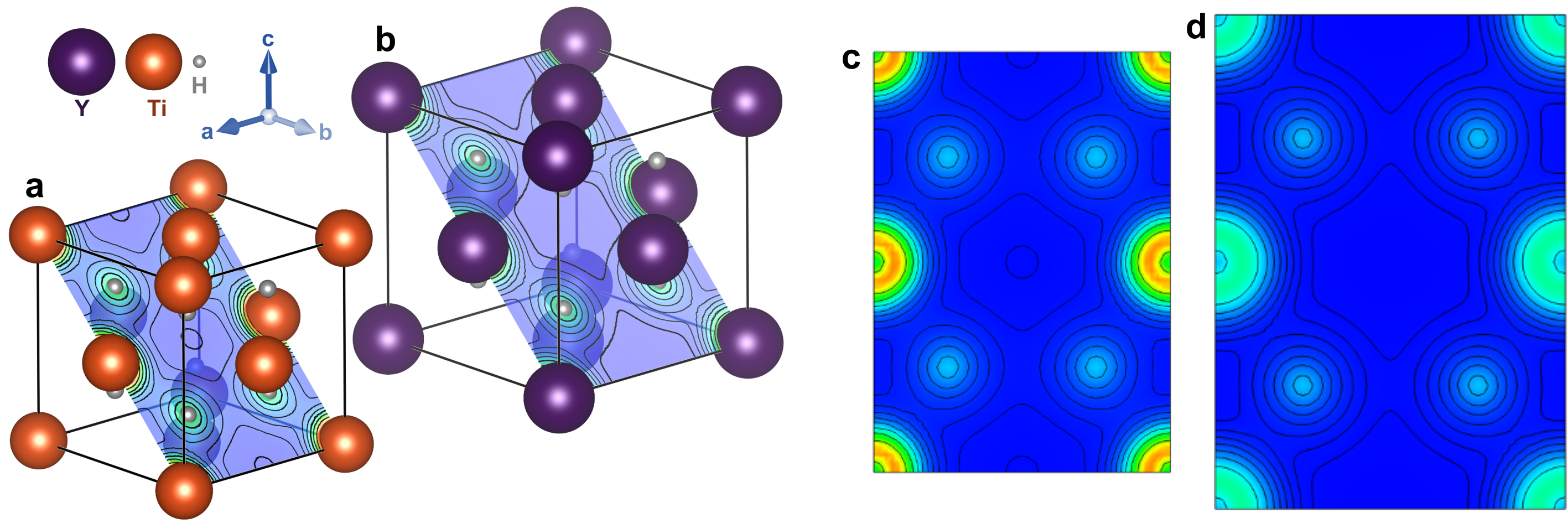}
    \caption{\textbf{Bond nature of metal hydrides} Using the relaxed TiH\textsubscript{2} and YH\textsubscript{2} structures (calculated with PBE), electron density maps were obtained, with \textbf{a} and \textbf{b} showing the three-dimensional (3D) density maps for TiH\textsubscript{2} and YH\textsubscript{2}, respectively. Two-dimensional (2D) electron density maps for TiH\textsubscript{2} and YH\textsubscript{2} are displayed in \textbf{c} and \textbf{d}, respectively. The 2D electron density maps are plotted along the (011) body diagonal lattice plane (1~\textit{d} from the origin) and on the same intensity scale to allow for a direct comparison. Note that the 3D maps are for reference only to show where the 2D maps have been extracted from (the colour scale does not match the 2D maps). Contour lines have been added to the 2D electron density maps. These contour lines have been plotted using the function $F(N) = A \times B^{N/step}$, where A = 1, B = 10, N\textsubscript{min} = -1, N\textsubscript{max} = 3, and step = 5. The maps are temperature scaled, with the blue regions indicating a low electron density and the red regions indicating a high electron density.}
    \label{fig:E_density}
\end{figure*}

The excellent agreement between experiment and theory in the electronic structure analysis allows further use of DFT to examine the bonding nature of the metal hydrides. The electron densities extracted from the DFT calculations of the two hydrides, depicted in Fig.~\ref{fig:E_density}, show a clearly localised density around the atoms, indicating an ionic/metallic bonding nature. This can be examined in more detail by considering the atomic charges and overlap (bond) populations (see Supplemental Material XIII at Ref.~\cite{SI} for these values). A similar approach was taken by Yang~\textit{et al.}\ to assess the effects of alloying on the chemical bonding of TiH\textsubscript{2}.~\cite{YANG2002109} The absolute values for both quantities are not themselves meaningful, particularly in the case of metallic systems, as illustrated by variations between absolute atomic charges calculated using different population analysis approaches (Mulliken, Hirshfeld and Bader). It is nonetheless possible to extract useful information about the trends.~\cite{Segall1996a,Segall1996b}

The overlap bond population is lower for the M-H bond than for the M-O bond in all cases (Ti, Y and PBE/PBE0 where applicable), with some variation between the different M-O populations. This signifies that both hydrides are more ionic than their corresponding oxides. This is further evidenced by the atomic charges, where for all three of the considered population analysis approaches, the effective valence is smaller for the hydrides than the corresponding oxide. This implies a more ionic bonding nature for the hydrides compared to the more covalent nature of the oxides.\par

Whilst the overall bonding nature of both hydrides is clear, a direct comparison between the two is complicated. Using the 2D density maps displayed in Figs.~\ref{fig:E_density}(c) and (d), a visual inspection of the densities suggests a more diffuse density on the metal atoms in YH\textsubscript{2} compared to TiH\textsubscript{2}, however, the differences are subtle. The overlap populations support this picture, being slightly smaller in TiH\textsubscript{2}, suggesting that TiH\textsubscript{2} is more ionic than YH\textsubscript{2}. However, the atomic charges indicate the opposite, with YH\textsubscript{2} having a smaller effective valence than TiH\textsubscript{2}. This difference is small for the Hirshfeld and Mulliken approaches but more significant for Bader. In other words, while the results clearly indicate that the hydrides exhibit an ionic bonding nature compared to the more covalent nature shown in the oxides, they are inconclusive about which of the two hydrides is more ionic. This is perhaps not surprising, given the small differences in electronegativity of the two metals (1.5 for Ti and 1.2 for Y). Using this as an indication to determine the ionic versus covalent bonding nature of the two systems,~\cite{ALLRED1961215, HUSAIN19891233} bonds in both hydrides can be classified as metallic or partial ionic and partial metallic bonds.~\cite{YOUNG2018} In contrast, the larger overlap population in YH\textsubscript{2} compared to TiH\textsubscript{2} suggests a stronger bond in the former, as indeed reflected by the obtained values of enthalpy of formation of the two compounds, larger in YH\textsubscript{2} than in TiH\textsubscript{2}.~\cite{Buschow_1982, CHERNIKOV1987441}\par

\subsection{Enthalpy of Formation}

The enthalpy of formation, $\Delta H_f$ is a highly relevant value for determining the hydrogen storage capability in a metal hydride.~\cite{GRAY_2021} The absorption/desorption plateau pressure $p_{H_2}$ for two-phase metal-hydrogen systems can be related to $\Delta H_f$ via the van~'t Hoff equation:~\cite{HUSTON1980435}

\begin{equation}
    \ln{p_{H_2}} = \frac{{\Delta}{H_f}}{RT} - \frac{{\Delta}S}{R},
\end{equation}
where $R$ is the universal gas constant (8.314~J~mol\textsuperscript{-1}~K\textsuperscript{-1}), $T$ is temperature, and $\Delta {S}$ is the change in entropy for H\textsubscript{2} absorption. Therefore, a larger negative enthalpy of formation indicates that a hydride phase can be formed at lower hydrogen partial pressures as desired. Griessen and Driessen proposed an empirical linear relationship between the enthalpy of formation of metal hydrides and characteristic energy, $\Delta E$ in the electronic band structure of the host metal:~\cite{Griessen_1984} 

\begin{equation}\label{Driessden}
    \Delta H_f = \frac{n_s}{2}(\alpha \Delta E + \beta),
\end{equation}
%
where $\Delta E=E_s-E_F$, $E_F$~=~0~eV, $E_s$ is the centre of the lowest conduction band of the host metal and equivalent to the energy at which the integrated density of states ($\int$DOS) of the host metal is equal to 0.5$n_s$, $n_s$ is the number of electrons per atom in the lowest \textit{s}-like conduction band of the host metal, $\alpha=29.62$ kJ/eV~mol~H, and $\beta=-135$~kJ/mol~H.~\cite{Griessen_1984} For alkali metals, $n_s$ is equal to one, but for all other metals (including Ti and Y), $n_s$ is equal to two. Therefore, for Y and Ti, $E_s$ is the energy at which the $\int$DOS = 1. The $\alpha$ and $\beta$ terms were derived by Griessen and Driessen by correlating the $\Delta E$ values determined by integrating the total density of states of the host metal, to the measured values of $\Delta H_f$ values found in the literature (see Fig.~4 in Ref.~\cite{Griessen_1984}). The authors claimed that the use of $\Delta E$ derived in this way was effective in reproducing the measured $\Delta H_f$ values. However, it is noted that there is a large degree of scatter in Fig.~4 of the paper used to determine the $\alpha$ and $\beta$ terms, as well as multiple different $\Delta{H_f}$ entries for the same metal, raising some concerns about the robustness of the model and the accuracy of the $\alpha$ and $\beta$ terms.\par

\begin{figure*}
\centering
    \includegraphics[keepaspectratio, width=0.9\linewidth]{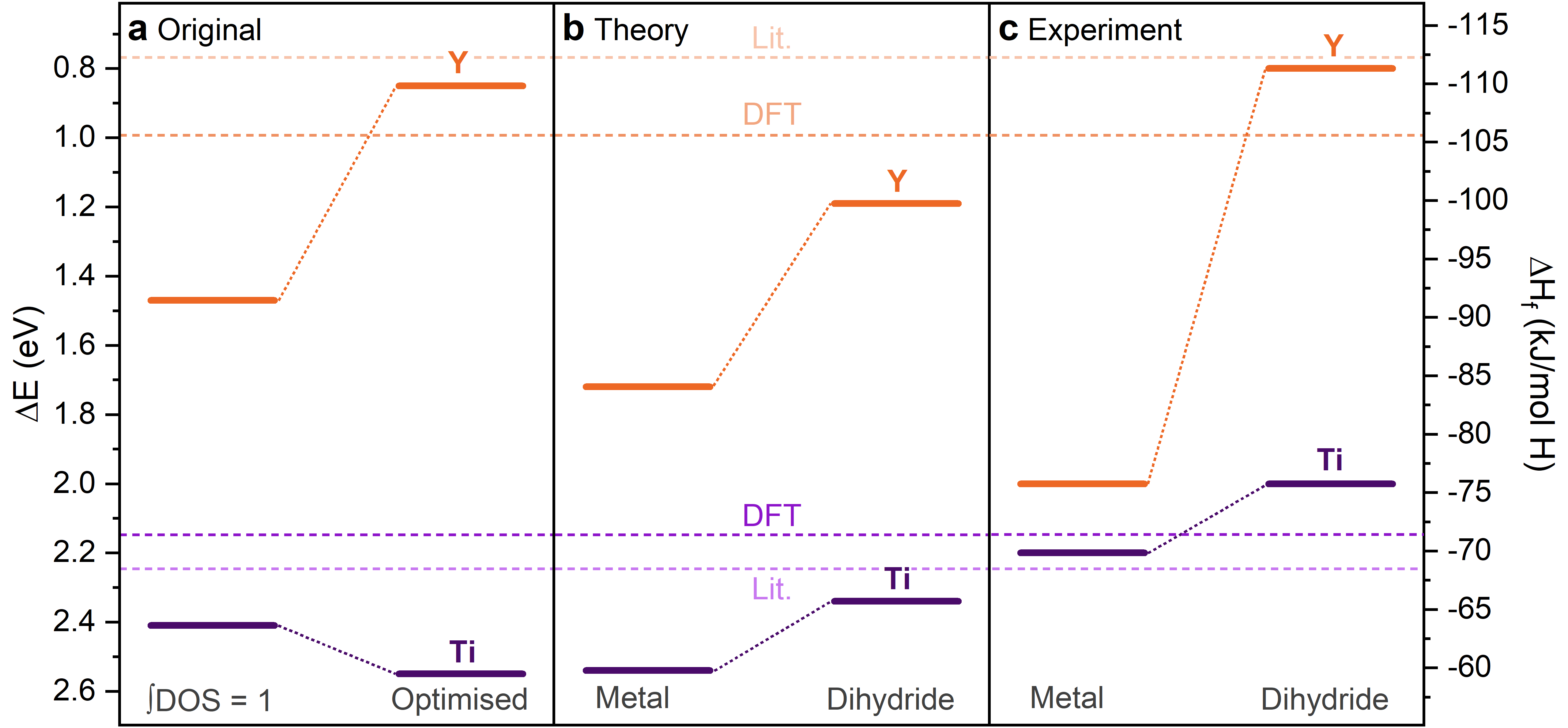}
    \caption{\textbf{Enthalpy of Formation} A comparison between the $\Delta E$ values obtained from various methods (left $y$-axis) and the resultant $\Delta H_f$ values calculated by inputting said $\Delta E$ values into Eqn.~\ref{Driessden} (right $y$-axis). \textbf{a} On the left-hand side, the $\Delta{E}$ values extracted from Fig.~4 in the study by Griessen and Driessen are displayed.~\cite{Griessen_1984} These values, termed ``Original'', were determined by the authors of Ref.~\cite{Griessen_1984} by calculating the energy where the integrated total density of states (DOS) of the host metal was equal to one. On the right-hand side of \textbf{a}, the optimised $\Delta{E}$ values for Ti and Y taken from Tab.~II from Ref.~\cite{Griessen_1984} are displayed. \textbf{b} $\Delta{E}$ values extracted from the PDOS of the metals and dihydrides calculated in this work, and termed ``Theory''. Abiding by the definitions of $\Delta{E}$ for the dihydrides, the position of the \textit{s} band closest to $E_F$ was used. For the metal, the main intensity \textit{s} state peak position was used to determine $\Delta{E}$. \textbf{c} $\Delta{E}$ values determined using the position of feature II in the valence band spectra collected with HAXPES, and termed ``Experiment''. For \textbf{b} and \textbf{c}, the left-hand side displays the metal value and the dihydride value on the right-hand side. Horizontal dashed reference lines are plotted to indicate the enthalpy of formation determined using DFT (darker line) and reported in the literature (Lit.) using experimental methods (lighter line).}
    \label{fig:deltaH_EF}
\end{figure*}

Relevant to this work, Griessen and Driessen used their model approach to derive the enthalpy of formation of TiH\textsubscript{2} and YH\textsubscript{2}. Using the calculated electronic band structures of Ti and Y metal available at the time of the publication of Ref.~\cite{Griessen_1984}, they determined $\Delta{E}$ for Ti and Y metal to be 2.41 and 1.47~eV, respectively. These values are displayed on the left-hand side of Fig.~\ref{fig:deltaH_EF}(a), and inputting them into Eqn.~\ref{Driessden} gives $\Delta{H_f}$ values of -63.6 and -91.5~kJ/mol H for TiH\textsubscript{2} and YH\textsubscript{2}, respectively (shown by viewing the right $y$-axis scale on Fig.~\ref{fig:deltaH_EF}). These $\Delta{H_f}$ values deviate significantly from experimental values reported in the literature, which are displayed in Fig.~\ref{fig:deltaH_EF} as light dashed lines ($\Delta H_{f, YH_2}$ = -112.25~kJ/mol~H (between 873-1073~K),~\cite{CHERNIKOV1987441}, $\Delta H_{f, YH_2}$ = -112.5~kJ/mol~H,~\cite{Buschow_1982} $\Delta H_{f, TiH_2}$ = -63.0~kJ/mol~H,~\cite{Buschow_1982} $\Delta H_{f, TiH_2}$ = -68.47~kJ/mol~H (at 737~K)~\cite{DANTZER1983913, JZhao_2008}). To compensate for this, the authors devised ``optimized'' $\Delta{E}$ values (see Tab. II in Ref.~\cite{Griessen_1984}) to reproduce a better match with experimental values of $\Delta{H_f}$. These optimised values are displayed on the right-hand side of Fig.~\ref{fig:deltaH_EF}(a). Using the optimised values a better match to the $\Delta{H_f}$ literature values was then obtained for YH\textsubscript{2} but not TiH\textsubscript{2}, and this is because of an error associated with the $\Delta{H_f}$ value of TiH\textsubscript{2} used by the authors, which does not match modern values.\par

The enthalpy of formation can be obtained directly from theoretical calculations, and DFT has been employed to calculate this property for metal dihydrides previously.~\cite{Miwa_2002, Smithson_2002, Wolverton_2004, Tao_2009, Ziani_2021} Here, the approach detailed in Ref.~\cite{Miwa_2002} was used to obtain values for titanium and yttrium dihydride. This approach required additional calculations of H\textsubscript{2} and bulk Ti and Y metal, with further details of the calculations and methods described in Supplemental Material XIV at Ref.~\cite{SI}. The resulting enthalpies of formation for titanium and yttrium dihydride are -71.4 and -105.6~kJ/mol H, respectively. Reasonable agreement with the literature is found, with the titanium dihydride DFT-determined enthalpy slightly overpredicted compared to the literature value and the opposite being true for the yttrium dihydride case. 

While DFT calculations are incomparably more accessible today than at the time when Griessen and Driessen published their model (1984), this direct calculation cannot link specific orbital character with the resulting values. Therefore, the theoretical PDOS and experimental HAXPES valence band spectra of the host metals and metal dihydrides were obtained to extract $\Delta{H_f}$ values independently. This provides the opportunity to expand on the model by Griessen and Driessen and to explore if the enthalpy of formation of metal dihydrides can be derived using data other than the integrated total density of states of the host metal. It is recalled and directly quoted from Ref.~\cite{Griessen_1984} that $E_s$ has ``primarily \textit{s} character with respect to the interstitial sites occupied by hydrogen atoms'', and ``$n_s$ represents the number of electrons per atom in the lowest \textit{s}-like conduction band of the host metal".~\cite{Griessen_1984} Based on these quotes, the first approach was to use the position of the main \textit{s} state intensity in our calculated PDOS of the two host metals as the value of $\Delta{E}$ (see Supplemental Material XV at Ref.~\cite{SI} for the unweighted PDOS of Ti and Y metal). Given that theory and experiment are referenced to an intrinsic $E_F$, $E_s$ and $\Delta E$ will be used interchangeably in this work. The values of $\Delta E$ from the metal PDOS spectra are displayed on the left-hand side of Fig.~\ref{fig:deltaH_EF}(b). This expands the application of Griessen and Driessen's model as they only had total DOS rather than individual PDOS available. However, comparing the resultant $\Delta{H_f}$ values to the values from the literature, it is clear that in both cases, the use of the main \textit{s} state intensity position significantly underestimates the values. This indicates that the model in its original form cannot be applied in this manner and instead requires optimisation of the $\alpha$ and $\beta$ terms. However, to generate optimised values of $\alpha$ and $\beta$, similar to the work in Ref.~\cite{Griessen_1984}, a larger dataset is required than the two metal systems studied in this work.\par 

Upon hydriding, the host metal \textit{s} band is known to be lowered due to the presence of hydrogen. For the case of a dihydride, this lowering is largely compensated by the increase in Coloumb energy due to the increased charge density around the protons.~\cite{Griessen_1984} However, alongside the lowering of the main metal \textit{s} band some 4-8~eV below $E_F$, conduction band states are also pulled below $E_F$.~\cite{Weaver1981} Both phenomena are apparent when comparing the PDOS spectra of the host metal and equivalent metal dihydride, which are included in Supplemental Material XV at Ref.~\cite{SI}. The position of the lowest lying \textit{s} band cannot be taken as the value of $E_s$ as it would give too large a $\Delta E$ value resulting in a positive $\Delta{H_f}$. Instead, if the position of the conduction band metal \textit{s} state pulled below $E_F$ is taken from the metal dihydride PDOS as $\Delta E$, Fig.~\ref{fig:deltaH_EF}(b) shows that the $\Delta E$ value decreases with respect to the $\Delta E$ value from the \textit{s} position in the metal PDOS. Consequently, when using these values in Eqn.~\ref{Driessden}, the estimated $\Delta{H_f}$ values for TiH\textsubscript{2} and YH\textsubscript{2} are -65.7 and -99.8~kJ/mol H, which improve on the values determined using the metal \textit{s} position from the metal PDOS. However, they do not provide a satisfactory agreement with the literature values. \par

Therefore, the HAXPES valence band of the metal and metal dihydrides was used and applied to the model. While the metal dihydride VB spectra are displayed in Fig.~\ref{fig:VB}, the Ti and Y metal VB spectra are shown in Supplemental Material XVI at Ref.~\cite{SI}. For Ti and Y metal, the \textit{s} state intensity will be enhanced with HAXPES, and by comparing the valence band spectra collected with HAXPES to those in literature collected with soft XPS or to PDOS calculations,~\cite{Ley_1977, Hochst1981, Fujimori_1984, Blaha_1988, TANAKA1990429, Mongstad2014TheOxide} feature II in both can be assumed to be \textit{s} dominated and therefore can be taken as the value for $E_s$. Taking these positions (as displayed in Fig.~\ref{fig:deltaH_EF}(c)) gives $\Delta E$ values of 2.2 and 2.0~eV for Ti and Y, respectively, which again causes the $\Delta{H_f}$ value for YH\textsubscript{2} to be significantly underestimated compared to the literature value. At the same time, a good agreement is found for the TiH\textsubscript{2} estimation. This indicates that feature II in the Y metal VB spectrum is incorrect and cannot be used as $E_s$. The primary source of error here is that Y metal is incredibly difficult to keep clean due to its strong potential to react with oxygen, and the VB spectrum is likely convoluted with oxygen-related states, leading to errors when selecting a feature for $E_s$.\par

Returning to the metal dihydride VB HAXPES spectra in Fig.~\ref{fig:VB}, the PDOS identified that feature II exhibited the strongest \textit{s} character. We find that using the position of feature II in the experimental valence band as the value of $E_s$ (YH\textsubscript{2-$\delta$} = 0.8, and TiH\textsubscript{2-$\delta$} = 2.0~eV) in Eqn.~\ref{Driessden}, one obtains $\Delta H_{f, YH_{2-\delta}}$ = -~111.3~kJ/mol~H and $\Delta H_{f, TiH_{2-\delta}}$ = -~75.8~kJ/mol~H. These values, displayed in Fig.~\ref{fig:deltaH_EF}(c), compared to all other approaches discussed above, provide the best agreement with the experimental values reported in the literature. Supplemental Material XVII at Ref.~\cite{SI} tabulates the $\Delta{H_f}$ values calculated using Eqn.~\ref{Driessden} and the various values of $\Delta E$.\par 

Although the model by Griessen and Driessen was explicitly defined as being based on the host metals' electronic structure in their original paper,~\cite{Griessen_1984} the agreement found here provides convincing evidence that from the experimental valence band of a metal dihydride, one can infer the enthalpy of formation, as there is a clear link between the position of the $s$-state dominated feature II and this critical property. To confirm this finding further, a library of metal dihydrides should be studied using HAXPES in the future, which will also allow the determination of new values of $\alpha$ and $\beta$.\par

\section{Conclusions}

Combining HAXPES core and valence photoelectron spectroscopy with theoretical spectra from DFT provides a powerful analytical approach to increase our understanding of metal hydrides. Photon energy-dependent HAXPES delivers a non-destructive depth profile of YH\textsubscript{2-$\delta$} and TiH\textsubscript{2-$\delta$} thin films. From core level analysis, it is shown that these samples can be modeled as a bilayer system consisting of a passive metal oxide overlayer protecting the underlying metal hydride. By tuning the photon energy across the hard X-ray regime from 2.4 to 7.2~keV, the metal-hydride-related states can be probed and enhanced while minimising any metal oxide surface contributions. By extracting quantitative information from the Ti~2\textit{p} and Y~3\textit{d} core level spectra and utilising the depth distribution model, quantitative estimates of the titanium and yttrium oxide layer thicknesses of 8.2$\pm$2 and 18.4$\pm$2~nm, respectively, were determined. Analysis of the valence spectra highlights the presence of metal hydride-related states near the $E_F$. By comparing the experimental spectra and theoretical cross-section weighted projected densities of states, the nature of hydride states close to $E_F$ could be clearly identified as a combination of \textit{d} and \textit{sp} contributions. Exploiting photoionisation cross-section effects with increasing photon energy, the importance of metal \textit{sp} states near $E_F$ alongside \textit{d} states were identified. Finally, the bond character of the two metal hydride systems was assessed, with results suggesting that both are more ionic than their oxide counterparts. Additionally, the enthalpies of formation of -~111.3~kJ/mol~H for YH\textsubscript{2-$\delta$} and -~75.8~kJ/mol~H for TiH\textsubscript{2-$\delta$} were directly obtained from valence spectra using the model proposed by Griessen and Driessen, in good agreement with values determined using thermodynamical experimental methods. This remarkable result directly links the electronic structure and macroscopic properties fundamental for characterising metal-hydrogen systems for energy storage applications.

\section{Methods}

\subsection{Synthesis}

YH\textsubscript{2-$\delta$} and TiH\textsubscript{2-$\delta$} thin films with a thickness of approximately 200~nm were prepared by reactive magnetron sputtering of a 2-in. metal target (MaTeck Germany, 99.99\% purity) in an Ar/H\textsubscript{2} atmosphere. Before deposition, the chamber was kept at a base pressure below $1\times10^{-6}$~mbar. During deposition, two independent mass flow controllers were used to define the composition of the gas mixture while the total deposition pressure was set to 3$\times$10\textsuperscript{-3}~mbar by means of a reducing valve mounted at the inlet of the pumping stage. An Ar/H\textsubscript{2} gas ratio of 7:1 and 7:2 was used for YH\textsubscript{2-$\delta$} and TiH\textsubscript{2-$\delta$} films, respectively. In both cases, plasma excitation was sustained with a total power of 200~W, supplied as direct current. All samples were grown on p-doped $<100>$ Si substrates without active heating and stored under an argon environment in a sealed container until the measurements were commenced.

\subsection{Characterisation methods}

Energy-dependent HAXPES experiments were conducted at the HAXPES end station of beamline P22 at PETRA III (DESY, Hamburg, Germany).~\cite{Schlueter2019ThePETRAIII} All samples were prepared within an argon-filled glove box and mounted on adhesive carbon tape. The samples were transferred into the HAXPES end station under an argon atmosphere to avoid exposure to air. Four photon energies were used; 2.4106, 3.2691, 6.0054, and 7.2310~keV, referred to as 2.4, 3.3, 6.0, and 7.2~keV, throughout the manuscript for simplicity. For all energies, a Si (111) double crystal monochromator (DCM) was used, and the following post-channel-cut monochromators were employed to achieve the final energy resolutions: Si (220) at 3.3~keV and Si (333) at 6.0 and 7.2~keV. No post-channel-cut monochromator was used at 2.4~keV. The achieved total energy resolutions as determined from taking the 16\% to 84\% width (16/84\% method) of the Fermi edge of a polycrystalline gold foil reference (see Supplemental Material XVIII at Ref.~\cite{SI} for the Fermi edge spectra) were 280~meV (2.4~keV), 243~meV (3.3~keV), 242~meV (6.0~keV), and 202~meV (7.2~keV). The error associated with these resolution values is $\pm$20~meV owing to the errors associated with using the 16/84\% method. Measurements were conducted with an X-ray incidence angle of 15$^{\circ}$ and near-normal emission geometry, and the core level and valence band spectra were collected at a pass energy of 30~eV and a step size of 50~meV. The end station is equipped with a SPECS Phoibos 225HV analyzer, providing a $\pm$30$^{\circ}$ wide acceptance angle, and a high dynamic range delay-line electron detector. The base pressure of the end station is 5$\times$10\textsuperscript{-10}~mbar. The core level and valence band spectra were aligned to the sample's intrinsic Fermi energy ($E_F$). Core level spectra were normalised to their respective areas, whereas the valence band spectra were normalised to the area of the metal hydride states adjacent to $E_F$. The spectral areas were determined in both cases after removing a Shirley-type background. The error associated with quoted BE values of the core lines is $\pm$0.2~eV, which accounts for both the total energy resolution of the measurements and the inherent error associated with peak fit analysis. The error associated with quoted relative atomic percentages is $\pm$0.5 at.\% owing to the error associated with peak fit analysis. HAXPES measurements on Ti and Y metal foils were conducted at beamline I09 of the Diamond Light Source (Harwell, UK). Details of these measurements can be found in Supplemental Material XVI at Ref.~\cite{SI}.

\subsection{Theoretical methods}\label{DFT}

Density functional theory (DFT)~\cite{Hohenberg1964, Kohn1965} calculations were performed using CASTEP~\cite{Clark2005}. Calculations employed the PBE exchange-correlation functional,~\cite{Perdew1996} and norm-conserving pseudopotentials with 12(11) valence electrons for Ti(Y), using a kinetic energy cut-off of 1100~eV. Calculations employed Monkhorst-Pack~\cite{Monkhorst1976} $k$-point grids of $3\times 3 \times 3$ for TiH$_2$, $3\times 3 \times 4$ for TiO$_2$, $2\times 2 \times 1$ for Ti$_2$O$_3$, $6 \times 6 \times 6$ for YH$_2$ and $2 \times 2 \times 2$ for Y$_2$O$_3$, while the projected densities of states (PDOS) for TiH$_2$, YH$_2$ and TiO$_2$ were calculated on a finer $12\times 12 \times 12$ $k$-point grid, and the PDOS for Ti$_2$O$_3$ was calculated on a $6\times 6 \times 2$ $k$-point grid. The $k$-point grid for Y$_2$O$_3$ was kept the same for the PDOS calculation. PDOS calculations were also performed using the hybrid PBE0 functional~\cite{Adamo1999} for the Ti-containing structures; the equivalent calculations for YH$_2$ and Y$_2$O$_3$ were prohibitively expensive due to the larger unit cell sizes. The cubic crystal structure was used for the calculations for both TiH\textsubscript{2} and YH\textsubscript{2}. An additional calculation was performed using the tetragonal crystal structure for TiH\textsubscript{2}, and very minor differences are observed compared to the calculations with the cubic crystal structure (See Supplemental Material IX at Ref.~\cite{SI} for the comparison between PDOS spectra of TiH\textsubscript{2} using the cubic and tetragonal crystal structures). Therefore, the results from the calculations using the cubic structure will only be discussed in this work. For all structures, both the atomic coordinates and cells were relaxed, imposing symmetry and using a maximum force tolerance of 0.02~eV/\AA. The PDOS, atomic charges, and bond (overlap) populations were calculated using a Mulliken-based approach~\cite{Mulliken1955,Segall1996a,Segall1996b}, while the atomic charges were also calculated using both Hirshfeld~\cite{Hirshfeld1977} and Bader~\cite{Bader1990,bader_code} population analysis approaches for comparison. We note that the spilling parameter for all systems was below 0.2\%, indicating that the atomic orbital basis set used to perform the projection was reasonably complete. Nonetheless, there are inherent uncertainties in projecting delocalised orbitals onto an atomic basis set, which also tend to be more severe for metallic systems, resulting in negative contributions to the PDOS, which are visible in the sum of the cross-section weighted PDOS and the difference plots shown in Fig.~\ref{fig:VB}. Gaussian smearing of 0.24 and 0.20~eV was applied to the PDOS to match the experimental broadening when measuring at 3.3~keV and 7.2~keV, respectively. See Supplemental Material XX at Ref.~\cite{SI} for a tabulation of the lattice parameters of the relaxed structures.\par

Post-processing was performed using OptaDOS.~\cite{Morris2014} The PDOS was further processed by applying Scofield photoionisation cross section weighting factors~\cite{Scofield1973, Kalha20} to each projected state and then aligning the PDOS to the theoretically calculated $E_F$. This provides a better comparison to the experiment, and the PDOS were summed to generate a simulated spectrum. See Supplemental Material XII at Ref.~\cite{SI} for the unweighted PDOS spectra. The Galore software package~\cite{Jackson2018} was used to interpolate the Scofield cross section tabulated data~\cite{Scofield1973, Kalha20} to determine the one-electron corrected photoionisation cross sections at the 3.3 and 7.2~keV photon excitation energies. The occupied Y~5\textit{s}/Ti~4\textit{s} and Y~4\textit{d}/Y~3\textit{d} one-electron corrected photoionisation cross sections were used to weight the \textit{s} and \textit{d} PDOS for Y/Ti. Given that the material system is metallic, it would be expected that \textit{p} conduction band states (i.e., \ Y~5\textit{p}/Ti~4\textit{p}) be pulled below the $E_F$ and be responsible for the \textit{p} state contribution observed in the valence band rather than the occupied Y~4\textit{p}/Ti~3\textit{p} shallow core level states. As no photoionisation cross sections are available for unoccupied states, the respective \textit{p} one-electron corrected cross sections were estimated by dividing the \textit{s} orbital cross section by a factor of two.~\cite{Mudd2014, regoutz2016a} Supplemental Material XXI at Ref.~\cite{SI} tabulates the one-electron corrected photoionisation cross section values used to weight the PDOS at both 3.3 and 7.2~keV photon energies.

\begin{acknowledgments}

CK acknowledges support from the Department of Chemistry, UCL. LER acknowledges support from an EPSRC Early Career Research Fellowship (EP/P033253/1). AR acknowledges the support from the Analytical Chemistry Trust Fund for her CAMS-UK Fellowship.
Parts of this research were carried out at beamline P22 at DESY, a member of the Helmholtz Association (HGF). The research leading to this result has been supported by the project CALIPSOplus under the Grant Agreement 730872 from the EU Framework Programme for Research and Innovation HORIZON 2020.
Funding for the HAXPES instrument at beamline P22 by the Federal Ministry of Education and Research (BMBF) under contracts 05KS7UM1 and 05K10UMA with Universit\"{a}t Mainz; 05KS7WW3, 05K10WW1 and 05K13WW1 with Universit\"{a}t W\"{u}rzburg is gratefully acknowledged. The authors would also like to acknowledge Dr. Volodymyr Baran for his assistance in preparing the samples in the glovebox at DESY. We acknowledge Diamond Light Source for time on Beamline I09 under Proposal NT29451-3.

All authors contributed to the concept and design of experiments. CK, GP, CS, AG, T-LL, PKT, and AR performed synchrotron radiation experiments, CK, AR, PB, YZ and FO analysed data, GC and BD provided samples, growth, and characterisation. LR performed theoretical calculations. CK, JO, FO, GP, and AR drafted the manuscript with input from all authors. 

\end{acknowledgments}


\bibliography{references_main.bib}

\begin{thebibliography}{107}%
\makeatletter
\providecommand \@ifxundefined [1]{%
 \@ifx{#1\undefined}
}%
\providecommand \@ifnum [1]{%
 \ifnum #1\expandafter \@firstoftwo
 \else \expandafter \@secondoftwo
 \fi
}%
\providecommand \@ifx [1]{%
 \ifx #1\expandafter \@firstoftwo
 \else \expandafter \@secondoftwo
 \fi
}%
\providecommand \natexlab [1]{#1}%
\providecommand \enquote  [1]{``#1''}%
\providecommand \bibnamefont  [1]{#1}%
\providecommand \bibfnamefont [1]{#1}%
\providecommand \citenamefont [1]{#1}%
\providecommand \href@noop [0]{\@secondoftwo}%
\providecommand \href [0]{\begingroup \@sanitize@url \@href}%
\providecommand \@href[1]{\@@startlink{#1}\@@href}%
\providecommand \@@href[1]{\endgroup#1\@@endlink}%
\providecommand \@sanitize@url [0]{\catcode `\\12\catcode `\$12\catcode
  `\&12\catcode `\#12\catcode `\^12\catcode `\_12\catcode `\%12\relax}%
\providecommand \@@startlink[1]{}%
\providecommand \@@endlink[0]{}%
\providecommand \url  [0]{\begingroup\@sanitize@url \@url }%
\providecommand \@url [1]{\endgroup\@href {#1}{\urlprefix }}%
\providecommand \urlprefix  [0]{URL }%
\providecommand \Eprint [0]{\href }%
\providecommand \doibase [0]{http://dx.doi.org/}%
\providecommand \selectlanguage [0]{\@gobble}%
\providecommand \bibinfo  [0]{\@secondoftwo}%
\providecommand \bibfield  [0]{\@secondoftwo}%
\providecommand \translation [1]{[#1]}%
\providecommand \BibitemOpen [0]{}%
\providecommand \bibitemStop [0]{}%
\providecommand \bibitemNoStop [0]{.\EOS\space}%
\providecommand \EOS [0]{\spacefactor3000\relax}%
\providecommand \BibitemShut  [1]{\csname bibitem#1\endcsname}%
\let\auto@bib@innerbib\@empty
\bibitem [{\citenamefont {Schlapbach}\ and\ \citenamefont
  {Z{\"{u}}ttel}(2001)}]{Schlappbach2001}%
  \BibitemOpen
  \bibfield  {author} {\bibinfo {author} {\bibfnamefont {L.}~\bibnamefont
  {Schlapbach}}\ and\ \bibinfo {author} {\bibfnamefont {A.}~\bibnamefont
  {Z{\"{u}}ttel}},\ }\href@noop {} {\bibfield  {journal} {\bibinfo  {journal}
  {Nature}\ }\textbf {\bibinfo {volume} {414}},\ \bibinfo {pages} {353}
  (\bibinfo {year} {2001})}\BibitemShut {NoStop}%
\bibitem [{\citenamefont {Joubert}\ \emph {et~al.}(2002)\citenamefont
  {Joubert}, \citenamefont {Latroche},\ and\ \citenamefont
  {Percheron-Gu{\'e}gan}}]{Joubert2002}%
  \BibitemOpen
  \bibfield  {author} {\bibinfo {author} {\bibfnamefont {J.~M.}\ \bibnamefont
  {Joubert}}, \bibinfo {author} {\bibfnamefont {M.}~\bibnamefont {Latroche}}, \
  and\ \bibinfo {author} {\bibfnamefont {A.}~\bibnamefont
  {Percheron-Gu{\'e}gan}},\ }\href {\doibase 10.1557/mrs2002.224} {\bibfield
  {journal} {\bibinfo  {journal} {MRS Bulletin}\ }\textbf {\bibinfo {volume}
  {27}},\ \bibinfo {pages} {694} (\bibinfo {year} {2002})}\BibitemShut
  {NoStop}%
\bibitem [{\citenamefont {Sakintuna}\ \emph {et~al.}(2007)\citenamefont
  {Sakintuna}, \citenamefont {Lamari-Darkrim},\ and\ \citenamefont
  {Hirscher}}]{Sakintuna2007}%
  \BibitemOpen
  \bibfield  {author} {\bibinfo {author} {\bibfnamefont {B.}~\bibnamefont
  {Sakintuna}}, \bibinfo {author} {\bibfnamefont {F.}~\bibnamefont
  {Lamari-Darkrim}}, \ and\ \bibinfo {author} {\bibfnamefont {M.}~\bibnamefont
  {Hirscher}},\ }\href@noop {} {\bibfield  {journal} {\bibinfo  {journal}
  {International Journal of Hydrogen Energy}\ }\textbf {\bibinfo {volume}
  {32}},\ \bibinfo {pages} {1121} (\bibinfo {year} {2007})}\BibitemShut
  {NoStop}%
\bibitem [{\citenamefont {{Bellosta von Colbe}}\ \emph
  {et~al.}(2019)\citenamefont {{Bellosta von Colbe}}, \citenamefont {Ares},
  \citenamefont {Barale}, \citenamefont {Baricco}, \citenamefont {Buckley},
  \citenamefont {Capurso}, \citenamefont {Gallandat}, \citenamefont {Grant},
  \citenamefont {Guzik}, \citenamefont {Jacob}, \citenamefont {Jensen},
  \citenamefont {Jensen}, \citenamefont {Jepsen}, \citenamefont {Klassen},
  \citenamefont {Lototskyy}, \citenamefont {Manickam}, \citenamefont {Montone},
  \citenamefont {Puszkiel}, \citenamefont {Sartori}, \citenamefont {Sheppard},
  \citenamefont {Stuart}, \citenamefont {Walker}, \citenamefont {Webb},
  \citenamefont {Yang}, \citenamefont {Yartys}, \citenamefont {Züttel},\ and\
  \citenamefont {Dornheim}}]{BELLOSTAVONCOLBE20197780}%
  \BibitemOpen
  \bibfield  {author} {\bibinfo {author} {\bibfnamefont {J.}~\bibnamefont
  {{Bellosta von Colbe}}}, \bibinfo {author} {\bibfnamefont {J.-R.}\
  \bibnamefont {Ares}}, \bibinfo {author} {\bibfnamefont {J.}~\bibnamefont
  {Barale}}, \bibinfo {author} {\bibfnamefont {M.}~\bibnamefont {Baricco}},
  \bibinfo {author} {\bibfnamefont {C.}~\bibnamefont {Buckley}}, \bibinfo
  {author} {\bibfnamefont {G.}~\bibnamefont {Capurso}}, \bibinfo {author}
  {\bibfnamefont {N.}~\bibnamefont {Gallandat}}, \bibinfo {author}
  {\bibfnamefont {D.~M.}\ \bibnamefont {Grant}}, \bibinfo {author}
  {\bibfnamefont {M.~N.}\ \bibnamefont {Guzik}}, \bibinfo {author}
  {\bibfnamefont {I.}~\bibnamefont {Jacob}}, \bibinfo {author} {\bibfnamefont
  {E.~H.}\ \bibnamefont {Jensen}}, \bibinfo {author} {\bibfnamefont
  {T.}~\bibnamefont {Jensen}}, \bibinfo {author} {\bibfnamefont
  {J.}~\bibnamefont {Jepsen}}, \bibinfo {author} {\bibfnamefont
  {T.}~\bibnamefont {Klassen}}, \bibinfo {author} {\bibfnamefont {M.~V.}\
  \bibnamefont {Lototskyy}}, \bibinfo {author} {\bibfnamefont {K.}~\bibnamefont
  {Manickam}}, \bibinfo {author} {\bibfnamefont {A.}~\bibnamefont {Montone}},
  \bibinfo {author} {\bibfnamefont {J.}~\bibnamefont {Puszkiel}}, \bibinfo
  {author} {\bibfnamefont {S.}~\bibnamefont {Sartori}}, \bibinfo {author}
  {\bibfnamefont {D.~A.}\ \bibnamefont {Sheppard}}, \bibinfo {author}
  {\bibfnamefont {A.}~\bibnamefont {Stuart}}, \bibinfo {author} {\bibfnamefont
  {G.}~\bibnamefont {Walker}}, \bibinfo {author} {\bibfnamefont {C.~J.}\
  \bibnamefont {Webb}}, \bibinfo {author} {\bibfnamefont {H.}~\bibnamefont
  {Yang}}, \bibinfo {author} {\bibfnamefont {V.}~\bibnamefont {Yartys}},
  \bibinfo {author} {\bibfnamefont {A.}~\bibnamefont {Züttel}}, \ and\
  \bibinfo {author} {\bibfnamefont {M.}~\bibnamefont {Dornheim}},\ }\href
  {\doibase https://doi.org/10.1016/j.ijhydene.2019.01.104} {\bibfield
  {journal} {\bibinfo  {journal} {International Journal of Hydrogen Energy}\
  }\textbf {\bibinfo {volume} {44}},\ \bibinfo {pages} {7780} (\bibinfo {year}
  {2019})}\BibitemShut {NoStop}%
\bibitem [{\citenamefont {Wu}\ \emph {et~al.}(2021)\citenamefont {Wu},
  \citenamefont {Guo}, \citenamefont {Yu}, \citenamefont {Jiang}, \citenamefont
  {Zhang}, \citenamefont {Qi}, \citenamefont {Fu}, \citenamefont {Xie},
  \citenamefont {Li}, \citenamefont {Zheng},\ and\ \citenamefont
  {Li}}]{Wu_2021}%
  \BibitemOpen
  \bibfield  {author} {\bibinfo {author} {\bibfnamefont {Y.}~\bibnamefont
  {Wu}}, \bibinfo {author} {\bibfnamefont {Y.}~\bibnamefont {Guo}}, \bibinfo
  {author} {\bibfnamefont {H.}~\bibnamefont {Yu}}, \bibinfo {author}
  {\bibfnamefont {X.}~\bibnamefont {Jiang}}, \bibinfo {author} {\bibfnamefont
  {Y.}~\bibnamefont {Zhang}}, \bibinfo {author} {\bibfnamefont
  {Y.}~\bibnamefont {Qi}}, \bibinfo {author} {\bibfnamefont {K.}~\bibnamefont
  {Fu}}, \bibinfo {author} {\bibfnamefont {L.}~\bibnamefont {Xie}}, \bibinfo
  {author} {\bibfnamefont {G.}~\bibnamefont {Li}}, \bibinfo {author}
  {\bibfnamefont {J.}~\bibnamefont {Zheng}}, \ and\ \bibinfo {author}
  {\bibfnamefont {X.}~\bibnamefont {Li}},\ }\href {\doibase
  10.31635/ccschem.020.202000255} {\bibfield  {journal} {\bibinfo  {journal}
  {CCS Chemistry}\ }\textbf {\bibinfo {volume} {3}},\ \bibinfo {pages} {974}
  (\bibinfo {year} {2021})}\BibitemShut {NoStop}%
\bibitem [{\citenamefont {Ngene}\ \emph {et~al.}(2017)\citenamefont {Ngene},
  \citenamefont {Longo}, \citenamefont {Mooij}, \citenamefont {Bras},\ and\
  \citenamefont {Dam}}]{Ngene2017}%
  \BibitemOpen
  \bibfield  {author} {\bibinfo {author} {\bibfnamefont {P.}~\bibnamefont
  {Ngene}}, \bibinfo {author} {\bibfnamefont {A.}~\bibnamefont {Longo}},
  \bibinfo {author} {\bibfnamefont {L.}~\bibnamefont {Mooij}}, \bibinfo
  {author} {\bibfnamefont {W.}~\bibnamefont {Bras}}, \ and\ \bibinfo {author}
  {\bibfnamefont {B.}~\bibnamefont {Dam}},\ }\href {\doibase
  10.1038/s41467-017-02043-9} {\bibfield  {journal} {\bibinfo  {journal}
  {Nature Communications}\ }\textbf {\bibinfo {volume} {8}},\ \bibinfo {pages}
  {1846} (\bibinfo {year} {2017})}\BibitemShut {NoStop}%
\bibitem [{\citenamefont {Modi}\ and\ \citenamefont
  {Aguey-Zinsou}(2021)}]{modi2021room}%
  \BibitemOpen
  \bibfield  {author} {\bibinfo {author} {\bibfnamefont {P.}~\bibnamefont
  {Modi}}\ and\ \bibinfo {author} {\bibfnamefont {K.-F.}\ \bibnamefont
  {Aguey-Zinsou}},\ }\href@noop {} {\bibfield  {journal} {\bibinfo  {journal}
  {Frontiers in Energy Research}\ }\textbf {\bibinfo {volume} {9}},\ \bibinfo
  {pages} {616115} (\bibinfo {year} {2021})}\BibitemShut {NoStop}%
\bibitem [{\citenamefont {Huiberts}\ \emph {et~al.}(1996)\citenamefont
  {Huiberts}, \citenamefont {Rector}, \citenamefont {Wijngaarden},
  \citenamefont {Jetten}, \citenamefont {{de Groot}}, \citenamefont {Dam},
  \citenamefont {Koeman}, \citenamefont {Griessen}, \citenamefont
  {Hjörvarsson}, \citenamefont {Olafsson},\ and\ \citenamefont
  {Cho}}]{HUIBERTS1996158}%
  \BibitemOpen
  \bibfield  {author} {\bibinfo {author} {\bibfnamefont {J.}~\bibnamefont
  {Huiberts}}, \bibinfo {author} {\bibfnamefont {J.}~\bibnamefont {Rector}},
  \bibinfo {author} {\bibfnamefont {R.}~\bibnamefont {Wijngaarden}}, \bibinfo
  {author} {\bibfnamefont {S.}~\bibnamefont {Jetten}}, \bibinfo {author}
  {\bibfnamefont {D.}~\bibnamefont {{de Groot}}}, \bibinfo {author}
  {\bibfnamefont {B.}~\bibnamefont {Dam}}, \bibinfo {author} {\bibfnamefont
  {N.}~\bibnamefont {Koeman}}, \bibinfo {author} {\bibfnamefont
  {R.}~\bibnamefont {Griessen}}, \bibinfo {author} {\bibfnamefont
  {B.}~\bibnamefont {Hjörvarsson}}, \bibinfo {author} {\bibfnamefont
  {S.}~\bibnamefont {Olafsson}}, \ and\ \bibinfo {author} {\bibfnamefont
  {Y.}~\bibnamefont {Cho}},\ }\href {\doibase
  https://doi.org/10.1016/0925-8388(96)02286-4} {\bibfield  {journal} {\bibinfo
   {journal} {Journal of Alloys and Compounds}\ }\textbf {\bibinfo {volume}
  {239}},\ \bibinfo {pages} {158} (\bibinfo {year} {1996})}\BibitemShut
  {NoStop}%
\bibitem [{\citenamefont {Palm}\ \emph {et~al.}(2018)\citenamefont {Palm},
  \citenamefont {Murray}, \citenamefont {Narayan},\ and\ \citenamefont
  {Munday}}]{Palm_2018}%
  \BibitemOpen
  \bibfield  {author} {\bibinfo {author} {\bibfnamefont {K.~J.}\ \bibnamefont
  {Palm}}, \bibinfo {author} {\bibfnamefont {J.~B.}\ \bibnamefont {Murray}},
  \bibinfo {author} {\bibfnamefont {T.~C.}\ \bibnamefont {Narayan}}, \ and\
  \bibinfo {author} {\bibfnamefont {J.~N.}\ \bibnamefont {Munday}},\ }\href
  {\doibase 10.1021/acsphotonics.8b01243} {\bibfield  {journal} {\bibinfo
  {journal} {ACS Photonics}\ }\textbf {\bibinfo {volume} {5}},\ \bibinfo
  {pages} {4677} (\bibinfo {year} {2018})}\BibitemShut {NoStop}%
\bibitem [{\citenamefont {Fujimori}\ and\ \citenamefont
  {Schlapbach}(1984)}]{Fujimori_1984}%
  \BibitemOpen
  \bibfield  {author} {\bibinfo {author} {\bibfnamefont {A.}~\bibnamefont
  {Fujimori}}\ and\ \bibinfo {author} {\bibfnamefont {L.}~\bibnamefont
  {Schlapbach}},\ }\href@noop {} {\bibfield  {journal} {\bibinfo  {journal}
  {Journal of Physics C: Solid State Physics}\ }\textbf {\bibinfo {volume}
  {17}},\ \bibinfo {pages} {341} (\bibinfo {year} {1984})}\BibitemShut
  {NoStop}%
\bibitem [{\citenamefont {Malinowski}(1983)}]{MALINOWSKI19831}%
  \BibitemOpen
  \bibfield  {author} {\bibinfo {author} {\bibfnamefont {M.}~\bibnamefont
  {Malinowski}},\ }\href {\doibase
  https://doi.org/10.1016/0022-5088(83)90243-6} {\bibfield  {journal} {\bibinfo
   {journal} {Journal of the Less Common Metals}\ }\textbf {\bibinfo {volume}
  {89}},\ \bibinfo {pages} {1} (\bibinfo {year} {1983})}\BibitemShut {NoStop}%
\bibitem [{\citenamefont {Liu}\ \emph {et~al.}(2019)\citenamefont {Liu},
  \citenamefont {Jeong}, \citenamefont {White}, \citenamefont {Feng},
  \citenamefont {Seon~Cho}, \citenamefont {Stavila}, \citenamefont {Allendorf},
  \citenamefont {Urban},\ and\ \citenamefont {Guo}}]{Liu_2019}%
  \BibitemOpen
  \bibfield  {author} {\bibinfo {author} {\bibfnamefont {Y.-S.}\ \bibnamefont
  {Liu}}, \bibinfo {author} {\bibfnamefont {S.}~\bibnamefont {Jeong}}, \bibinfo
  {author} {\bibfnamefont {J.~L.}\ \bibnamefont {White}}, \bibinfo {author}
  {\bibfnamefont {X.}~\bibnamefont {Feng}}, \bibinfo {author} {\bibfnamefont
  {E.}~\bibnamefont {Seon~Cho}}, \bibinfo {author} {\bibfnamefont
  {V.}~\bibnamefont {Stavila}}, \bibinfo {author} {\bibfnamefont {M.~D.}\
  \bibnamefont {Allendorf}}, \bibinfo {author} {\bibfnamefont {J.~J.}\
  \bibnamefont {Urban}}, \ and\ \bibinfo {author} {\bibfnamefont
  {J.}~\bibnamefont {Guo}},\ }\href {\doibase
  https://doi.org/10.1002/cphc.201801185} {\bibfield  {journal} {\bibinfo
  {journal} {ChemPhysChem}\ }\textbf {\bibinfo {volume} {20}},\ \bibinfo
  {pages} {1261} (\bibinfo {year} {2019})}\BibitemShut {NoStop}%
\bibitem [{\citenamefont {King}\ \emph {et~al.}(2021)\citenamefont {King},
  \citenamefont {Picozzi}, \citenamefont {Egdell},\ and\ \citenamefont
  {Panaccione}}]{king_2021}%
  \BibitemOpen
  \bibfield  {author} {\bibinfo {author} {\bibfnamefont {P.~D.~C.}\
  \bibnamefont {King}}, \bibinfo {author} {\bibfnamefont {S.}~\bibnamefont
  {Picozzi}}, \bibinfo {author} {\bibfnamefont {R.~G.}\ \bibnamefont {Egdell}},
  \ and\ \bibinfo {author} {\bibfnamefont {G.}~\bibnamefont {Panaccione}},\
  }\href {\doibase 10.1021/acs.chemrev.0c00616} {\bibfield  {journal} {\bibinfo
   {journal} {Chemical Reviews}\ }\textbf {\bibinfo {volume} {121}},\ \bibinfo
  {pages} {2816} (\bibinfo {year} {2021})}\BibitemShut {NoStop}%
\bibitem [{\citenamefont {Weaver}\ \emph {et~al.}(1979)\citenamefont {Weaver},
  \citenamefont {Peterson},\ and\ \citenamefont {Benbow}}]{Weaver1979}%
  \BibitemOpen
  \bibfield  {author} {\bibinfo {author} {\bibfnamefont {J.~H.}\ \bibnamefont
  {Weaver}}, \bibinfo {author} {\bibfnamefont {D.~T.}\ \bibnamefont
  {Peterson}}, \ and\ \bibinfo {author} {\bibfnamefont {R.~L.}\ \bibnamefont
  {Benbow}},\ }\href@noop {} {\bibfield  {journal} {\bibinfo  {journal}
  {Physical Review B}\ }\textbf {\bibinfo {volume} {20}},\ \bibinfo {pages}
  {5301} (\bibinfo {year} {1979})}\BibitemShut {NoStop}%
\bibitem [{\citenamefont {Weaver}\ \emph {et~al.}(1981)\citenamefont {Weaver},
  \citenamefont {Peterman}, \citenamefont {Peterson},\ and\ \citenamefont
  {Franciosi}}]{Weaver1981}%
  \BibitemOpen
  \bibfield  {author} {\bibinfo {author} {\bibfnamefont {J.~H.}\ \bibnamefont
  {Weaver}}, \bibinfo {author} {\bibfnamefont {D.~J.}\ \bibnamefont
  {Peterman}}, \bibinfo {author} {\bibfnamefont {D.~T.}\ \bibnamefont
  {Peterson}}, \ and\ \bibinfo {author} {\bibfnamefont {A.}~\bibnamefont
  {Franciosi}},\ }\href@noop {} {\bibfield  {journal} {\bibinfo  {journal}
  {Physical Review B}\ }\textbf {\bibinfo {volume} {23}},\ \bibinfo {pages}
  {1692} (\bibinfo {year} {1981})}\BibitemShut {NoStop}%
\bibitem [{\citenamefont {Schlapbach}\ and\ \citenamefont
  {Osterwalder}(1982)}]{SCHLAPBACH1982271}%
  \BibitemOpen
  \bibfield  {author} {\bibinfo {author} {\bibfnamefont {L.}~\bibnamefont
  {Schlapbach}}\ and\ \bibinfo {author} {\bibfnamefont {J.}~\bibnamefont
  {Osterwalder}},\ }\href {\doibase
  https://doi.org/10.1016/0038-1098(82)90541-5} {\bibfield  {journal} {\bibinfo
   {journal} {Solid State Communications}\ }\textbf {\bibinfo {volume} {42}},\
  \bibinfo {pages} {271} (\bibinfo {year} {1982})}\BibitemShut {NoStop}%
\bibitem [{\citenamefont {Butera}\ \emph {et~al.}(1983)\citenamefont {Butera},
  \citenamefont {Weaver}, \citenamefont {Peterman}, \citenamefont {Franciosi},\
  and\ \citenamefont {Peterson}}]{Butera1983}%
  \BibitemOpen
  \bibfield  {author} {\bibinfo {author} {\bibfnamefont {R.~A.}\ \bibnamefont
  {Butera}}, \bibinfo {author} {\bibfnamefont {J.~H.}\ \bibnamefont {Weaver}},
  \bibinfo {author} {\bibfnamefont {D.~J.}\ \bibnamefont {Peterman}}, \bibinfo
  {author} {\bibfnamefont {A.}~\bibnamefont {Franciosi}}, \ and\ \bibinfo
  {author} {\bibfnamefont {D.~T.}\ \bibnamefont {Peterson}},\ }\href {\doibase
  10.1063/1.446046} {\bibfield  {journal} {\bibinfo  {journal} {The Journal of
  Chemical Physics}\ }\textbf {\bibinfo {volume} {79}},\ \bibinfo {pages}
  {2395} (\bibinfo {year} {1983})}\BibitemShut {NoStop}%
\bibitem [{\citenamefont {Osterwalder}(1985)}]{Osterwalder1985OnHydrides}%
  \BibitemOpen
  \bibfield  {author} {\bibinfo {author} {\bibfnamefont {J.}~\bibnamefont
  {Osterwalder}},\ }\href {\doibase 10.1007/BF01307765} {\bibfield  {journal}
  {\bibinfo  {journal} {Zeitschrift f{\"{u}}r Physik B Condensed Matter}\
  }\textbf {\bibinfo {volume} {61}},\ \bibinfo {pages} {113} (\bibinfo {year}
  {1985})}\BibitemShut {NoStop}%
\bibitem [{\citenamefont {Riesterer}(1987)}]{Riesterer1987}%
  \BibitemOpen
  \bibfield  {author} {\bibinfo {author} {\bibfnamefont {T.}~\bibnamefont
  {Riesterer}},\ }\href@noop {} {\bibfield  {journal} {\bibinfo  {journal}
  {Zeitschrift f{\"{u}}r Physik B Condensed Matter}\ }\textbf {\bibinfo
  {volume} {66}},\ \bibinfo {pages} {441} (\bibinfo {year} {1987})}\BibitemShut
  {NoStop}%
\bibitem [{\citenamefont {Hayoz}\ \emph {et~al.}(2000)\citenamefont {Hayoz},
  \citenamefont {Pillo}, \citenamefont {Bovet}, \citenamefont {Z{\"{u}}ttel},
  \citenamefont {Guthrie}, \citenamefont {Pastore}, \citenamefont
  {Schlapbach},\ and\ \citenamefont {Aebi}}]{Hayoz_2000}%
  \BibitemOpen
  \bibfield  {author} {\bibinfo {author} {\bibfnamefont {J.}~\bibnamefont
  {Hayoz}}, \bibinfo {author} {\bibfnamefont {T.}~\bibnamefont {Pillo}},
  \bibinfo {author} {\bibfnamefont {M.}~\bibnamefont {Bovet}}, \bibinfo
  {author} {\bibfnamefont {A.}~\bibnamefont {Z{\"{u}}ttel}}, \bibinfo {author}
  {\bibfnamefont {S.}~\bibnamefont {Guthrie}}, \bibinfo {author} {\bibfnamefont
  {G.}~\bibnamefont {Pastore}}, \bibinfo {author} {\bibfnamefont
  {L.}~\bibnamefont {Schlapbach}}, \ and\ \bibinfo {author} {\bibfnamefont
  {P.}~\bibnamefont {Aebi}},\ }\href {\doibase 10.1116/1.1286073} {\bibfield
  {journal} {\bibinfo  {journal} {Journal of Vacuum Science {\&} Technology A:
  Vacuum, Surfaces, and Films}\ }\textbf {\bibinfo {volume} {18}},\ \bibinfo
  {pages} {2417} (\bibinfo {year} {2000})}\BibitemShut {NoStop}%
\bibitem [{\citenamefont {Hayoz}\ \emph {et~al.}(2001)\citenamefont {Hayoz},
  \citenamefont {Schoenes}, \citenamefont {Schlapbach},\ and\ \citenamefont
  {Aebi}}]{Hayoz2001SwitchableStudy}%
  \BibitemOpen
  \bibfield  {author} {\bibinfo {author} {\bibfnamefont {J.}~\bibnamefont
  {Hayoz}}, \bibinfo {author} {\bibfnamefont {J.}~\bibnamefont {Schoenes}},
  \bibinfo {author} {\bibfnamefont {L.}~\bibnamefont {Schlapbach}}, \ and\
  \bibinfo {author} {\bibfnamefont {P.}~\bibnamefont {Aebi}},\ }\href {\doibase
  10.1063/1.1405835} {\bibfield  {journal} {\bibinfo  {journal} {Journal of
  Applied Physics}\ }\textbf {\bibinfo {volume} {90}},\ \bibinfo {pages} {3925}
  (\bibinfo {year} {2001})}\BibitemShut {NoStop}%
\bibitem [{\citenamefont {Hayoz}\ \emph {et~al.}(2002)\citenamefont {Hayoz},
  \citenamefont {Koitzsch}, \citenamefont {Popovi{\'{c}}}, \citenamefont
  {Bovet}, \citenamefont {Naumovi{\'{c}}},\ and\ \citenamefont
  {Aebi}}]{Hayoz2002ANGLE-SCANNED:}%
  \BibitemOpen
  \bibfield  {author} {\bibinfo {author} {\bibfnamefont {J.}~\bibnamefont
  {Hayoz}}, \bibinfo {author} {\bibfnamefont {C.}~\bibnamefont {Koitzsch}},
  \bibinfo {author} {\bibfnamefont {D.}~\bibnamefont {Popovi{\'{c}}}}, \bibinfo
  {author} {\bibfnamefont {M.}~\bibnamefont {Bovet}}, \bibinfo {author}
  {\bibfnamefont {D.}~\bibnamefont {Naumovi{\'{c}}}}, \ and\ \bibinfo {author}
  {\bibfnamefont {P.}~\bibnamefont {Aebi}},\ }\href {\doibase
  10.1142/S0218625X02001719} {\bibfield  {journal} {\bibinfo  {journal}
  {Surface Review and Letters}\ }\textbf {\bibinfo {volume} {9}},\ \bibinfo
  {pages} {235} (\bibinfo {year} {2002})}\BibitemShut {NoStop}%
\bibitem [{\citenamefont {Hayoz}\ \emph {et~al.}(2003)\citenamefont {Hayoz},
  \citenamefont {Koitzsch}, \citenamefont {Bovet}, \citenamefont
  {Naumovi{\'{c}}}, \citenamefont {Schlapbach},\ and\ \citenamefont
  {Aebi}}]{Hayoz_2003}%
  \BibitemOpen
  \bibfield  {author} {\bibinfo {author} {\bibfnamefont {J.}~\bibnamefont
  {Hayoz}}, \bibinfo {author} {\bibfnamefont {C.}~\bibnamefont {Koitzsch}},
  \bibinfo {author} {\bibfnamefont {M.}~\bibnamefont {Bovet}}, \bibinfo
  {author} {\bibfnamefont {D.}~\bibnamefont {Naumovi{\'{c}}}}, \bibinfo
  {author} {\bibfnamefont {L.}~\bibnamefont {Schlapbach}}, \ and\ \bibinfo
  {author} {\bibfnamefont {P.}~\bibnamefont {Aebi}},\ }\href {\doibase
  10.1103/PhysRevLett.90.196804} {\bibfield  {journal} {\bibinfo  {journal}
  {Physical Review Letters}\ }\textbf {\bibinfo {volume} {90}},\ \bibinfo
  {pages} {4} (\bibinfo {year} {2003})}\BibitemShut {NoStop}%
\bibitem [{\citenamefont {Uno}\ \emph {et~al.}(2004)\citenamefont {Uno},
  \citenamefont {Yamada}, \citenamefont {Maruyama}, \citenamefont {Muta},\ and\
  \citenamefont {Yamanaka}}]{UNO2004101}%
  \BibitemOpen
  \bibfield  {author} {\bibinfo {author} {\bibfnamefont {M.}~\bibnamefont
  {Uno}}, \bibinfo {author} {\bibfnamefont {K.}~\bibnamefont {Yamada}},
  \bibinfo {author} {\bibfnamefont {T.}~\bibnamefont {Maruyama}}, \bibinfo
  {author} {\bibfnamefont {H.}~\bibnamefont {Muta}}, \ and\ \bibinfo {author}
  {\bibfnamefont {S.}~\bibnamefont {Yamanaka}},\ }\href {\doibase
  https://doi.org/10.1016/j.jallcom.2003.07.006} {\bibfield  {journal}
  {\bibinfo  {journal} {Journal of Alloys and Compounds}\ }\textbf {\bibinfo
  {volume} {366}},\ \bibinfo {pages} {101} (\bibinfo {year}
  {2004})}\BibitemShut {NoStop}%
\bibitem [{\citenamefont {Kato}\ \emph {et~al.}(2012)\citenamefont {Kato},
  \citenamefont {Borgschulte}, \citenamefont {Ferri}, \citenamefont {Bielmann},
  \citenamefont {Crivello}, \citenamefont {Wiedenmann}, \citenamefont
  {Parlinska-Wojtan}, \citenamefont {Rossbach}, \citenamefont {Lu},
  \citenamefont {Remhof},\ and\ \citenamefont {Züttel}}]{kato2012co}%
  \BibitemOpen
  \bibfield  {author} {\bibinfo {author} {\bibfnamefont {S.}~\bibnamefont
  {Kato}}, \bibinfo {author} {\bibfnamefont {A.}~\bibnamefont {Borgschulte}},
  \bibinfo {author} {\bibfnamefont {D.}~\bibnamefont {Ferri}}, \bibinfo
  {author} {\bibfnamefont {M.}~\bibnamefont {Bielmann}}, \bibinfo {author}
  {\bibfnamefont {J.-C.}\ \bibnamefont {Crivello}}, \bibinfo {author}
  {\bibfnamefont {D.}~\bibnamefont {Wiedenmann}}, \bibinfo {author}
  {\bibfnamefont {M.}~\bibnamefont {Parlinska-Wojtan}}, \bibinfo {author}
  {\bibfnamefont {P.}~\bibnamefont {Rossbach}}, \bibinfo {author}
  {\bibfnamefont {Y.}~\bibnamefont {Lu}}, \bibinfo {author} {\bibfnamefont
  {A.}~\bibnamefont {Remhof}}, \ and\ \bibinfo {author} {\bibfnamefont
  {A.}~\bibnamefont {Züttel}},\ }\href@noop {} {\bibfield  {journal} {\bibinfo
   {journal} {Physical Chemistry Chemical Physics}\ }\textbf {\bibinfo {volume}
  {14}},\ \bibinfo {pages} {5518} (\bibinfo {year} {2012})}\BibitemShut
  {NoStop}%
\bibitem [{\citenamefont {Mongstad}\ \emph {et~al.}(2014)\citenamefont
  {Mongstad}, \citenamefont {Th{\o}gersen}, \citenamefont {Subrahmanyam},\ and\
  \citenamefont {Karazhanov}}]{Mongstad2014TheOxide}%
  \BibitemOpen
  \bibfield  {author} {\bibinfo {author} {\bibfnamefont {T.}~\bibnamefont
  {Mongstad}}, \bibinfo {author} {\bibfnamefont {A.}~\bibnamefont
  {Th{\o}gersen}}, \bibinfo {author} {\bibfnamefont {A.}~\bibnamefont
  {Subrahmanyam}}, \ and\ \bibinfo {author} {\bibfnamefont {S.}~\bibnamefont
  {Karazhanov}},\ }\href {\doibase 10.1016/j.solmat.2014.05.037} {\bibfield
  {journal} {\bibinfo  {journal} {Solar Energy Materials and Solar Cells}\
  }\textbf {\bibinfo {volume} {128}},\ \bibinfo {pages} {270} (\bibinfo {year}
  {2014})}\BibitemShut {NoStop}%
\bibitem [{\citenamefont {Billeter}\ \emph {et~al.}(2021)\citenamefont
  {Billeter}, \citenamefont {{\L}odziana},\ and\ \citenamefont
  {Borgschulte}}]{billeter2021surface}%
  \BibitemOpen
  \bibfield  {author} {\bibinfo {author} {\bibfnamefont {E.}~\bibnamefont
  {Billeter}}, \bibinfo {author} {\bibfnamefont {Z.}~\bibnamefont
  {{\L}odziana}}, \ and\ \bibinfo {author} {\bibfnamefont {A.}~\bibnamefont
  {Borgschulte}},\ }\href@noop {} {\bibfield  {journal} {\bibinfo  {journal}
  {The Journal of Physical Chemistry C}\ }\textbf {\bibinfo {volume} {125}},\
  \bibinfo {pages} {25339} (\bibinfo {year} {2021})}\BibitemShut {NoStop}%
\bibitem [{\citenamefont {Panaccione}\ \emph {et~al.}(2005)\citenamefont
  {Panaccione}, \citenamefont {Cautero}, \citenamefont {Cautero}, \citenamefont
  {Fondacaro}, \citenamefont {Grioni}, \citenamefont {Lacovig}, \citenamefont
  {Monaco}, \citenamefont {Offi}, \citenamefont {Paolicelli}, \citenamefont
  {Sacchi}, \citenamefont {Stojic}, \citenamefont {Stefani}, \citenamefont
  {Tommasini},\ and\ \citenamefont {Torelli}}]{Panaccione2005}%
  \BibitemOpen
  \bibfield  {author} {\bibinfo {author} {\bibfnamefont {G.}~\bibnamefont
  {Panaccione}}, \bibinfo {author} {\bibfnamefont {G.}~\bibnamefont {Cautero}},
  \bibinfo {author} {\bibfnamefont {M.}~\bibnamefont {Cautero}}, \bibinfo
  {author} {\bibfnamefont {A.}~\bibnamefont {Fondacaro}}, \bibinfo {author}
  {\bibfnamefont {M.}~\bibnamefont {Grioni}}, \bibinfo {author} {\bibfnamefont
  {P.}~\bibnamefont {Lacovig}}, \bibinfo {author} {\bibfnamefont
  {G.}~\bibnamefont {Monaco}}, \bibinfo {author} {\bibfnamefont
  {F.}~\bibnamefont {Offi}}, \bibinfo {author} {\bibfnamefont {G.}~\bibnamefont
  {Paolicelli}}, \bibinfo {author} {\bibfnamefont {M.}~\bibnamefont {Sacchi}},
  \bibinfo {author} {\bibfnamefont {N.}~\bibnamefont {Stojic}}, \bibinfo
  {author} {\bibfnamefont {G.}~\bibnamefont {Stefani}}, \bibinfo {author}
  {\bibfnamefont {R.}~\bibnamefont {Tommasini}}, \ and\ \bibinfo {author}
  {\bibfnamefont {P.}~\bibnamefont {Torelli}},\ }\href@noop {} {\bibfield
  {journal} {\bibinfo  {journal} {Journal of Physics: Condensed Matter}\
  }\textbf {\bibinfo {volume} {17}},\ \bibinfo {pages} {2671} (\bibinfo {year}
  {2005})}\BibitemShut {NoStop}%
\bibitem [{\citenamefont {Sacchi}\ \emph {et~al.}(2005)\citenamefont {Sacchi},
  \citenamefont {Offi}, \citenamefont {Torelli}, \citenamefont {Fondacaro},
  \citenamefont {Spezzani}, \citenamefont {Cautero}, \citenamefont {Cautero},
  \citenamefont {Huotari}, \citenamefont {Grioni}, \citenamefont {Delaunay},
  \citenamefont {Fabrizioli}, \citenamefont {Vank{\'{o}}}, \citenamefont
  {Monaco}, \citenamefont {Paolicelli}, \citenamefont {Stefani},\ and\
  \citenamefont {Panaccione}}]{Sacchi2005}%
  \BibitemOpen
  \bibfield  {author} {\bibinfo {author} {\bibfnamefont {M.}~\bibnamefont
  {Sacchi}}, \bibinfo {author} {\bibfnamefont {F.}~\bibnamefont {Offi}},
  \bibinfo {author} {\bibfnamefont {P.}~\bibnamefont {Torelli}}, \bibinfo
  {author} {\bibfnamefont {A.}~\bibnamefont {Fondacaro}}, \bibinfo {author}
  {\bibfnamefont {C.}~\bibnamefont {Spezzani}}, \bibinfo {author}
  {\bibfnamefont {M.}~\bibnamefont {Cautero}}, \bibinfo {author} {\bibfnamefont
  {G.}~\bibnamefont {Cautero}}, \bibinfo {author} {\bibfnamefont
  {S.}~\bibnamefont {Huotari}}, \bibinfo {author} {\bibfnamefont
  {M.}~\bibnamefont {Grioni}}, \bibinfo {author} {\bibfnamefont
  {R.}~\bibnamefont {Delaunay}}, \bibinfo {author} {\bibfnamefont
  {M.}~\bibnamefont {Fabrizioli}}, \bibinfo {author} {\bibfnamefont
  {G.}~\bibnamefont {Vank{\'{o}}}}, \bibinfo {author} {\bibfnamefont
  {G.}~\bibnamefont {Monaco}}, \bibinfo {author} {\bibfnamefont
  {G.}~\bibnamefont {Paolicelli}}, \bibinfo {author} {\bibfnamefont
  {G.}~\bibnamefont {Stefani}}, \ and\ \bibinfo {author} {\bibfnamefont
  {G.}~\bibnamefont {Panaccione}},\ }\href {\doibase
  10.1103/PhysRevB.71.155117} {\bibfield  {journal} {\bibinfo  {journal}
  {Physical Review B}\ }\textbf {\bibinfo {volume} {71}},\ \bibinfo {pages} {1}
  (\bibinfo {year} {2005})}\BibitemShut {NoStop}%
\bibitem [{\citenamefont {Fadley}(2005)}]{Fadley2005}%
  \BibitemOpen
  \bibfield  {author} {\bibinfo {author} {\bibfnamefont {C.~S.}\ \bibnamefont
  {Fadley}},\ }\href {\doibase https://doi.org/10.1016/j.nima.2005.05.009}
  {\bibfield  {journal} {\bibinfo  {journal} {Nuclear Instruments and Methods
  in Physics Research Section A: Accelerators, Spectrometers, Detectors and
  Associated Equipment}\ }\textbf {\bibinfo {volume} {547}},\ \bibinfo {pages}
  {24} (\bibinfo {year} {2005})}\BibitemShut {NoStop}%
\bibitem [{\citenamefont {Payne}\ \emph {et~al.}(2009)\citenamefont {Payne},
  \citenamefont {Paolicelli}, \citenamefont {Offi}, \citenamefont {Panaccione},
  \citenamefont {Lacovig}, \citenamefont {Beamson}, \citenamefont {Fondacaro},
  \citenamefont {Monaco}, \citenamefont {Vanko},\ and\ \citenamefont
  {Egdell}}]{PAYNE200926}%
  \BibitemOpen
  \bibfield  {author} {\bibinfo {author} {\bibfnamefont {D.}~\bibnamefont
  {Payne}}, \bibinfo {author} {\bibfnamefont {G.}~\bibnamefont {Paolicelli}},
  \bibinfo {author} {\bibfnamefont {F.}~\bibnamefont {Offi}}, \bibinfo {author}
  {\bibfnamefont {G.}~\bibnamefont {Panaccione}}, \bibinfo {author}
  {\bibfnamefont {P.}~\bibnamefont {Lacovig}}, \bibinfo {author} {\bibfnamefont
  {G.}~\bibnamefont {Beamson}}, \bibinfo {author} {\bibfnamefont
  {A.}~\bibnamefont {Fondacaro}}, \bibinfo {author} {\bibfnamefont
  {G.}~\bibnamefont {Monaco}}, \bibinfo {author} {\bibfnamefont
  {G.}~\bibnamefont {Vanko}}, \ and\ \bibinfo {author} {\bibfnamefont
  {R.}~\bibnamefont {Egdell}},\ }\href {\doibase
  https://doi.org/10.1016/j.elspec.2008.10.002} {\bibfield  {journal} {\bibinfo
   {journal} {Journal of Electron Spectroscopy and Related Phenomena}\ }\textbf
  {\bibinfo {volume} {169}},\ \bibinfo {pages} {26} (\bibinfo {year}
  {2009})}\BibitemShut {NoStop}%
\bibitem [{\citenamefont {Claessen}\ \emph {et~al.}(2009)\citenamefont
  {Claessen}, \citenamefont {Sing}, \citenamefont {Paul}, \citenamefont
  {Berner}, \citenamefont {Wetscherek}, \citenamefont {Müller},\ and\
  \citenamefont {Drube}}]{Claessen_2009}%
  \BibitemOpen
  \bibfield  {author} {\bibinfo {author} {\bibfnamefont {R.}~\bibnamefont
  {Claessen}}, \bibinfo {author} {\bibfnamefont {M.}~\bibnamefont {Sing}},
  \bibinfo {author} {\bibfnamefont {M.}~\bibnamefont {Paul}}, \bibinfo {author}
  {\bibfnamefont {G.}~\bibnamefont {Berner}}, \bibinfo {author} {\bibfnamefont
  {A.}~\bibnamefont {Wetscherek}}, \bibinfo {author} {\bibfnamefont
  {A.}~\bibnamefont {Müller}}, \ and\ \bibinfo {author} {\bibfnamefont
  {W.}~\bibnamefont {Drube}},\ }\href {\doibase 10.1088/1367-2630/11/12/125007}
  {\bibfield  {journal} {\bibinfo  {journal} {New Journal of Physics}\ }\textbf
  {\bibinfo {volume} {11}},\ \bibinfo {pages} {125007} (\bibinfo {year}
  {2009})}\BibitemShut {NoStop}%
\bibitem [{\citenamefont {Mudd}\ \emph {et~al.}(2014)\citenamefont {Mudd},
  \citenamefont {Lee}, \citenamefont {Mu\~noz Sanjos\'e}, \citenamefont
  {Z\'u\~niga P\'erez}, \citenamefont {Payne}, \citenamefont {Egdell},\ and\
  \citenamefont {McConville}}]{Mudd2014}%
  \BibitemOpen
  \bibfield  {author} {\bibinfo {author} {\bibfnamefont {J.~J.}\ \bibnamefont
  {Mudd}}, \bibinfo {author} {\bibfnamefont {T.-L.}\ \bibnamefont {Lee}},
  \bibinfo {author} {\bibfnamefont {V.}~\bibnamefont {Mu\~noz Sanjos\'e}},
  \bibinfo {author} {\bibfnamefont {J.}~\bibnamefont {Z\'u\~niga P\'erez}},
  \bibinfo {author} {\bibfnamefont {D.~J.}\ \bibnamefont {Payne}}, \bibinfo
  {author} {\bibfnamefont {R.~G.}\ \bibnamefont {Egdell}}, \ and\ \bibinfo
  {author} {\bibfnamefont {C.~F.}\ \bibnamefont {McConville}},\ }\href
  {\doibase 10.1103/PhysRevB.89.165305} {\bibfield  {journal} {\bibinfo
  {journal} {Physical Review B}\ }\textbf {\bibinfo {volume} {89}},\ \bibinfo
  {pages} {165305} (\bibinfo {year} {2014})}\BibitemShut {NoStop}%
\bibitem [{\citenamefont {Woicik}(2016)}]{woicik}%
  \BibitemOpen
  \bibinfo {editor} {\bibfnamefont {J.}~\bibnamefont {Woicik}},\ ed.,\
  \href@noop {} {\emph {\bibinfo {title} {Hard X-ray Photoelectron Spectroscopy
  (HAXPES), Springer Series in Surface Sciences}}},\ Vol.~\bibinfo {volume}
  {59}\ (\bibinfo  {publisher} {Springer International Publishing},\ \bibinfo
  {address} {Switzerland},\ \bibinfo {year} {2016})\BibitemShut {NoStop}%
\bibitem [{\citenamefont {Kalha}\ \emph {et~al.}(2021)\citenamefont {Kalha},
  \citenamefont {Fernando}, \citenamefont {Bhatt}, \citenamefont {Johansson},
  \citenamefont {Lindblad}, \citenamefont {Rensmo}, \citenamefont {Medina},
  \citenamefont {Lindblad}, \citenamefont {Siol}, \citenamefont {Jeurgens},
  \citenamefont {Cancellieri}, \citenamefont {Rossnagel}, \citenamefont
  {Medjanik}, \citenamefont {Sch{\"{o}}nhense}, \citenamefont {Simon},
  \citenamefont {Gray}, \citenamefont {Nem{\v{s}}{\'{a}}k}, \citenamefont
  {L{\"{o}}mker}, \citenamefont {Schlueter},\ and\ \citenamefont
  {Regoutz}}]{HAXPES_Big_Boy}%
  \BibitemOpen
  \bibfield  {author} {\bibinfo {author} {\bibfnamefont {C.}~\bibnamefont
  {Kalha}}, \bibinfo {author} {\bibfnamefont {N.~K.}\ \bibnamefont {Fernando}},
  \bibinfo {author} {\bibfnamefont {P.}~\bibnamefont {Bhatt}}, \bibinfo
  {author} {\bibfnamefont {F.~O.~L.}\ \bibnamefont {Johansson}}, \bibinfo
  {author} {\bibfnamefont {A.}~\bibnamefont {Lindblad}}, \bibinfo {author}
  {\bibfnamefont {H.}~\bibnamefont {Rensmo}}, \bibinfo {author} {\bibfnamefont
  {L.~Z.}\ \bibnamefont {Medina}}, \bibinfo {author} {\bibfnamefont
  {R.}~\bibnamefont {Lindblad}}, \bibinfo {author} {\bibfnamefont
  {S.}~\bibnamefont {Siol}}, \bibinfo {author} {\bibfnamefont {L.~P.~H.}\
  \bibnamefont {Jeurgens}}, \bibinfo {author} {\bibfnamefont {C.}~\bibnamefont
  {Cancellieri}}, \bibinfo {author} {\bibfnamefont {K.}~\bibnamefont
  {Rossnagel}}, \bibinfo {author} {\bibfnamefont {K.}~\bibnamefont {Medjanik}},
  \bibinfo {author} {\bibfnamefont {G.}~\bibnamefont {Sch{\"{o}}nhense}},
  \bibinfo {author} {\bibfnamefont {M.}~\bibnamefont {Simon}}, \bibinfo
  {author} {\bibfnamefont {A.~X.}\ \bibnamefont {Gray}}, \bibinfo {author}
  {\bibfnamefont {S.}~\bibnamefont {Nem{\v{s}}{\'{a}}k}}, \bibinfo {author}
  {\bibfnamefont {P.}~\bibnamefont {L{\"{o}}mker}}, \bibinfo {author}
  {\bibfnamefont {C.}~\bibnamefont {Schlueter}}, \ and\ \bibinfo {author}
  {\bibfnamefont {A.}~\bibnamefont {Regoutz}},\ }\href {\doibase
  10.1088/1361-648x/abeacd} {\bibfield  {journal} {\bibinfo  {journal} {Journal
  of Physics: Condensed Matter}\ }\textbf {\bibinfo {volume} {33}},\ \bibinfo
  {pages} {233001} (\bibinfo {year} {2021})}\BibitemShut {NoStop}%
\bibitem [{\citenamefont {Griessen}\ and\ \citenamefont
  {Driessen}(1984)}]{Griessen_1984}%
  \BibitemOpen
  \bibfield  {author} {\bibinfo {author} {\bibfnamefont {R.}~\bibnamefont
  {Griessen}}\ and\ \bibinfo {author} {\bibfnamefont {A.}~\bibnamefont
  {Driessen}},\ }\href {\doibase 10.1103/PhysRevB.30.4372} {\bibfield
  {journal} {\bibinfo  {journal} {Physical Review B}\ }\textbf {\bibinfo
  {volume} {30}},\ \bibinfo {pages} {4372} (\bibinfo {year}
  {1984})}\BibitemShut {NoStop}%
\bibitem [{SI()}]{SI}%
  \BibitemOpen
  \href@noop {} {}\bibinfo {note} {See Supplemental Material at [URL will be
  inserted by publisher] for the collected survey spectra, details on the peak
  fitting procedure and DDF methodology, information on the resolution of the
  measurements, additional core level and valence band spectra, spectra
  collected after air exposure, the unweighted PDOS of both the metals and
  dihydrides, the complete cross-section weighted dihydride PDOS at both 3.3
  and 7.2~keV photon energies, additional computational details, summary of the
  methods used to derive the enthalpy of formation from theory and experiment,
  and HAXPES collected Ti and Y metal valence band spectra.}\BibitemShut
  {Stop}%
\bibitem [{\citenamefont {Tanuma}\ \emph {et~al.}(2011)\citenamefont {Tanuma},
  \citenamefont {Powell},\ and\ \citenamefont {Penn}}]{Tanuma_2011}%
  \BibitemOpen
  \bibfield  {author} {\bibinfo {author} {\bibfnamefont {S.}~\bibnamefont
  {Tanuma}}, \bibinfo {author} {\bibfnamefont {C.~J.}\ \bibnamefont {Powell}},
  \ and\ \bibinfo {author} {\bibfnamefont {D.~R.}\ \bibnamefont {Penn}},\
  }\href {\doibase https://doi.org/10.1002/sia.3522} {\bibfield  {journal}
  {\bibinfo  {journal} {Surface and Interface Analysis}\ }\textbf {\bibinfo
  {volume} {43}},\ \bibinfo {pages} {689} (\bibinfo {year} {2011})}\BibitemShut
  {NoStop}%
\bibitem [{\citenamefont {Lamartine}\ \emph {et~al.}(1980)\citenamefont
  {Lamartine}, \citenamefont {Haas},\ and\ \citenamefont
  {Solomon}}]{LAMARTINE1980537}%
  \BibitemOpen
  \bibfield  {author} {\bibinfo {author} {\bibfnamefont {B.}~\bibnamefont
  {Lamartine}}, \bibinfo {author} {\bibfnamefont {T.}~\bibnamefont {Haas}}, \
  and\ \bibinfo {author} {\bibfnamefont {J.}~\bibnamefont {Solomon}},\ }\href
  {\doibase https://doi.org/10.1016/0378-5963(80)90097-5} {\bibfield  {journal}
  {\bibinfo  {journal} {Applications of Surface Science}\ }\textbf {\bibinfo
  {volume} {4}},\ \bibinfo {pages} {537} (\bibinfo {year} {1980})}\BibitemShut
  {NoStop}%
\bibitem [{\citenamefont {Ma}\ \emph {et~al.}(2009)\citenamefont {Ma},
  \citenamefont {Kang}, \citenamefont {Dai}, \citenamefont {Liang},
  \citenamefont {Fang}, \citenamefont {Wang}, \citenamefont {Wang},\ and\
  \citenamefont {Cheng}}]{MA20092250}%
  \BibitemOpen
  \bibfield  {author} {\bibinfo {author} {\bibfnamefont {L.-P.}\ \bibnamefont
  {Ma}}, \bibinfo {author} {\bibfnamefont {X.-D.}\ \bibnamefont {Kang}},
  \bibinfo {author} {\bibfnamefont {H.-B.}\ \bibnamefont {Dai}}, \bibinfo
  {author} {\bibfnamefont {Y.}~\bibnamefont {Liang}}, \bibinfo {author}
  {\bibfnamefont {Z.-Z.}\ \bibnamefont {Fang}}, \bibinfo {author}
  {\bibfnamefont {P.-J.}\ \bibnamefont {Wang}}, \bibinfo {author}
  {\bibfnamefont {P.}~\bibnamefont {Wang}}, \ and\ \bibinfo {author}
  {\bibfnamefont {H.-M.}\ \bibnamefont {Cheng}},\ }\href {\doibase
  https://doi.org/10.1016/j.actamat.2009.01.025} {\bibfield  {journal}
  {\bibinfo  {journal} {Acta Materialia}\ }\textbf {\bibinfo {volume} {57}},\
  \bibinfo {pages} {2250} (\bibinfo {year} {2009})}\BibitemShut {NoStop}%
\bibitem [{\citenamefont {Ren}\ \emph {et~al.}(2014)\citenamefont {Ren},
  \citenamefont {Wang}, \citenamefont {Liu},\ and\ \citenamefont
  {Ohachi}}]{Ren_2014}%
  \BibitemOpen
  \bibfield  {author} {\bibinfo {author} {\bibfnamefont {N.}~\bibnamefont
  {Ren}}, \bibinfo {author} {\bibfnamefont {G.}~\bibnamefont {Wang}}, \bibinfo
  {author} {\bibfnamefont {H.}~\bibnamefont {Liu}}, \ and\ \bibinfo {author}
  {\bibfnamefont {T.}~\bibnamefont {Ohachi}},\ }\href {\doibase
  10.1016/j.materresbull.2013.11.002} {\bibfield  {journal} {\bibinfo
  {journal} {Materials Research Bulletin}\ }\textbf {\bibinfo {volume} {50}},\
  \bibinfo {pages} {379} (\bibinfo {year} {2014})}\BibitemShut {NoStop}%
\bibitem [{\citenamefont {Biwer}\ and\ \citenamefont
  {Bernasek}(1986)}]{BIWER1986207}%
  \BibitemOpen
  \bibfield  {author} {\bibinfo {author} {\bibfnamefont {B.}~\bibnamefont
  {Biwer}}\ and\ \bibinfo {author} {\bibfnamefont {S.}~\bibnamefont
  {Bernasek}},\ }\href {\doibase https://doi.org/10.1016/0039-6028(86)90795-8}
  {\bibfield  {journal} {\bibinfo  {journal} {Surface Science}\ }\textbf
  {\bibinfo {volume} {167}},\ \bibinfo {pages} {207} (\bibinfo {year}
  {1986})}\BibitemShut {NoStop}%
\bibitem [{\citenamefont {Sleigh}\ \emph {et~al.}(1996)\citenamefont {Sleigh},
  \citenamefont {Pijpers}, \citenamefont {Jaspers}, \citenamefont {Coussens},\
  and\ \citenamefont {Meier}}]{SLEIGH199641}%
  \BibitemOpen
  \bibfield  {author} {\bibinfo {author} {\bibfnamefont {C.}~\bibnamefont
  {Sleigh}}, \bibinfo {author} {\bibfnamefont {A.}~\bibnamefont {Pijpers}},
  \bibinfo {author} {\bibfnamefont {A.}~\bibnamefont {Jaspers}}, \bibinfo
  {author} {\bibfnamefont {B.}~\bibnamefont {Coussens}}, \ and\ \bibinfo
  {author} {\bibfnamefont {R.~J.}\ \bibnamefont {Meier}},\ }\href {\doibase
  https://doi.org/10.1016/0368-2048(95)02392-5} {\bibfield  {journal} {\bibinfo
   {journal} {Journal of Electron Spectroscopy and Related Phenomena}\ }\textbf
  {\bibinfo {volume} {77}},\ \bibinfo {pages} {41} (\bibinfo {year}
  {1996})}\BibitemShut {NoStop}%
\bibitem [{\citenamefont {Cornelius}\ \emph {et~al.}(2019)\citenamefont
  {Cornelius}, \citenamefont {Colombi}, \citenamefont {Nafezarefi},
  \citenamefont {Schreuders}, \citenamefont {Heller}, \citenamefont {Munnik},\
  and\ \citenamefont {Dam}}]{Cornelius_2019}%
  \BibitemOpen
  \bibfield  {author} {\bibinfo {author} {\bibfnamefont {S.}~\bibnamefont
  {Cornelius}}, \bibinfo {author} {\bibfnamefont {G.}~\bibnamefont {Colombi}},
  \bibinfo {author} {\bibfnamefont {F.}~\bibnamefont {Nafezarefi}}, \bibinfo
  {author} {\bibfnamefont {H.}~\bibnamefont {Schreuders}}, \bibinfo {author}
  {\bibfnamefont {R.}~\bibnamefont {Heller}}, \bibinfo {author} {\bibfnamefont
  {F.}~\bibnamefont {Munnik}}, \ and\ \bibinfo {author} {\bibfnamefont
  {B.}~\bibnamefont {Dam}},\ }\href {\doibase 10.1021/acs.jpclett.9b00088}
  {\bibfield  {journal} {\bibinfo  {journal} {The Journal of Physical Chemistry
  Letters}\ }\textbf {\bibinfo {volume} {10}},\ \bibinfo {pages} {1342}
  (\bibinfo {year} {2019})}\BibitemShut {NoStop}%
\bibitem [{\citenamefont {Reichl}\ and\ \citenamefont
  {Gaukler}(1986)}]{REICHL1986196}%
  \BibitemOpen
  \bibfield  {author} {\bibinfo {author} {\bibfnamefont {R.}~\bibnamefont
  {Reichl}}\ and\ \bibinfo {author} {\bibfnamefont {K.}~\bibnamefont
  {Gaukler}},\ }\href {\doibase https://doi.org/10.1016/0169-4332(86)90005-X}
  {\bibfield  {journal} {\bibinfo  {journal} {Applied Surface Science}\
  }\textbf {\bibinfo {volume} {26}},\ \bibinfo {pages} {196} (\bibinfo {year}
  {1986})}\BibitemShut {NoStop}%
\bibitem [{\citenamefont {Barreca}\ \emph {et~al.}(2001)\citenamefont
  {Barreca}, \citenamefont {Battiston}, \citenamefont {Berto}, \citenamefont
  {Gerbasi},\ and\ \citenamefont {Tondello}}]{Barreca_2001}%
  \BibitemOpen
  \bibfield  {author} {\bibinfo {author} {\bibfnamefont {D.}~\bibnamefont
  {Barreca}}, \bibinfo {author} {\bibfnamefont {G.~A.}\ \bibnamefont
  {Battiston}}, \bibinfo {author} {\bibfnamefont {D.}~\bibnamefont {Berto}},
  \bibinfo {author} {\bibfnamefont {R.}~\bibnamefont {Gerbasi}}, \ and\
  \bibinfo {author} {\bibfnamefont {E.}~\bibnamefont {Tondello}},\ }\href
  {\doibase 10.1116/11.20020404} {\bibfield  {journal} {\bibinfo  {journal}
  {Surface Science Spectra}\ }\textbf {\bibinfo {volume} {8}},\ \bibinfo
  {pages} {234} (\bibinfo {year} {2001})}\BibitemShut {NoStop}%
\bibitem [{\citenamefont {Gougousi}\ and\ \citenamefont
  {Chen}(2008)}]{GOUGOUSI20086197}%
  \BibitemOpen
  \bibfield  {author} {\bibinfo {author} {\bibfnamefont {T.}~\bibnamefont
  {Gougousi}}\ and\ \bibinfo {author} {\bibfnamefont {Z.}~\bibnamefont
  {Chen}},\ }\href {\doibase https://doi.org/10.1016/j.tsf.2007.11.104}
  {\bibfield  {journal} {\bibinfo  {journal} {Thin Solid Films}\ }\textbf
  {\bibinfo {volume} {516}},\ \bibinfo {pages} {6197} (\bibinfo {year}
  {2008})}\BibitemShut {NoStop}%
\bibitem [{\citenamefont {Mitrovic}\ \emph {et~al.}(2014)\citenamefont
  {Mitrovic}, \citenamefont {Althobaiti}, \citenamefont {Weerakkody},
  \citenamefont {Dhanak}, \citenamefont {Linhart}, \citenamefont {Veal},
  \citenamefont {Sedghi}, \citenamefont {Hall}, \citenamefont {Chalker},
  \citenamefont {Tsoutsou},\ and\ \citenamefont
  {Dimoulas}}]{Mitrovic_Veal_2014}%
  \BibitemOpen
  \bibfield  {author} {\bibinfo {author} {\bibfnamefont {I.~Z.}\ \bibnamefont
  {Mitrovic}}, \bibinfo {author} {\bibfnamefont {M.}~\bibnamefont
  {Althobaiti}}, \bibinfo {author} {\bibfnamefont {A.~D.}\ \bibnamefont
  {Weerakkody}}, \bibinfo {author} {\bibfnamefont {V.~R.}\ \bibnamefont
  {Dhanak}}, \bibinfo {author} {\bibfnamefont {W.~M.}\ \bibnamefont {Linhart}},
  \bibinfo {author} {\bibfnamefont {T.~D.}\ \bibnamefont {Veal}}, \bibinfo
  {author} {\bibfnamefont {N.}~\bibnamefont {Sedghi}}, \bibinfo {author}
  {\bibfnamefont {S.}~\bibnamefont {Hall}}, \bibinfo {author} {\bibfnamefont
  {P.~R.}\ \bibnamefont {Chalker}}, \bibinfo {author} {\bibfnamefont
  {D.}~\bibnamefont {Tsoutsou}}, \ and\ \bibinfo {author} {\bibfnamefont
  {A.}~\bibnamefont {Dimoulas}},\ }\href {\doibase 10.1063/1.4868091}
  {\bibfield  {journal} {\bibinfo  {journal} {Journal of Applied Physics}\
  }\textbf {\bibinfo {volume} {115}},\ \bibinfo {pages} {114102} (\bibinfo
  {year} {2014})}\BibitemShut {NoStop}%
\bibitem [{\citenamefont {Pouilleau}\ \emph {et~al.}(1997)\citenamefont
  {Pouilleau}, \citenamefont {Devilliers}, \citenamefont {Garrido},
  \citenamefont {Durand-Vidal},\ and\ \citenamefont
  {Mahé}}]{POUILLEAU1997235}%
  \BibitemOpen
  \bibfield  {author} {\bibinfo {author} {\bibfnamefont {J.}~\bibnamefont
  {Pouilleau}}, \bibinfo {author} {\bibfnamefont {D.}~\bibnamefont
  {Devilliers}}, \bibinfo {author} {\bibfnamefont {F.}~\bibnamefont {Garrido}},
  \bibinfo {author} {\bibfnamefont {S.}~\bibnamefont {Durand-Vidal}}, \ and\
  \bibinfo {author} {\bibfnamefont {E.}~\bibnamefont {Mahé}},\ }\href
  {\doibase https://doi.org/10.1016/S0921-5107(97)00043-3} {\bibfield
  {journal} {\bibinfo  {journal} {Materials Science and Engineering: B}\
  }\textbf {\bibinfo {volume} {47}},\ \bibinfo {pages} {235} (\bibinfo {year}
  {1997})}\BibitemShut {NoStop}%
\bibitem [{\citenamefont {Diebold}(2003)}]{DIEBOLD200353}%
  \BibitemOpen
  \bibfield  {author} {\bibinfo {author} {\bibfnamefont {U.}~\bibnamefont
  {Diebold}},\ }\href {\doibase https://doi.org/10.1016/S0167-5729(02)00100-0}
  {\bibfield  {journal} {\bibinfo  {journal} {Surface Science Reports}\
  }\textbf {\bibinfo {volume} {48}},\ \bibinfo {pages} {53} (\bibinfo {year}
  {2003})}\BibitemShut {NoStop}%
\bibitem [{\citenamefont {Huang}\ \emph {et~al.}(2009)\citenamefont {Huang},
  \citenamefont {Peng},\ and\ \citenamefont {Ohuchi}}]{HUANG20092825}%
  \BibitemOpen
  \bibfield  {author} {\bibinfo {author} {\bibfnamefont {L.}~\bibnamefont
  {Huang}}, \bibinfo {author} {\bibfnamefont {F.}~\bibnamefont {Peng}}, \ and\
  \bibinfo {author} {\bibfnamefont {F.~S.}\ \bibnamefont {Ohuchi}},\ }\href
  {\doibase https://doi.org/10.1016/j.susc.2009.07.030} {\bibfield  {journal}
  {\bibinfo  {journal} {Surface Science}\ }\textbf {\bibinfo {volume} {603}},\
  \bibinfo {pages} {2825} (\bibinfo {year} {2009})}\BibitemShut {NoStop}%
\bibitem [{\citenamefont {Berens}\ \emph {et~al.}(2020)\citenamefont {Berens},
  \citenamefont {Bichelmaier}, \citenamefont {Fernando}, \citenamefont
  {Thakur}, \citenamefont {Lee}, \citenamefont {Mascheck}, \citenamefont
  {Wiell}, \citenamefont {Eriksson}, \citenamefont {Kahk}, \citenamefont
  {Lischner}, \citenamefont {Mistry}, \citenamefont {Aichinger}, \citenamefont
  {Pobegen},\ and\ \citenamefont {Regoutz}}]{Berens_2020}%
  \BibitemOpen
  \bibfield  {author} {\bibinfo {author} {\bibfnamefont {J.}~\bibnamefont
  {Berens}}, \bibinfo {author} {\bibfnamefont {S.}~\bibnamefont {Bichelmaier}},
  \bibinfo {author} {\bibfnamefont {N.~K.}\ \bibnamefont {Fernando}}, \bibinfo
  {author} {\bibfnamefont {P.~K.}\ \bibnamefont {Thakur}}, \bibinfo {author}
  {\bibfnamefont {T.-L.}\ \bibnamefont {Lee}}, \bibinfo {author} {\bibfnamefont
  {M.}~\bibnamefont {Mascheck}}, \bibinfo {author} {\bibfnamefont
  {T.}~\bibnamefont {Wiell}}, \bibinfo {author} {\bibfnamefont {S.~K.}\
  \bibnamefont {Eriksson}}, \bibinfo {author} {\bibfnamefont {J.~M.}\
  \bibnamefont {Kahk}}, \bibinfo {author} {\bibfnamefont {J.}~\bibnamefont
  {Lischner}}, \bibinfo {author} {\bibfnamefont {M.~V.}\ \bibnamefont
  {Mistry}}, \bibinfo {author} {\bibfnamefont {T.}~\bibnamefont {Aichinger}},
  \bibinfo {author} {\bibfnamefont {G.}~\bibnamefont {Pobegen}}, \ and\
  \bibinfo {author} {\bibfnamefont {A.}~\bibnamefont {Regoutz}},\ }\href
  {\doibase 10.1088/2515-7655/ab8c5e} {\bibfield  {journal} {\bibinfo
  {journal} {Journal of Physics: Energy}\ }\textbf {\bibinfo {volume} {2}},\
  \bibinfo {pages} {035001} (\bibinfo {year} {2020})}\BibitemShut {NoStop}%
\bibitem [{\citenamefont {Shinotsuka}\ \emph {et~al.}(2015)\citenamefont
  {Shinotsuka}, \citenamefont {Tanuma}, \citenamefont {Powell},\ and\
  \citenamefont {Penn}}]{Shinotsuka_2015}%
  \BibitemOpen
  \bibfield  {author} {\bibinfo {author} {\bibfnamefont {H.}~\bibnamefont
  {Shinotsuka}}, \bibinfo {author} {\bibfnamefont {S.}~\bibnamefont {Tanuma}},
  \bibinfo {author} {\bibfnamefont {C.~J.}\ \bibnamefont {Powell}}, \ and\
  \bibinfo {author} {\bibfnamefont {D.~R.}\ \bibnamefont {Penn}},\ }\href
  {\doibase https://doi.org/10.1002/sia.5789} {\bibfield  {journal} {\bibinfo
  {journal} {Surface and Interface Analysis}\ }\textbf {\bibinfo {volume}
  {47}},\ \bibinfo {pages} {871} (\bibinfo {year} {2015})}\BibitemShut
  {NoStop}%
\bibitem [{\citenamefont {Bosseboeuf}\ \emph {et~al.}(2019)\citenamefont
  {Bosseboeuf}, \citenamefont {Lemettre}, \citenamefont {Wu}, \citenamefont
  {Moulin}, \citenamefont {Coste}, \citenamefont {Bessouet}, \citenamefont
  {Hammami}, \citenamefont {Renard},\ and\ \citenamefont
  {Vincent}}]{bosseboeuf2019effect}%
  \BibitemOpen
  \bibfield  {author} {\bibinfo {author} {\bibfnamefont {A.}~\bibnamefont
  {Bosseboeuf}}, \bibinfo {author} {\bibfnamefont {S.}~\bibnamefont
  {Lemettre}}, \bibinfo {author} {\bibfnamefont {M.}~\bibnamefont {Wu}},
  \bibinfo {author} {\bibfnamefont {J.}~\bibnamefont {Moulin}}, \bibinfo
  {author} {\bibfnamefont {P.}~\bibnamefont {Coste}}, \bibinfo {author}
  {\bibfnamefont {C.}~\bibnamefont {Bessouet}}, \bibinfo {author}
  {\bibfnamefont {S.}~\bibnamefont {Hammami}}, \bibinfo {author} {\bibfnamefont
  {C.}~\bibnamefont {Renard}}, \ and\ \bibinfo {author} {\bibfnamefont
  {L.}~\bibnamefont {Vincent}},\ }\href@noop {} {\bibfield  {journal} {\bibinfo
   {journal} {Sensors and Materials}\ }\textbf {\bibinfo {volume} {31}},\
  \bibinfo {pages} {2825} (\bibinfo {year} {2019})}\BibitemShut {NoStop}%
\bibitem [{\citenamefont {Bessouet}\ \emph {et~al.}(2021)\citenamefont
  {Bessouet}, \citenamefont {Lemettre}, \citenamefont {Kutyla}, \citenamefont
  {Bosseboeuf}, \citenamefont {Coste}, \citenamefont {Sauvage}, \citenamefont
  {Lecoq}, \citenamefont {Wendling}, \citenamefont {Bellamy}, \citenamefont
  {Jagtap}, \citenamefont {Escoubas}, \citenamefont {Guichet}, \citenamefont
  {Thomas},\ and\ \citenamefont {Moulin}}]{Bessouet_2021}%
  \BibitemOpen
  \bibfield  {author} {\bibinfo {author} {\bibfnamefont {C.}~\bibnamefont
  {Bessouet}}, \bibinfo {author} {\bibfnamefont {S.}~\bibnamefont {Lemettre}},
  \bibinfo {author} {\bibfnamefont {C.}~\bibnamefont {Kutyla}}, \bibinfo
  {author} {\bibfnamefont {A.}~\bibnamefont {Bosseboeuf}}, \bibinfo {author}
  {\bibfnamefont {P.}~\bibnamefont {Coste}}, \bibinfo {author} {\bibfnamefont
  {T.}~\bibnamefont {Sauvage}}, \bibinfo {author} {\bibfnamefont
  {H.}~\bibnamefont {Lecoq}}, \bibinfo {author} {\bibfnamefont
  {O.}~\bibnamefont {Wendling}}, \bibinfo {author} {\bibfnamefont
  {A.}~\bibnamefont {Bellamy}}, \bibinfo {author} {\bibfnamefont
  {P.}~\bibnamefont {Jagtap}}, \bibinfo {author} {\bibfnamefont
  {S.}~\bibnamefont {Escoubas}}, \bibinfo {author} {\bibfnamefont
  {C.}~\bibnamefont {Guichet}}, \bibinfo {author} {\bibfnamefont
  {O.}~\bibnamefont {Thomas}}, \ and\ \bibinfo {author} {\bibfnamefont
  {J.}~\bibnamefont {Moulin}},\ }\href {\doibase 10.1116/6.0001084} {\bibfield
  {journal} {\bibinfo  {journal} {Journal of Vacuum Science \& Technology B}\
  }\textbf {\bibinfo {volume} {39}},\ \bibinfo {pages} {054202} (\bibinfo
  {year} {2021})}\BibitemShut {NoStop}%
\bibitem [{\citenamefont {Offi}\ \emph {et~al.}(2021)\citenamefont {Offi},
  \citenamefont {Yamauchi}, \citenamefont {Picozzi}, \citenamefont
  {Lollobrigida}, \citenamefont {Verna}, \citenamefont {Schlueter},
  \citenamefont {Lee}, \citenamefont {Regoutz}, \citenamefont {Payne},
  \citenamefont {Petrov}, \citenamefont {Vinai}, \citenamefont {Pierantozzi},
  \citenamefont {Pincelli}, \citenamefont {Panaccione},\ and\ \citenamefont
  {Borgatti}}]{Offi2021}%
  \BibitemOpen
  \bibfield  {author} {\bibinfo {author} {\bibfnamefont {F.}~\bibnamefont
  {Offi}}, \bibinfo {author} {\bibfnamefont {K.}~\bibnamefont {Yamauchi}},
  \bibinfo {author} {\bibfnamefont {S.}~\bibnamefont {Picozzi}}, \bibinfo
  {author} {\bibfnamefont {V.}~\bibnamefont {Lollobrigida}}, \bibinfo {author}
  {\bibfnamefont {A.}~\bibnamefont {Verna}}, \bibinfo {author} {\bibfnamefont
  {C.}~\bibnamefont {Schlueter}}, \bibinfo {author} {\bibfnamefont {T.-L.}\
  \bibnamefont {Lee}}, \bibinfo {author} {\bibfnamefont {A.}~\bibnamefont
  {Regoutz}}, \bibinfo {author} {\bibfnamefont {D.}~\bibnamefont {Payne}},
  \bibinfo {author} {\bibfnamefont {A.}~\bibnamefont {Petrov}}, \bibinfo
  {author} {\bibfnamefont {G.}~\bibnamefont {Vinai}}, \bibinfo {author}
  {\bibfnamefont {G.}~\bibnamefont {Pierantozzi}}, \bibinfo {author}
  {\bibfnamefont {T.}~\bibnamefont {Pincelli}}, \bibinfo {author}
  {\bibfnamefont {G.}~\bibnamefont {Panaccione}}, \ and\ \bibinfo {author}
  {\bibfnamefont {F.}~\bibnamefont {Borgatti}},\ }\href@noop {} {\bibfield
  {journal} {\bibinfo  {journal} {Physical Review Materials}\ }\textbf
  {\bibinfo {volume} {5}},\ \bibinfo {pages} {104403} (\bibinfo {year}
  {2021})}\BibitemShut {NoStop}%
\bibitem [{\citenamefont {Zhang}\ and\ \citenamefont
  {Xiao}(1998)}]{Zhang_1998}%
  \BibitemOpen
  \bibfield  {author} {\bibinfo {author} {\bibfnamefont {S.}~\bibnamefont
  {Zhang}}\ and\ \bibinfo {author} {\bibfnamefont {R.}~\bibnamefont {Xiao}},\
  }\href {\doibase 10.1063/1.366615} {\bibfield  {journal} {\bibinfo  {journal}
  {Journal of Applied Physics}\ }\textbf {\bibinfo {volume} {83}},\ \bibinfo
  {pages} {3842} (\bibinfo {year} {1998})}\BibitemShut {NoStop}%
\bibitem [{\citenamefont {Tanuma}\ \emph {et~al.}(2003)\citenamefont {Tanuma},
  \citenamefont {Powell},\ and\ \citenamefont {Penn}}]{Tanuma_2003}%
  \BibitemOpen
  \bibfield  {author} {\bibinfo {author} {\bibfnamefont {S.}~\bibnamefont
  {Tanuma}}, \bibinfo {author} {\bibfnamefont {C.~J.}\ \bibnamefont {Powell}},
  \ and\ \bibinfo {author} {\bibfnamefont {D.~R.}\ \bibnamefont {Penn}},\
  }\href {\doibase https://doi.org/10.1002/sia.1526} {\bibfield  {journal}
  {\bibinfo  {journal} {Surface and Interface Analysis}\ }\textbf {\bibinfo
  {volume} {35}},\ \bibinfo {pages} {268} (\bibinfo {year} {2003})}\BibitemShut
  {NoStop}%
\bibitem [{\citenamefont {Wang}\ \emph {et~al.}(2009)\citenamefont {Wang},
  \citenamefont {Badylevich}, \citenamefont {Afanas’ev}, \citenamefont
  {Stesmans}, \citenamefont {Adelmann}, \citenamefont {Van~Elshocht},
  \citenamefont {Kittl}, \citenamefont {Lukosius}, \citenamefont {Walczyk},\
  and\ \citenamefont {Wenger}}]{Wang_2009}%
  \BibitemOpen
  \bibfield  {author} {\bibinfo {author} {\bibfnamefont {W.~C.}\ \bibnamefont
  {Wang}}, \bibinfo {author} {\bibfnamefont {M.}~\bibnamefont {Badylevich}},
  \bibinfo {author} {\bibfnamefont {V.~V.}\ \bibnamefont {Afanas’ev}},
  \bibinfo {author} {\bibfnamefont {A.}~\bibnamefont {Stesmans}}, \bibinfo
  {author} {\bibfnamefont {C.}~\bibnamefont {Adelmann}}, \bibinfo {author}
  {\bibfnamefont {S.}~\bibnamefont {Van~Elshocht}}, \bibinfo {author}
  {\bibfnamefont {J.~A.}\ \bibnamefont {Kittl}}, \bibinfo {author}
  {\bibfnamefont {M.}~\bibnamefont {Lukosius}}, \bibinfo {author}
  {\bibfnamefont {C.}~\bibnamefont {Walczyk}}, \ and\ \bibinfo {author}
  {\bibfnamefont {C.}~\bibnamefont {Wenger}},\ }\href {\doibase
  10.1063/1.3236536} {\bibfield  {journal} {\bibinfo  {journal} {Applied
  Physics Letters}\ }\textbf {\bibinfo {volume} {95}},\ \bibinfo {pages}
  {132903} (\bibinfo {year} {2009})}\BibitemShut {NoStop}%
\bibitem [{\citenamefont {Mudavakkat}\ \emph {et~al.}(2012)\citenamefont
  {Mudavakkat}, \citenamefont {Atuchin}, \citenamefont {Kruchinin},
  \citenamefont {Kayani},\ and\ \citenamefont {Ramana}}]{MUDAVAKKAT2012893}%
  \BibitemOpen
  \bibfield  {author} {\bibinfo {author} {\bibfnamefont {V.}~\bibnamefont
  {Mudavakkat}}, \bibinfo {author} {\bibfnamefont {V.}~\bibnamefont {Atuchin}},
  \bibinfo {author} {\bibfnamefont {V.}~\bibnamefont {Kruchinin}}, \bibinfo
  {author} {\bibfnamefont {A.}~\bibnamefont {Kayani}}, \ and\ \bibinfo {author}
  {\bibfnamefont {C.}~\bibnamefont {Ramana}},\ }\href {\doibase
  https://doi.org/10.1016/j.optmat.2011.11.027} {\bibfield  {journal} {\bibinfo
   {journal} {Optical Materials}\ }\textbf {\bibinfo {volume} {34}},\ \bibinfo
  {pages} {893} (\bibinfo {year} {2012})}\BibitemShut {NoStop}%
\bibitem [{\citenamefont {Weaver}\ and\ \citenamefont
  {Peterson}(1980)}]{WEAVER1980207}%
  \BibitemOpen
  \bibfield  {author} {\bibinfo {author} {\bibfnamefont {J.}~\bibnamefont
  {Weaver}}\ and\ \bibinfo {author} {\bibfnamefont {D.}~\bibnamefont
  {Peterson}},\ }\href {\doibase https://doi.org/10.1016/0022-5088(80)90091-0}
  {\bibfield  {journal} {\bibinfo  {journal} {Journal of the Less Common
  Metals}\ }\textbf {\bibinfo {volume} {74}},\ \bibinfo {pages} {207} (\bibinfo
  {year} {1980})}\BibitemShut {NoStop}%
\bibitem [{\citenamefont {Barrett}\ and\ \citenamefont
  {Jordan}(1987)}]{Barrett1987}%
  \BibitemOpen
  \bibfield  {author} {\bibinfo {author} {\bibfnamefont {S.~D.}\ \bibnamefont
  {Barrett}}\ and\ \bibinfo {author} {\bibfnamefont {R.~G.}\ \bibnamefont
  {Jordan}},\ }\href {\doibase 10.1007/BF01305429} {\bibfield  {journal}
  {\bibinfo  {journal} {Zeitschrift f{\"u}r Physik B Condensed Matter}\
  }\textbf {\bibinfo {volume} {66}},\ \bibinfo {pages} {375} (\bibinfo {year}
  {1987})}\BibitemShut {NoStop}%
\bibitem [{\citenamefont {Baptist}\ \emph {et~al.}(1988)\citenamefont
  {Baptist}, \citenamefont {Pellissier},\ and\ \citenamefont
  {Chauvet}}]{Baptist1988}%
  \BibitemOpen
  \bibfield  {author} {\bibinfo {author} {\bibfnamefont {R.}~\bibnamefont
  {Baptist}}, \bibinfo {author} {\bibfnamefont {A.}~\bibnamefont {Pellissier}},
  \ and\ \bibinfo {author} {\bibfnamefont {G.}~\bibnamefont {Chauvet}},\
  }\href@noop {} {\bibfield  {journal} {\bibinfo  {journal} {Zeitschrift
  f{\"{u}}r Physik B Condensed Matter}\ }\textbf {\bibinfo {volume} {73}},\
  \bibinfo {pages} {107} (\bibinfo {year} {1988})}\BibitemShut {NoStop}%
\bibitem [{\citenamefont {Budke}\ \emph {et~al.}(2008)\citenamefont {Budke},
  \citenamefont {Correa},\ and\ \citenamefont {Donath}}]{Budke_2008}%
  \BibitemOpen
  \bibfield  {author} {\bibinfo {author} {\bibfnamefont {M.}~\bibnamefont
  {Budke}}, \bibinfo {author} {\bibfnamefont {J.~S.}\ \bibnamefont {Correa}}, \
  and\ \bibinfo {author} {\bibfnamefont {M.}~\bibnamefont {Donath}},\
  }\href@noop {} {\bibfield  {journal} {\bibinfo  {journal} {Physical Review
  B}\ }\textbf {\bibinfo {volume} {77}} (\bibinfo {year} {2008})}\BibitemShut
  {NoStop}%
\bibitem [{\citenamefont {Scanlon}\ \emph {et~al.}(2013)\citenamefont
  {Scanlon}, \citenamefont {Dunnill}, \citenamefont {Buckeridge}, \citenamefont
  {Shevlin}, \citenamefont {Logsdail}, \citenamefont {Woodley}, \citenamefont
  {Catlow}, \citenamefont {Powell}, \citenamefont {Palgrave}, \citenamefont
  {Parkin}, \citenamefont {Watson}, \citenamefont {Keal}, \citenamefont
  {Sherwood}, \citenamefont {Walsh},\ and\ \citenamefont
  {Sokol}}]{Scanlon2013}%
  \BibitemOpen
  \bibfield  {author} {\bibinfo {author} {\bibfnamefont {D.~O.}\ \bibnamefont
  {Scanlon}}, \bibinfo {author} {\bibfnamefont {C.~W.}\ \bibnamefont
  {Dunnill}}, \bibinfo {author} {\bibfnamefont {J.}~\bibnamefont {Buckeridge}},
  \bibinfo {author} {\bibfnamefont {S.~A.}\ \bibnamefont {Shevlin}}, \bibinfo
  {author} {\bibfnamefont {A.~J.}\ \bibnamefont {Logsdail}}, \bibinfo {author}
  {\bibfnamefont {S.~M.}\ \bibnamefont {Woodley}}, \bibinfo {author}
  {\bibfnamefont {C.~R.~A.}\ \bibnamefont {Catlow}}, \bibinfo {author}
  {\bibfnamefont {M.~J.}\ \bibnamefont {Powell}}, \bibinfo {author}
  {\bibfnamefont {R.~G.}\ \bibnamefont {Palgrave}}, \bibinfo {author}
  {\bibfnamefont {I.~P.}\ \bibnamefont {Parkin}}, \bibinfo {author}
  {\bibfnamefont {G.~W.}\ \bibnamefont {Watson}}, \bibinfo {author}
  {\bibfnamefont {T.~W.}\ \bibnamefont {Keal}}, \bibinfo {author}
  {\bibfnamefont {P.}~\bibnamefont {Sherwood}}, \bibinfo {author}
  {\bibfnamefont {A.}~\bibnamefont {Walsh}}, \ and\ \bibinfo {author}
  {\bibfnamefont {A.~A.}\ \bibnamefont {Sokol}},\ }\href {\doibase
  10.1038/nmat3697} {\bibfield  {journal} {\bibinfo  {journal} {Nature
  Materials}\ }\textbf {\bibinfo {volume} {12}},\ \bibinfo {pages} {798}
  (\bibinfo {year} {2013})}\BibitemShut {NoStop}%
\bibitem [{\citenamefont {Gupta}(1979)}]{Gupta_1979}%
  \BibitemOpen
  \bibfield  {author} {\bibinfo {author} {\bibfnamefont {M.}~\bibnamefont
  {Gupta}},\ }\href@noop {} {\bibfield  {journal} {\bibinfo  {journal} {Solid
  State Communications}\ }\textbf {\bibinfo {volume} {29}},\ \bibinfo {pages}
  {47} (\bibinfo {year} {1979})}\BibitemShut {NoStop}%
\bibitem [{\citenamefont {Smithson}\ \emph {et~al.}(2002)\citenamefont
  {Smithson}, \citenamefont {Marianetti}, \citenamefont {Morgan}, \citenamefont
  {Van~der Ven}, \citenamefont {Predith},\ and\ \citenamefont
  {Ceder}}]{Smithson_2002}%
  \BibitemOpen
  \bibfield  {author} {\bibinfo {author} {\bibfnamefont {H.}~\bibnamefont
  {Smithson}}, \bibinfo {author} {\bibfnamefont {C.~A.}\ \bibnamefont
  {Marianetti}}, \bibinfo {author} {\bibfnamefont {D.}~\bibnamefont {Morgan}},
  \bibinfo {author} {\bibfnamefont {A.}~\bibnamefont {Van~der Ven}}, \bibinfo
  {author} {\bibfnamefont {A.}~\bibnamefont {Predith}}, \ and\ \bibinfo
  {author} {\bibfnamefont {G.}~\bibnamefont {Ceder}},\ }\href {\doibase
  10.1103/PhysRevB.66.144107} {\bibfield  {journal} {\bibinfo  {journal}
  {Physical Review B}\ }\textbf {\bibinfo {volume} {66}},\ \bibinfo {pages}
  {144107} (\bibinfo {year} {2002})}\BibitemShut {NoStop}%
\bibitem [{\citenamefont {Peterman}\ \emph {et~al.}(1979)\citenamefont
  {Peterman}, \citenamefont {Harmon}, \citenamefont {Marchiando},\ and\
  \citenamefont {Weaver}}]{Peterman_1979}%
  \BibitemOpen
  \bibfield  {author} {\bibinfo {author} {\bibfnamefont {D.~J.}\ \bibnamefont
  {Peterman}}, \bibinfo {author} {\bibfnamefont {B.~N.}\ \bibnamefont
  {Harmon}}, \bibinfo {author} {\bibfnamefont {J.}~\bibnamefont {Marchiando}},
  \ and\ \bibinfo {author} {\bibfnamefont {J.~H.}\ \bibnamefont {Weaver}},\
  }\href {\doibase 10.1103/PhysRevB.19.4867} {\bibfield  {journal} {\bibinfo
  {journal} {Phys. Rev. B}\ }\textbf {\bibinfo {volume} {19}},\ \bibinfo
  {pages} {4867} (\bibinfo {year} {1979})}\BibitemShut {NoStop}%
\bibitem [{\citenamefont {Wolf}\ and\ \citenamefont
  {Herzig}(2000)}]{Wolf2000First-principlesBonding}%
  \BibitemOpen
  \bibfield  {author} {\bibinfo {author} {\bibfnamefont {W.}~\bibnamefont
  {Wolf}}\ and\ \bibinfo {author} {\bibfnamefont {P.}~\bibnamefont {Herzig}},\
  }\href@noop {} {\bibfield  {journal} {\bibinfo  {journal} {Journal of
  Physics: Condensed Matter}\ }\textbf {\bibinfo {volume} {12}},\ \bibinfo
  {pages} {4535} (\bibinfo {year} {2000})}\BibitemShut {NoStop}%
\bibitem [{\citenamefont {Mehta}\ \emph {et~al.}(2021)\citenamefont {Mehta},
  \citenamefont {Vogel}, \citenamefont {Shivprasad}, \citenamefont {Luther},
  \citenamefont {Andersson}, \citenamefont {Rao}, \citenamefont {Kotlyar},
  \citenamefont {Clausen},\ and\ \citenamefont {Cooper}}]{Mehta2021AHydride}%
  \BibitemOpen
  \bibfield  {author} {\bibinfo {author} {\bibfnamefont {V.~K.}\ \bibnamefont
  {Mehta}}, \bibinfo {author} {\bibfnamefont {S.~C.}\ \bibnamefont {Vogel}},
  \bibinfo {author} {\bibfnamefont {A.~P.}\ \bibnamefont {Shivprasad}},
  \bibinfo {author} {\bibfnamefont {E.~P.}\ \bibnamefont {Luther}}, \bibinfo
  {author} {\bibfnamefont {D.~A.}\ \bibnamefont {Andersson}}, \bibinfo {author}
  {\bibfnamefont {D.~V.}\ \bibnamefont {Rao}}, \bibinfo {author} {\bibfnamefont
  {D.}~\bibnamefont {Kotlyar}}, \bibinfo {author} {\bibfnamefont
  {B.}~\bibnamefont {Clausen}}, \ and\ \bibinfo {author} {\bibfnamefont
  {M.~W.}\ \bibnamefont {Cooper}},\ }\href@noop {} {\bibfield  {journal}
  {\bibinfo  {journal} {Journal of Nuclear Materials}\ }\textbf {\bibinfo
  {volume} {547}},\ \bibinfo {pages} {152837} (\bibinfo {year}
  {2021})}\BibitemShut {NoStop}%
\bibitem [{\citenamefont {{Luiggi A.}}(2021)}]{LUIGGIA2021102639}%
  \BibitemOpen
  \bibfield  {author} {\bibinfo {author} {\bibfnamefont {N.~J.}\ \bibnamefont
  {{Luiggi A.}}},\ }\href {\doibase
  https://doi.org/10.1016/j.mtcomm.2021.102639} {\bibfield  {journal} {\bibinfo
   {journal} {Materials Today Communications}\ }\textbf {\bibinfo {volume}
  {28}},\ \bibinfo {pages} {102639} (\bibinfo {year} {2021})}\BibitemShut
  {NoStop}%
\bibitem [{\citenamefont {Yang}\ \emph {et~al.}(2002)\citenamefont {Yang},
  \citenamefont {Wang}, \citenamefont {Zhao}, \citenamefont {Wang},
  \citenamefont {Ye},\ and\ \citenamefont {Wang}}]{YANG2002109}%
  \BibitemOpen
  \bibfield  {author} {\bibinfo {author} {\bibfnamefont {R.}~\bibnamefont
  {Yang}}, \bibinfo {author} {\bibfnamefont {Y.}~\bibnamefont {Wang}}, \bibinfo
  {author} {\bibfnamefont {Y.}~\bibnamefont {Zhao}}, \bibinfo {author}
  {\bibfnamefont {L.}~\bibnamefont {Wang}}, \bibinfo {author} {\bibfnamefont
  {H.}~\bibnamefont {Ye}}, \ and\ \bibinfo {author} {\bibfnamefont
  {C.}~\bibnamefont {Wang}},\ }\href {\doibase
  https://doi.org/10.1016/S1359-6454(01)00317-2} {\bibfield  {journal}
  {\bibinfo  {journal} {Acta Materialia}\ }\textbf {\bibinfo {volume} {50}},\
  \bibinfo {pages} {109} (\bibinfo {year} {2002})}\BibitemShut {NoStop}%
\bibitem [{\citenamefont {Segall}\ \emph
  {et~al.}(1996{\natexlab{a}})\citenamefont {Segall}, \citenamefont {Shah},
  \citenamefont {Pickard},\ and\ \citenamefont {Payne}}]{Segall1996a}%
  \BibitemOpen
  \bibfield  {author} {\bibinfo {author} {\bibfnamefont {M.~D.}\ \bibnamefont
  {Segall}}, \bibinfo {author} {\bibfnamefont {R.}~\bibnamefont {Shah}},
  \bibinfo {author} {\bibfnamefont {C.~J.}\ \bibnamefont {Pickard}}, \ and\
  \bibinfo {author} {\bibfnamefont {M.~C.}\ \bibnamefont {Payne}},\ }\href
  {\doibase 10.1103/PhysRevB.54.16317} {\bibfield  {journal} {\bibinfo
  {journal} {Phys. Rev. B}\ }\textbf {\bibinfo {volume} {54}},\ \bibinfo
  {pages} {16317} (\bibinfo {year} {1996}{\natexlab{a}})}\BibitemShut {NoStop}%
\bibitem [{\citenamefont {Segall}\ \emph
  {et~al.}(1996{\natexlab{b}})\citenamefont {Segall}, \citenamefont {Pickard},
  \citenamefont {Shah},\ and\ \citenamefont {Payne}}]{Segall1996b}%
  \BibitemOpen
  \bibfield  {author} {\bibinfo {author} {\bibfnamefont {M.~D.}\ \bibnamefont
  {Segall}}, \bibinfo {author} {\bibfnamefont {C.~J.}\ \bibnamefont {Pickard}},
  \bibinfo {author} {\bibfnamefont {R.}~\bibnamefont {Shah}}, \ and\ \bibinfo
  {author} {\bibfnamefont {M.~C.}\ \bibnamefont {Payne}},\ }\href {\doibase
  10.1080/002689796173912} {\bibfield  {journal} {\bibinfo  {journal}
  {Molecular Physics}\ }\textbf {\bibinfo {volume} {89}},\ \bibinfo {pages}
  {571} (\bibinfo {year} {1996}{\natexlab{b}})}\BibitemShut {NoStop}%
\bibitem [{\citenamefont {Allred}(1961)}]{ALLRED1961215}%
  \BibitemOpen
  \bibfield  {author} {\bibinfo {author} {\bibfnamefont {A.}~\bibnamefont
  {Allred}},\ }\href {\doibase https://doi.org/10.1016/0022-1902(61)80142-5}
  {\bibfield  {journal} {\bibinfo  {journal} {Journal of Inorganic and Nuclear
  Chemistry}\ }\textbf {\bibinfo {volume} {17}},\ \bibinfo {pages} {215}
  (\bibinfo {year} {1961})}\BibitemShut {NoStop}%
\bibitem [{\citenamefont {Husain}\ \emph {et~al.}(1989)\citenamefont {Husain},
  \citenamefont {Batra},\ and\ \citenamefont {Srivastava}}]{HUSAIN19891233}%
  \BibitemOpen
  \bibfield  {author} {\bibinfo {author} {\bibfnamefont {M.}~\bibnamefont
  {Husain}}, \bibinfo {author} {\bibfnamefont {A.}~\bibnamefont {Batra}}, \
  and\ \bibinfo {author} {\bibfnamefont {K.}~\bibnamefont {Srivastava}},\
  }\href {\doibase https://doi.org/10.1016/S0277-5387(00)81146-8} {\bibfield
  {journal} {\bibinfo  {journal} {Polyhedron}\ }\textbf {\bibinfo {volume}
  {8}},\ \bibinfo {pages} {1233} (\bibinfo {year} {1989})}\BibitemShut
  {NoStop}%
\bibitem [{\citenamefont {Young}(2018)}]{YOUNG2018}%
  \BibitemOpen
  \bibfield  {author} {\bibinfo {author} {\bibfnamefont {K.}~\bibnamefont
  {Young}},\ }in\ \href {\doibase
  https://doi.org/10.1016/B978-0-12-409547-2.05894-7} {\emph {\bibinfo
  {booktitle} {Reference Module in Chemistry, Molecular Sciences and Chemical
  Engineering}}}\ (\bibinfo  {publisher} {Elsevier},\ \bibinfo {year}
  {2018})\BibitemShut {NoStop}%
\bibitem [{\citenamefont {Buschow}\ \emph {et~al.}(1982)\citenamefont
  {Buschow}, \citenamefont {Bouten},\ and\ \citenamefont
  {Miedema}}]{Buschow_1982}%
  \BibitemOpen
  \bibfield  {author} {\bibinfo {author} {\bibfnamefont {K.~H.~J.}\
  \bibnamefont {Buschow}}, \bibinfo {author} {\bibfnamefont {P.~C.~P.}\
  \bibnamefont {Bouten}}, \ and\ \bibinfo {author} {\bibfnamefont {A.~R.}\
  \bibnamefont {Miedema}},\ }\href {\doibase 10.1088/0034-4885/45/9/001}
  {\bibfield  {journal} {\bibinfo  {journal} {Reports on Progress in Physics}\
  }\textbf {\bibinfo {volume} {45}},\ \bibinfo {pages} {937} (\bibinfo {year}
  {1982})}\BibitemShut {NoStop}%
\bibitem [{\citenamefont {Chernikov}\ \emph {et~al.}(1987)\citenamefont
  {Chernikov}, \citenamefont {Savin}, \citenamefont {Fadeev}, \citenamefont
  {Landin},\ and\ \citenamefont {Izhvanov}}]{CHERNIKOV1987441}%
  \BibitemOpen
  \bibfield  {author} {\bibinfo {author} {\bibfnamefont {A.}~\bibnamefont
  {Chernikov}}, \bibinfo {author} {\bibfnamefont {V.}~\bibnamefont {Savin}},
  \bibinfo {author} {\bibfnamefont {V.}~\bibnamefont {Fadeev}}, \bibinfo
  {author} {\bibfnamefont {N.}~\bibnamefont {Landin}}, \ and\ \bibinfo {author}
  {\bibfnamefont {L.}~\bibnamefont {Izhvanov}},\ }\href {\doibase
  https://doi.org/10.1016/0022-5088(87)90139-1} {\bibfield  {journal} {\bibinfo
   {journal} {Journal of the Less Common Metals}\ }\textbf {\bibinfo {volume}
  {130}},\ \bibinfo {pages} {441} (\bibinfo {year} {1987})}\BibitemShut
  {NoStop}%
\bibitem [{\citenamefont {Gray}(2021)}]{GRAY_2021}%
  \BibitemOpen
  \bibfield  {author} {\bibinfo {author} {\bibfnamefont {E.~M.}\ \bibnamefont
  {Gray}},\ }\href {\doibase https://doi.org/10.1016/j.ijhydene.2021.02.025}
  {\bibfield  {journal} {\bibinfo  {journal} {International Journal of Hydrogen
  Energy}\ }\textbf {\bibinfo {volume} {46}},\ \bibinfo {pages} {15702}
  (\bibinfo {year} {2021})}\BibitemShut {NoStop}%
\bibitem [{\citenamefont {Huston}\ and\ \citenamefont
  {Sandrock}(1980)}]{HUSTON1980435}%
  \BibitemOpen
  \bibfield  {author} {\bibinfo {author} {\bibfnamefont {E.}~\bibnamefont
  {Huston}}\ and\ \bibinfo {author} {\bibfnamefont {G.}~\bibnamefont
  {Sandrock}},\ }\href {\doibase https://doi.org/10.1016/0022-5088(80)90182-4}
  {\bibfield  {journal} {\bibinfo  {journal} {Journal of the Less Common
  Metals}\ }\textbf {\bibinfo {volume} {74}},\ \bibinfo {pages} {435} (\bibinfo
  {year} {1980})}\BibitemShut {NoStop}%
\bibitem [{\citenamefont {Dantzer}(1983)}]{DANTZER1983913}%
  \BibitemOpen
  \bibfield  {author} {\bibinfo {author} {\bibfnamefont {P.}~\bibnamefont
  {Dantzer}},\ }\href {\doibase https://doi.org/10.1016/0022-3697(83)90130-0}
  {\bibfield  {journal} {\bibinfo  {journal} {Journal of Physics and Chemistry
  of Solids}\ }\textbf {\bibinfo {volume} {44}},\ \bibinfo {pages} {913}
  (\bibinfo {year} {1983})}\BibitemShut {NoStop}%
\bibitem [{\citenamefont {wei Zhao}\ \emph {et~al.}(2008)\citenamefont {wei
  Zhao}, \citenamefont {Ding}, \citenamefont {feng Tian}, \citenamefont {juan
  Zhao},\ and\ \citenamefont {liang Hou}}]{JZhao_2008}%
  \BibitemOpen
  \bibfield  {author} {\bibinfo {author} {\bibfnamefont {J.}~\bibnamefont {wei
  Zhao}}, \bibinfo {author} {\bibfnamefont {H.}~\bibnamefont {Ding}}, \bibinfo
  {author} {\bibfnamefont {X.}~\bibnamefont {feng Tian}}, \bibinfo {author}
  {\bibfnamefont {W.}~\bibnamefont {juan Zhao}}, \ and\ \bibinfo {author}
  {\bibfnamefont {H.}~\bibnamefont {liang Hou}},\ }\href {\doibase
  10.1088/1674-0068/21/06/569-574} {\bibfield  {journal} {\bibinfo  {journal}
  {Chinese Journal of Chemical Physics}\ }\textbf {\bibinfo {volume} {21}},\
  \bibinfo {pages} {569} (\bibinfo {year} {2008})}\BibitemShut {NoStop}%
\bibitem [{\citenamefont {Miwa}\ and\ \citenamefont
  {Fukumoto}(2002)}]{Miwa_2002}%
  \BibitemOpen
  \bibfield  {author} {\bibinfo {author} {\bibfnamefont {K.}~\bibnamefont
  {Miwa}}\ and\ \bibinfo {author} {\bibfnamefont {A.}~\bibnamefont
  {Fukumoto}},\ }\href {\doibase 10.1103/PhysRevB.65.155114} {\bibfield
  {journal} {\bibinfo  {journal} {Physical Review B}\ }\textbf {\bibinfo
  {volume} {65}},\ \bibinfo {pages} {155114} (\bibinfo {year}
  {2002})}\BibitemShut {NoStop}%
\bibitem [{\citenamefont {Wolverton}\ \emph {et~al.}(2004)\citenamefont
  {Wolverton}, \citenamefont {Ozoli\ifmmode \mbox{\c{n}}\else
  \c{n}\fi{}\ifmmode~\check{s}\else \v{s}\fi{}},\ and\ \citenamefont
  {Asta}}]{Wolverton_2004}%
  \BibitemOpen
  \bibfield  {author} {\bibinfo {author} {\bibfnamefont {C.}~\bibnamefont
  {Wolverton}}, \bibinfo {author} {\bibfnamefont {V.}~\bibnamefont
  {Ozoli\ifmmode \mbox{\c{n}}\else \c{n}\fi{}\ifmmode~\check{s}\else
  \v{s}\fi{}}}, \ and\ \bibinfo {author} {\bibfnamefont {M.}~\bibnamefont
  {Asta}},\ }\href {\doibase 10.1103/PhysRevB.69.144109} {\bibfield  {journal}
  {\bibinfo  {journal} {Physical Review B}\ }\textbf {\bibinfo {volume} {69}},\
  \bibinfo {pages} {144109} (\bibinfo {year} {2004})}\BibitemShut {NoStop}%
\bibitem [{\citenamefont {Tao}\ \emph {et~al.}(2009)\citenamefont {Tao},
  \citenamefont {Notten}, \citenamefont {van Santen},\ and\ \citenamefont
  {Jansen}}]{Tao_2009}%
  \BibitemOpen
  \bibfield  {author} {\bibinfo {author} {\bibfnamefont {S.~X.}\ \bibnamefont
  {Tao}}, \bibinfo {author} {\bibfnamefont {P.~H.~L.}\ \bibnamefont {Notten}},
  \bibinfo {author} {\bibfnamefont {R.~A.}\ \bibnamefont {van Santen}}, \ and\
  \bibinfo {author} {\bibfnamefont {A.~P.~J.}\ \bibnamefont {Jansen}},\ }\href
  {\doibase 10.1103/PhysRevB.79.144121} {\bibfield  {journal} {\bibinfo
  {journal} {Physical Review B}\ }\textbf {\bibinfo {volume} {79}},\ \bibinfo
  {pages} {144121} (\bibinfo {year} {2009})}\BibitemShut {NoStop}%
\bibitem [{\citenamefont {Ziani}\ and\ \citenamefont
  {Gueddim}(2021)}]{Ziani_2021}%
  \BibitemOpen
  \bibfield  {author} {\bibinfo {author} {\bibfnamefont {H.}~\bibnamefont
  {Ziani}}\ and\ \bibinfo {author} {\bibfnamefont {A.}~\bibnamefont
  {Gueddim}},\ }in\ \href {\doibase 10.1109/SIENR50924.2021.9631902} {\emph
  {\bibinfo {booktitle} {2020 6th International Symposium on New and Renewable
  Energy (SIENR)}}}\ (\bibinfo {year} {2021})\ pp.\ \bibinfo {pages}
  {1--4}\BibitemShut {NoStop}%
\bibitem [{\citenamefont {Ley}\ \emph {et~al.}(1977)\citenamefont {Ley},
  \citenamefont {Dabbousi}, \citenamefont {Kowalczyk}, \citenamefont
  {McFeely},\ and\ \citenamefont {Shirley}}]{Ley_1977}%
  \BibitemOpen
  \bibfield  {author} {\bibinfo {author} {\bibfnamefont {L.}~\bibnamefont
  {Ley}}, \bibinfo {author} {\bibfnamefont {O.~B.}\ \bibnamefont {Dabbousi}},
  \bibinfo {author} {\bibfnamefont {S.~P.}\ \bibnamefont {Kowalczyk}}, \bibinfo
  {author} {\bibfnamefont {F.~R.}\ \bibnamefont {McFeely}}, \ and\ \bibinfo
  {author} {\bibfnamefont {D.~A.}\ \bibnamefont {Shirley}},\ }\href {\doibase
  10.1103/PhysRevB.16.5372} {\bibfield  {journal} {\bibinfo  {journal}
  {Physical Review B}\ }\textbf {\bibinfo {volume} {16}},\ \bibinfo {pages}
  {5372} (\bibinfo {year} {1977})}\BibitemShut {NoStop}%
\bibitem [{\citenamefont {H{\"o}chst}\ \emph {et~al.}(1981)\citenamefont
  {H{\"o}chst}, \citenamefont {Steiner}, \citenamefont {Reiter},\ and\
  \citenamefont {H{\"u}fner}}]{Hochst1981}%
  \BibitemOpen
  \bibfield  {author} {\bibinfo {author} {\bibfnamefont {H.}~\bibnamefont
  {H{\"o}chst}}, \bibinfo {author} {\bibfnamefont {P.}~\bibnamefont {Steiner}},
  \bibinfo {author} {\bibfnamefont {G.}~\bibnamefont {Reiter}}, \ and\ \bibinfo
  {author} {\bibfnamefont {S.}~\bibnamefont {H{\"u}fner}},\ }\href {\doibase
  10.1007/BF01422023} {\bibfield  {journal} {\bibinfo  {journal} {Zeitschrift
  f{\"u}r Physik B Condensed Matter}\ }\textbf {\bibinfo {volume} {42}},\
  \bibinfo {pages} {199} (\bibinfo {year} {1981})}\BibitemShut {NoStop}%
\bibitem [{\citenamefont {Blaha}\ \emph {et~al.}(1988)\citenamefont {Blaha},
  \citenamefont {Schwarz},\ and\ \citenamefont {Dederichs}}]{Blaha_1988}%
  \BibitemOpen
  \bibfield  {author} {\bibinfo {author} {\bibfnamefont {P.}~\bibnamefont
  {Blaha}}, \bibinfo {author} {\bibfnamefont {K.}~\bibnamefont {Schwarz}}, \
  and\ \bibinfo {author} {\bibfnamefont {P.~H.}\ \bibnamefont {Dederichs}},\
  }\href {\doibase 10.1103/PhysRevB.38.9368} {\bibfield  {journal} {\bibinfo
  {journal} {Phys. Rev. B}\ }\textbf {\bibinfo {volume} {38}},\ \bibinfo
  {pages} {9368} (\bibinfo {year} {1988})}\BibitemShut {NoStop}%
\bibitem [{\citenamefont {Tanaka}\ \emph {et~al.}(1990)\citenamefont {Tanaka},
  \citenamefont {Ushida}, \citenamefont {Sumiyama},\ and\ \citenamefont
  {Nakamura}}]{TANAKA1990429}%
  \BibitemOpen
  \bibfield  {author} {\bibinfo {author} {\bibfnamefont {K.}~\bibnamefont
  {Tanaka}}, \bibinfo {author} {\bibfnamefont {M.}~\bibnamefont {Ushida}},
  \bibinfo {author} {\bibfnamefont {K.}~\bibnamefont {Sumiyama}}, \ and\
  \bibinfo {author} {\bibfnamefont {Y.}~\bibnamefont {Nakamura}},\ }\href@noop
  {} {\bibfield  {journal} {\bibinfo  {journal} {Journal of Non-Crystalline
  Solids}\ }\textbf {\bibinfo {volume} {117-118}},\ \bibinfo {pages} {429}
  (\bibinfo {year} {1990})}\BibitemShut {NoStop}%
\bibitem [{\citenamefont {Schlueter}\ \emph {et~al.}(2019)\citenamefont
  {Schlueter}, \citenamefont {Gloskovskii}, \citenamefont {Ederer},
  \citenamefont {Schostak}, \citenamefont {Piec}, \citenamefont {Sarkar},
  \citenamefont {Matveyev}, \citenamefont {L{\"{o}}mker}, \citenamefont {Sing},
  \citenamefont {Claessen}, \citenamefont {Wiemann}, \citenamefont {Schneider},
  \citenamefont {Medjanik}, \citenamefont {Sch{\"{o}}nhense}, \citenamefont
  {Amann}, \citenamefont {Nilsson},\ and\ \citenamefont
  {Drube}}]{Schlueter2019ThePETRAIII}%
  \BibitemOpen
  \bibfield  {author} {\bibinfo {author} {\bibfnamefont {C.}~\bibnamefont
  {Schlueter}}, \bibinfo {author} {\bibfnamefont {A.}~\bibnamefont
  {Gloskovskii}}, \bibinfo {author} {\bibfnamefont {K.}~\bibnamefont {Ederer}},
  \bibinfo {author} {\bibfnamefont {I.}~\bibnamefont {Schostak}}, \bibinfo
  {author} {\bibfnamefont {S.}~\bibnamefont {Piec}}, \bibinfo {author}
  {\bibfnamefont {I.}~\bibnamefont {Sarkar}}, \bibinfo {author} {\bibfnamefont
  {Y.}~\bibnamefont {Matveyev}}, \bibinfo {author} {\bibfnamefont
  {P.}~\bibnamefont {L{\"{o}}mker}}, \bibinfo {author} {\bibfnamefont
  {M.}~\bibnamefont {Sing}}, \bibinfo {author} {\bibfnamefont {R.}~\bibnamefont
  {Claessen}}, \bibinfo {author} {\bibfnamefont {C.}~\bibnamefont {Wiemann}},
  \bibinfo {author} {\bibfnamefont {C.~M.}\ \bibnamefont {Schneider}}, \bibinfo
  {author} {\bibfnamefont {K.}~\bibnamefont {Medjanik}}, \bibinfo {author}
  {\bibfnamefont {G.}~\bibnamefont {Sch{\"{o}}nhense}}, \bibinfo {author}
  {\bibfnamefont {P.}~\bibnamefont {Amann}}, \bibinfo {author} {\bibfnamefont
  {A.}~\bibnamefont {Nilsson}}, \ and\ \bibinfo {author} {\bibfnamefont
  {W.}~\bibnamefont {Drube}},\ }\href@noop {} {\bibfield  {journal} {\bibinfo
  {journal} {AIP Conference Proceedings}\ }\textbf {\bibinfo {volume} {2054}}
  (\bibinfo {year} {2019})}\BibitemShut {NoStop}%
\bibitem [{\citenamefont {Hohenberg}\ and\ \citenamefont
  {Kohn}(1964)}]{Hohenberg1964}%
  \BibitemOpen
  \bibfield  {author} {\bibinfo {author} {\bibfnamefont {P.}~\bibnamefont
  {Hohenberg}}\ and\ \bibinfo {author} {\bibfnamefont {W.}~\bibnamefont
  {Kohn}},\ }\href@noop {} {\bibfield  {journal} {\bibinfo  {journal} {Physical
  Review}\ }\textbf {\bibinfo {volume} {136}},\ \bibinfo {pages} {864}
  (\bibinfo {year} {1964})}\BibitemShut {NoStop}%
\bibitem [{\citenamefont {Kohn}\ and\ \citenamefont {Sham}(1965)}]{Kohn1965}%
  \BibitemOpen
  \bibfield  {author} {\bibinfo {author} {\bibfnamefont {W.}~\bibnamefont
  {Kohn}}\ and\ \bibinfo {author} {\bibfnamefont {L.~J.}\ \bibnamefont
  {Sham}},\ }\href@noop {} {\bibfield  {journal} {\bibinfo  {journal} {Physical
  Review}\ }\textbf {\bibinfo {volume} {140}},\ \bibinfo {pages} {1133}
  (\bibinfo {year} {1965})}\BibitemShut {NoStop}%
\bibitem [{\citenamefont {Clark}\ \emph {et~al.}(2005)\citenamefont {Clark},
  \citenamefont {Segall}, \citenamefont {Pickard~Ii}, \citenamefont {Hasnip},
  \citenamefont {Probert}, \citenamefont {Refson},\ and\ \citenamefont
  {Payne}}]{Clark2005}%
  \BibitemOpen
  \bibfield  {author} {\bibinfo {author} {\bibfnamefont {S.~J.}\ \bibnamefont
  {Clark}}, \bibinfo {author} {\bibfnamefont {M.~D.}\ \bibnamefont {Segall}},
  \bibinfo {author} {\bibfnamefont {C.~J.}\ \bibnamefont {Pickard~Ii}},
  \bibinfo {author} {\bibfnamefont {P.~J.}\ \bibnamefont {Hasnip}}, \bibinfo
  {author} {\bibfnamefont {M.~I.~J.}\ \bibnamefont {Probert}}, \bibinfo
  {author} {\bibfnamefont {K.}~\bibnamefont {Refson}}, \ and\ \bibinfo {author}
  {\bibfnamefont {M.~C.}\ \bibnamefont {Payne}},\ }\href
  {www.accelrys.com/references/castep/.} {\bibfield  {journal} {\bibinfo
  {journal} {Zeitschrift f{\"{u}}r Kristallographie - Crystalline Materials}\
  }\textbf {\bibinfo {volume} {220}},\ \bibinfo {pages} {567} (\bibinfo {year}
  {2005})}\BibitemShut {NoStop}%
\bibitem [{\citenamefont {Perdew}\ \emph {et~al.}(1996)\citenamefont {Perdew},
  \citenamefont {Burke},\ and\ \citenamefont {Ernzerhof}}]{Perdew1996}%
  \BibitemOpen
  \bibfield  {author} {\bibinfo {author} {\bibfnamefont {J.~P.}\ \bibnamefont
  {Perdew}}, \bibinfo {author} {\bibfnamefont {K.}~\bibnamefont {Burke}}, \
  and\ \bibinfo {author} {\bibfnamefont {M.}~\bibnamefont {Ernzerhof}},\
  }\href@noop {} {\bibfield  {journal} {\bibinfo  {journal} {Physical Review
  Letters}\ }\textbf {\bibinfo {volume} {77}},\ \bibinfo {pages} {3865}
  (\bibinfo {year} {1996})}\BibitemShut {NoStop}%
\bibitem [{\citenamefont {Monkhorst}\ and\ \citenamefont
  {Pack}(1976)}]{Monkhorst1976}%
  \BibitemOpen
  \bibfield  {author} {\bibinfo {author} {\bibfnamefont {H.~J.}\ \bibnamefont
  {Monkhorst}}\ and\ \bibinfo {author} {\bibfnamefont {J.~D.}\ \bibnamefont
  {Pack}},\ }\href@noop {} {\bibfield  {journal} {\bibinfo  {journal} {Physical
  Review B}\ }\textbf {\bibinfo {volume} {13}},\ \bibinfo {pages} {5188}
  (\bibinfo {year} {1976})}\BibitemShut {NoStop}%
\bibitem [{\citenamefont {Adamo}\ and\ \citenamefont
  {Barone}(1999)}]{Adamo1999}%
  \BibitemOpen
  \bibfield  {author} {\bibinfo {author} {\bibfnamefont {C.}~\bibnamefont
  {Adamo}}\ and\ \bibinfo {author} {\bibfnamefont {V.}~\bibnamefont {Barone}},\
  }\href {\doibase 10.1063/1.478522} {\bibfield  {journal} {\bibinfo  {journal}
  {Journal of Chemical Physics}\ }\textbf {\bibinfo {volume} {110}},\ \bibinfo
  {pages} {6158} (\bibinfo {year} {1999})}\BibitemShut {NoStop}%
\bibitem [{\citenamefont {Mulliken}(1955)}]{Mulliken1955}%
  \BibitemOpen
  \bibfield  {author} {\bibinfo {author} {\bibfnamefont {R.~S.}\ \bibnamefont
  {Mulliken}},\ }\href {\doibase 10.1063/1.1740588} {\bibfield  {journal}
  {\bibinfo  {journal} {The Journal of Chemical Physics}\ }\textbf {\bibinfo
  {volume} {23}},\ \bibinfo {pages} {1833} (\bibinfo {year}
  {1955})}\BibitemShut {NoStop}%
\bibitem [{\citenamefont {Hirshfeld}(1977)}]{Hirshfeld1977}%
  \BibitemOpen
  \bibfield  {author} {\bibinfo {author} {\bibfnamefont {F.~L.}\ \bibnamefont
  {Hirshfeld}},\ }\href {\doibase 10.1007/BF00549096} {\bibfield  {journal}
  {\bibinfo  {journal} {Theoretica Chimica Acta}\ }\textbf {\bibinfo {volume}
  {44}},\ \bibinfo {pages} {129} (\bibinfo {year} {1977})}\BibitemShut
  {NoStop}%
\bibitem [{\citenamefont {Bader}(1990)}]{Bader1990}%
  \BibitemOpen
  \bibfield  {author} {\bibinfo {author} {\bibfnamefont {R.~F.~W.}\
  \bibnamefont {Bader}},\ }\href@noop {} {\emph {\bibinfo {title} {Atoms in
  Molecules}}}\ (\bibinfo  {publisher} {Oxford University Press: New York},\
  \bibinfo {year} {1990})\BibitemShut {NoStop}%
\bibitem [{\citenamefont {Arnaldsson}\ \emph {et~al.}(2022)\citenamefont
  {Arnaldsson}, \citenamefont {Tang}, \citenamefont {Chill}, \citenamefont
  {Chai}, \citenamefont {Anselm},\ and\ \citenamefont
  {Henkelman}}]{bader_code}%
  \BibitemOpen
  \bibfield  {author} {\bibinfo {author} {\bibfnamefont {A.}~\bibnamefont
  {Arnaldsson}}, \bibinfo {author} {\bibfnamefont {W.}~\bibnamefont {Tang}},
  \bibinfo {author} {\bibfnamefont {S.}~\bibnamefont {Chill}}, \bibinfo
  {author} {\bibfnamefont {W.}~\bibnamefont {Chai}}, \bibinfo {author}
  {\bibfnamefont {R.}~\bibnamefont {Anselm}}, \ and\ \bibinfo {author}
  {\bibfnamefont {G.}~\bibnamefont {Henkelman}},\ }\href
  {http://theory.cm.utexas.edu/henkelman/code/bader} {\enquote {\bibinfo
  {title} {Bader charge analysis code},}\ } (\bibinfo {year}
  {2022})\BibitemShut {NoStop}%
\bibitem [{\citenamefont {Morris}\ \emph {et~al.}(2014)\citenamefont {Morris},
  \citenamefont {Nicholls}, \citenamefont {Pickard},\ and\ \citenamefont
  {Yates}}]{Morris2014}%
  \BibitemOpen
  \bibfield  {author} {\bibinfo {author} {\bibfnamefont {A.~J.}\ \bibnamefont
  {Morris}}, \bibinfo {author} {\bibfnamefont {R.~J.}\ \bibnamefont
  {Nicholls}}, \bibinfo {author} {\bibfnamefont {C.~J.}\ \bibnamefont
  {Pickard}}, \ and\ \bibinfo {author} {\bibfnamefont {J.~R.}\ \bibnamefont
  {Yates}},\ }\href {\doibase 10.1016/j.cpc.2014.02.013} {\bibfield  {journal}
  {\bibinfo  {journal} {Computer Physics Communications}\ }\textbf {\bibinfo
  {volume} {185}},\ \bibinfo {pages} {1477} (\bibinfo {year}
  {2014})}\BibitemShut {NoStop}%
\bibitem [{\citenamefont {Scofield}(1973)}]{Scofield1973}%
  \BibitemOpen
  \bibfield  {author} {\bibinfo {author} {\bibfnamefont {J.~H.}\ \bibnamefont
  {Scofield}},\ }\href@noop {} {\emph {\bibinfo {title} {{Theoretical
  Photoionization Cross Sections from 1 to 1500 keV}}}},\ \bibinfo {type}
  {Tech. Rep.}\ (\bibinfo  {institution} {Lawrence Livermore Laboratory},\
  \bibinfo {year} {1973})\BibitemShut {NoStop}%
\bibitem [{\citenamefont {Kalha}\ \emph {et~al.}(2020)\citenamefont {Kalha},
  \citenamefont {Fernando},\ and\ \citenamefont {Regoutz}}]{Kalha20}%
  \BibitemOpen
  \bibfield  {author} {\bibinfo {author} {\bibfnamefont {C.}~\bibnamefont
  {Kalha}}, \bibinfo {author} {\bibfnamefont {N.}~\bibnamefont {Fernando}}, \
  and\ \bibinfo {author} {\bibfnamefont {A.}~\bibnamefont {Regoutz}},\ }\href
  {https://figshare.com/articles/dataset/Digitisation_of_Scofield_Photoionisation_Cross_Section_Tabulated_Data/12967079}
  {\enquote {\bibinfo {title} {{Digitisation of Scofield Photoionisation Cross
  Section Tabulated Data}},}\ } (\bibinfo {year} {2020}),\ \bibinfo {note}
  {10.6084/m9.figshare.12967079.v1}\BibitemShut {NoStop}%
\bibitem [{\citenamefont {Jackson}\ \emph {et~al.}(2018)\citenamefont
  {Jackson}, \citenamefont {Ganose}, \citenamefont {Regoutz}, \citenamefont
  {Egdell},\ and\ \citenamefont {Scanlon}}]{Jackson2018}%
  \BibitemOpen
  \bibfield  {author} {\bibinfo {author} {\bibfnamefont {A.~J.}\ \bibnamefont
  {Jackson}}, \bibinfo {author} {\bibfnamefont {A.~M.}\ \bibnamefont {Ganose}},
  \bibinfo {author} {\bibfnamefont {A.}~\bibnamefont {Regoutz}}, \bibinfo
  {author} {\bibfnamefont {R.~G.}\ \bibnamefont {Egdell}}, \ and\ \bibinfo
  {author} {\bibfnamefont {D.~O.}\ \bibnamefont {Scanlon}},\ }\href {\doibase
  10.21105/joss.00773} {\bibfield  {journal} {\bibinfo  {journal} {Journal of
  Open Source Software}\ }\textbf {\bibinfo {volume} {3}},\ \bibinfo {pages}
  {773} (\bibinfo {year} {2018})}\BibitemShut {NoStop}%
\bibitem [{\citenamefont {Regoutz}(2016)}]{regoutz2016a}%
  \BibitemOpen
  \bibfield  {author} {\bibinfo {author} {\bibfnamefont {A.}~\bibnamefont
  {Regoutz}},\ }\emph {\bibinfo {title} {Structural and electronic properties
  of metal oxides}},\ \href@noop {} {Ph.D. thesis},\ \bibinfo  {school} {Oxford
  University, UK} (\bibinfo {year} {2016})\BibitemShut {NoStop}%
\end{thebibliography}%


%


\begin{thebibliography}{37}%
\makeatletter
\providecommand \@ifxundefined [1]{%
 \@ifx{#1\undefined}
}%
\providecommand \@ifnum [1]{%
 \ifnum #1\expandafter \@firstoftwo
 \else \expandafter \@secondoftwo
 \fi
}%
\providecommand \@ifx [1]{%
 \ifx #1\expandafter \@firstoftwo
 \else \expandafter \@secondoftwo
 \fi
}%
\providecommand \natexlab [1]{#1}%
\providecommand \enquote  [1]{``#1''}%
\providecommand \bibnamefont  [1]{#1}%
\providecommand \bibfnamefont [1]{#1}%
\providecommand \citenamefont [1]{#1}%
\providecommand \href@noop [0]{\@secondoftwo}%
\providecommand \href [0]{\begingroup \@sanitize@url \@href}%
\providecommand \@href[1]{\@@startlink{#1}\@@href}%
\providecommand \@@href[1]{\endgroup#1\@@endlink}%
\providecommand \@sanitize@url [0]{\catcode `\\12\catcode `\$12\catcode
  `\&12\catcode `\#12\catcode `\^12\catcode `\_12\catcode `\%12\relax}%
\providecommand \@@startlink[1]{}%
\providecommand \@@endlink[0]{}%
\providecommand \url  [0]{\begingroup\@sanitize@url \@url }%
\providecommand \@url [1]{\endgroup\@href {#1}{\urlprefix }}%
\providecommand \urlprefix  [0]{URL }%
\providecommand \Eprint [0]{\href }%
\providecommand \doibase [0]{http://dx.doi.org/}%
\providecommand \selectlanguage [0]{\@gobble}%
\providecommand \bibinfo  [0]{\@secondoftwo}%
\providecommand \bibfield  [0]{\@secondoftwo}%
\providecommand \translation [1]{[#1]}%
\providecommand \BibitemOpen [0]{}%
\providecommand \bibitemStop [0]{}%
\providecommand \bibitemNoStop [0]{.\EOS\space}%
\providecommand \EOS [0]{\spacefactor3000\relax}%
\providecommand \BibitemShut  [1]{\csname bibitem#1\endcsname}%
\let\auto@bib@innerbib\@empty
\bibitem [{\citenamefont {Fujimori}\ and\ \citenamefont
  {Schlapbach}(1984)}]{Fujimori_1984}%
  \BibitemOpen
  \bibfield  {author} {\bibinfo {author} {\bibfnamefont {A.}~\bibnamefont
  {Fujimori}}\ and\ \bibinfo {author} {\bibfnamefont {L.}~\bibnamefont
  {Schlapbach}},\ }\href@noop {} {\bibfield  {journal} {\bibinfo  {journal}
  {Journal of Physics C: Solid State Physics}\ }\textbf {\bibinfo {volume}
  {17}},\ \bibinfo {pages} {341} (\bibinfo {year} {1984})}\BibitemShut
  {NoStop}%
\bibitem [{\citenamefont {Lamartine}\ \emph {et~al.}(1980)\citenamefont
  {Lamartine}, \citenamefont {Haas},\ and\ \citenamefont
  {Solomon}}]{LAMARTINE1980537}%
  \BibitemOpen
  \bibfield  {author} {\bibinfo {author} {\bibfnamefont {B.}~\bibnamefont
  {Lamartine}}, \bibinfo {author} {\bibfnamefont {T.}~\bibnamefont {Haas}}, \
  and\ \bibinfo {author} {\bibfnamefont {J.}~\bibnamefont {Solomon}},\ }\href
  {\doibase https://doi.org/10.1016/0378-5963(80)90097-5} {\bibfield  {journal}
  {\bibinfo  {journal} {Applications of Surface Science}\ }\textbf {\bibinfo
  {volume} {4}},\ \bibinfo {pages} {537} (\bibinfo {year} {1980})}\BibitemShut
  {NoStop}%
\bibitem [{\citenamefont {Hayoz}\ \emph {et~al.}(2000)\citenamefont {Hayoz},
  \citenamefont {Pillo}, \citenamefont {Bovet}, \citenamefont {Z{\"{u}}ttel},
  \citenamefont {Guthrie}, \citenamefont {Pastore}, \citenamefont
  {Schlapbach},\ and\ \citenamefont {Aebi}}]{Hayoz_2000}%
  \BibitemOpen
  \bibfield  {author} {\bibinfo {author} {\bibfnamefont {J.}~\bibnamefont
  {Hayoz}}, \bibinfo {author} {\bibfnamefont {T.}~\bibnamefont {Pillo}},
  \bibinfo {author} {\bibfnamefont {M.}~\bibnamefont {Bovet}}, \bibinfo
  {author} {\bibfnamefont {A.}~\bibnamefont {Z{\"{u}}ttel}}, \bibinfo {author}
  {\bibfnamefont {S.}~\bibnamefont {Guthrie}}, \bibinfo {author} {\bibfnamefont
  {G.}~\bibnamefont {Pastore}}, \bibinfo {author} {\bibfnamefont
  {L.}~\bibnamefont {Schlapbach}}, \ and\ \bibinfo {author} {\bibfnamefont
  {P.}~\bibnamefont {Aebi}},\ }\href {\doibase 10.1116/1.1286073} {\bibfield
  {journal} {\bibinfo  {journal} {Journal of Vacuum Science {\&} Technology A:
  Vacuum, Surfaces, and Films}\ }\textbf {\bibinfo {volume} {18}},\ \bibinfo
  {pages} {2417} (\bibinfo {year} {2000})}\BibitemShut {NoStop}%
\bibitem [{\citenamefont {Mongstad}\ \emph {et~al.}(2014)\citenamefont
  {Mongstad}, \citenamefont {Th{\o}gersen}, \citenamefont {Subrahmanyam},\ and\
  \citenamefont {Karazhanov}}]{Mongstad2014TheOxide}%
  \BibitemOpen
  \bibfield  {author} {\bibinfo {author} {\bibfnamefont {T.}~\bibnamefont
  {Mongstad}}, \bibinfo {author} {\bibfnamefont {A.}~\bibnamefont
  {Th{\o}gersen}}, \bibinfo {author} {\bibfnamefont {A.}~\bibnamefont
  {Subrahmanyam}}, \ and\ \bibinfo {author} {\bibfnamefont {S.}~\bibnamefont
  {Karazhanov}},\ }\href {\doibase 10.1016/j.solmat.2014.05.037} {\bibfield
  {journal} {\bibinfo  {journal} {Solar Energy Materials and Solar Cells}\
  }\textbf {\bibinfo {volume} {128}},\ \bibinfo {pages} {270} (\bibinfo {year}
  {2014})}\BibitemShut {NoStop}%
\bibitem [{\citenamefont {Majumdar}\ and\ \citenamefont
  {Chatterjee}(1991)}]{Majumdar_1991}%
  \BibitemOpen
  \bibfield  {author} {\bibinfo {author} {\bibfnamefont {D.}~\bibnamefont
  {Majumdar}}\ and\ \bibinfo {author} {\bibfnamefont {D.}~\bibnamefont
  {Chatterjee}},\ }\href {\doibase 10.1063/1.349611} {\bibfield  {journal}
  {\bibinfo  {journal} {Journal of Applied Physics}\ }\textbf {\bibinfo
  {volume} {70}},\ \bibinfo {pages} {988} (\bibinfo {year} {1991})}\BibitemShut
  {NoStop}%
\bibitem [{\citenamefont {Uwamino}\ \emph {et~al.}(1984)\citenamefont
  {Uwamino}, \citenamefont {Ishizuka},\ and\ \citenamefont
  {Yamatera}}]{UWAMINO198467}%
  \BibitemOpen
  \bibfield  {author} {\bibinfo {author} {\bibfnamefont {Y.}~\bibnamefont
  {Uwamino}}, \bibinfo {author} {\bibfnamefont {T.}~\bibnamefont {Ishizuka}}, \
  and\ \bibinfo {author} {\bibfnamefont {H.}~\bibnamefont {Yamatera}},\ }\href
  {\doibase https://doi.org/10.1016/0368-2048(84)80060-2} {\bibfield  {journal}
  {\bibinfo  {journal} {Journal of Electron Spectroscopy and Related
  Phenomena}\ }\textbf {\bibinfo {volume} {34}},\ \bibinfo {pages} {67}
  (\bibinfo {year} {1984})}\BibitemShut {NoStop}%
\bibitem [{\citenamefont {Mitrovic}\ \emph {et~al.}(2014)\citenamefont
  {Mitrovic}, \citenamefont {Althobaiti}, \citenamefont {Weerakkody},
  \citenamefont {Dhanak}, \citenamefont {Linhart}, \citenamefont {Veal},
  \citenamefont {Sedghi}, \citenamefont {Hall}, \citenamefont {Chalker},
  \citenamefont {Tsoutsou},\ and\ \citenamefont
  {Dimoulas}}]{Mitrovic_Veal_2014}%
  \BibitemOpen
  \bibfield  {author} {\bibinfo {author} {\bibfnamefont {I.~Z.}\ \bibnamefont
  {Mitrovic}}, \bibinfo {author} {\bibfnamefont {M.}~\bibnamefont
  {Althobaiti}}, \bibinfo {author} {\bibfnamefont {A.~D.}\ \bibnamefont
  {Weerakkody}}, \bibinfo {author} {\bibfnamefont {V.~R.}\ \bibnamefont
  {Dhanak}}, \bibinfo {author} {\bibfnamefont {W.~M.}\ \bibnamefont {Linhart}},
  \bibinfo {author} {\bibfnamefont {T.~D.}\ \bibnamefont {Veal}}, \bibinfo
  {author} {\bibfnamefont {N.}~\bibnamefont {Sedghi}}, \bibinfo {author}
  {\bibfnamefont {S.}~\bibnamefont {Hall}}, \bibinfo {author} {\bibfnamefont
  {P.~R.}\ \bibnamefont {Chalker}}, \bibinfo {author} {\bibfnamefont
  {D.}~\bibnamefont {Tsoutsou}}, \ and\ \bibinfo {author} {\bibfnamefont
  {A.}~\bibnamefont {Dimoulas}},\ }\href {\doibase 10.1063/1.4868091}
  {\bibfield  {journal} {\bibinfo  {journal} {Journal of Applied Physics}\
  }\textbf {\bibinfo {volume} {115}},\ \bibinfo {pages} {114102} (\bibinfo
  {year} {2014})}\BibitemShut {NoStop}%
\bibitem [{\citenamefont {Ingo}\ \emph {et~al.}(1990)\citenamefont {Ingo},
  \citenamefont {Paparazzo}, \citenamefont {Bagnarelli},\ and\ \citenamefont
  {Zacchetti}}]{Ingo_1990}%
  \BibitemOpen
  \bibfield  {author} {\bibinfo {author} {\bibfnamefont {G.~M.}\ \bibnamefont
  {Ingo}}, \bibinfo {author} {\bibfnamefont {E.}~\bibnamefont {Paparazzo}},
  \bibinfo {author} {\bibfnamefont {O.}~\bibnamefont {Bagnarelli}}, \ and\
  \bibinfo {author} {\bibfnamefont {N.}~\bibnamefont {Zacchetti}},\ }\href
  {\doibase https://doi.org/10.1002/sia.7401601107} {\bibfield  {journal}
  {\bibinfo  {journal} {Surface and Interface Analysis}\ }\textbf {\bibinfo
  {volume} {16}},\ \bibinfo {pages} {515} (\bibinfo {year} {1990})}\BibitemShut
  {NoStop}%
\bibitem [{\citenamefont {Barreca}\ \emph {et~al.}(2001)\citenamefont
  {Barreca}, \citenamefont {Battiston}, \citenamefont {Berto}, \citenamefont
  {Gerbasi},\ and\ \citenamefont {Tondello}}]{Barreca_2001}%
  \BibitemOpen
  \bibfield  {author} {\bibinfo {author} {\bibfnamefont {D.}~\bibnamefont
  {Barreca}}, \bibinfo {author} {\bibfnamefont {G.~A.}\ \bibnamefont
  {Battiston}}, \bibinfo {author} {\bibfnamefont {D.}~\bibnamefont {Berto}},
  \bibinfo {author} {\bibfnamefont {R.}~\bibnamefont {Gerbasi}}, \ and\
  \bibinfo {author} {\bibfnamefont {E.}~\bibnamefont {Tondello}},\ }\href
  {\doibase 10.1116/11.20020404} {\bibfield  {journal} {\bibinfo  {journal}
  {Surface Science Spectra}\ }\textbf {\bibinfo {volume} {8}},\ \bibinfo
  {pages} {234} (\bibinfo {year} {2001})}\BibitemShut {NoStop}%
\bibitem [{\citenamefont {Reichl}\ and\ \citenamefont
  {Gaukler}(1986)}]{REICHL1986196}%
  \BibitemOpen
  \bibfield  {author} {\bibinfo {author} {\bibfnamefont {R.}~\bibnamefont
  {Reichl}}\ and\ \bibinfo {author} {\bibfnamefont {K.}~\bibnamefont
  {Gaukler}},\ }\href {\doibase https://doi.org/10.1016/0169-4332(86)90005-X}
  {\bibfield  {journal} {\bibinfo  {journal} {Applied Surface Science}\
  }\textbf {\bibinfo {volume} {26}},\ \bibinfo {pages} {196} (\bibinfo {year}
  {1986})}\BibitemShut {NoStop}%
\bibitem [{\citenamefont {Biwer}\ and\ \citenamefont
  {Bernasek}(1986)}]{BIWER1986207}%
  \BibitemOpen
  \bibfield  {author} {\bibinfo {author} {\bibfnamefont {B.}~\bibnamefont
  {Biwer}}\ and\ \bibinfo {author} {\bibfnamefont {S.}~\bibnamefont
  {Bernasek}},\ }\href {\doibase https://doi.org/10.1016/0039-6028(86)90795-8}
  {\bibfield  {journal} {\bibinfo  {journal} {Surface Science}\ }\textbf
  {\bibinfo {volume} {167}},\ \bibinfo {pages} {207} (\bibinfo {year}
  {1986})}\BibitemShut {NoStop}%
\bibitem [{\citenamefont {Sleigh}\ \emph {et~al.}(1996)\citenamefont {Sleigh},
  \citenamefont {Pijpers}, \citenamefont {Jaspers}, \citenamefont {Coussens},\
  and\ \citenamefont {Meier}}]{SLEIGH199641}%
  \BibitemOpen
  \bibfield  {author} {\bibinfo {author} {\bibfnamefont {C.}~\bibnamefont
  {Sleigh}}, \bibinfo {author} {\bibfnamefont {A.}~\bibnamefont {Pijpers}},
  \bibinfo {author} {\bibfnamefont {A.}~\bibnamefont {Jaspers}}, \bibinfo
  {author} {\bibfnamefont {B.}~\bibnamefont {Coussens}}, \ and\ \bibinfo
  {author} {\bibfnamefont {R.~J.}\ \bibnamefont {Meier}},\ }\href {\doibase
  https://doi.org/10.1016/0368-2048(95)02392-5} {\bibfield  {journal} {\bibinfo
   {journal} {Journal of Electron Spectroscopy and Related Phenomena}\ }\textbf
  {\bibinfo {volume} {77}},\ \bibinfo {pages} {41} (\bibinfo {year}
  {1996})}\BibitemShut {NoStop}%
\bibitem [{\citenamefont {Kaciulis}\ \emph {et~al.}(2018)\citenamefont
  {Kaciulis}, \citenamefont {Soltani}, \citenamefont {Mezzi}, \citenamefont
  {Montanari}, \citenamefont {Lapi}, \citenamefont {Richetta}, \citenamefont
  {Varone},\ and\ \citenamefont {Barbieri}}]{Kaciulis_2018}%
  \BibitemOpen
  \bibfield  {author} {\bibinfo {author} {\bibfnamefont {S.}~\bibnamefont
  {Kaciulis}}, \bibinfo {author} {\bibfnamefont {P.}~\bibnamefont {Soltani}},
  \bibinfo {author} {\bibfnamefont {A.}~\bibnamefont {Mezzi}}, \bibinfo
  {author} {\bibfnamefont {R.}~\bibnamefont {Montanari}}, \bibinfo {author}
  {\bibfnamefont {G.}~\bibnamefont {Lapi}}, \bibinfo {author} {\bibfnamefont
  {M.}~\bibnamefont {Richetta}}, \bibinfo {author} {\bibfnamefont
  {A.}~\bibnamefont {Varone}}, \ and\ \bibinfo {author} {\bibfnamefont
  {G.}~\bibnamefont {Barbieri}},\ }\href {\doibase
  https://doi.org/10.1002/sia.6428} {\bibfield  {journal} {\bibinfo  {journal}
  {Surface and Interface Analysis}\ }\textbf {\bibinfo {volume} {50}},\
  \bibinfo {pages} {1195} (\bibinfo {year} {2018})}\BibitemShut {NoStop}%
\bibitem [{\citenamefont {Ren}\ \emph {et~al.}(2014)\citenamefont {Ren},
  \citenamefont {Wang}, \citenamefont {Liu},\ and\ \citenamefont
  {Ohachi}}]{Ren_2014}%
  \BibitemOpen
  \bibfield  {author} {\bibinfo {author} {\bibfnamefont {N.}~\bibnamefont
  {Ren}}, \bibinfo {author} {\bibfnamefont {G.}~\bibnamefont {Wang}}, \bibinfo
  {author} {\bibfnamefont {H.}~\bibnamefont {Liu}}, \ and\ \bibinfo {author}
  {\bibfnamefont {T.}~\bibnamefont {Ohachi}},\ }\href {\doibase
  10.1016/j.materresbull.2013.11.002} {\bibfield  {journal} {\bibinfo
  {journal} {Materials Research Bulletin}\ }\textbf {\bibinfo {volume} {50}},\
  \bibinfo {pages} {379} (\bibinfo {year} {2014})}\BibitemShut {NoStop}%
\bibitem [{\citenamefont {Ma}\ \emph {et~al.}(2009)\citenamefont {Ma},
  \citenamefont {Kang}, \citenamefont {Dai}, \citenamefont {Liang},
  \citenamefont {Fang}, \citenamefont {Wang}, \citenamefont {Wang},\ and\
  \citenamefont {Cheng}}]{MA20092250}%
  \BibitemOpen
  \bibfield  {author} {\bibinfo {author} {\bibfnamefont {L.-P.}\ \bibnamefont
  {Ma}}, \bibinfo {author} {\bibfnamefont {X.-D.}\ \bibnamefont {Kang}},
  \bibinfo {author} {\bibfnamefont {H.-B.}\ \bibnamefont {Dai}}, \bibinfo
  {author} {\bibfnamefont {Y.}~\bibnamefont {Liang}}, \bibinfo {author}
  {\bibfnamefont {Z.-Z.}\ \bibnamefont {Fang}}, \bibinfo {author}
  {\bibfnamefont {P.-J.}\ \bibnamefont {Wang}}, \bibinfo {author}
  {\bibfnamefont {P.}~\bibnamefont {Wang}}, \ and\ \bibinfo {author}
  {\bibfnamefont {H.-M.}\ \bibnamefont {Cheng}},\ }\href {\doibase
  https://doi.org/10.1016/j.actamat.2009.01.025} {\bibfield  {journal}
  {\bibinfo  {journal} {Acta Materialia}\ }\textbf {\bibinfo {volume} {57}},\
  \bibinfo {pages} {2250} (\bibinfo {year} {2009})}\BibitemShut {NoStop}%
\bibitem [{\citenamefont {Gonzalez-Elipe}\ \emph {et~al.}(1989)\citenamefont
  {Gonzalez-Elipe}, \citenamefont {Munuera}, \citenamefont {Espinos},\ and\
  \citenamefont {Sanz}}]{Gonzalez_1989}%
  \BibitemOpen
  \bibfield  {author} {\bibinfo {author} {\bibfnamefont {A.~R.}\ \bibnamefont
  {Gonzalez-Elipe}}, \bibinfo {author} {\bibfnamefont {G.}~\bibnamefont
  {Munuera}}, \bibinfo {author} {\bibfnamefont {J.~P.}\ \bibnamefont
  {Espinos}}, \ and\ \bibinfo {author} {\bibfnamefont {J.~M.}\ \bibnamefont
  {Sanz}},\ }\href@noop {} {\bibfield  {journal} {\bibinfo  {journal} {Surface
  Science}\ }\textbf {\bibinfo {volume} {220}},\ \bibinfo {pages} {368}
  (\bibinfo {year} {1989})}\BibitemShut {NoStop}%
\bibitem [{\citenamefont {Diebold}(2003)}]{DIEBOLD200353}%
  \BibitemOpen
  \bibfield  {author} {\bibinfo {author} {\bibfnamefont {U.}~\bibnamefont
  {Diebold}},\ }\href {\doibase https://doi.org/10.1016/S0167-5729(02)00100-0}
  {\bibfield  {journal} {\bibinfo  {journal} {Surface Science Reports}\
  }\textbf {\bibinfo {volume} {48}},\ \bibinfo {pages} {53} (\bibinfo {year}
  {2003})}\BibitemShut {NoStop}%
\bibitem [{\citenamefont {Riesterer}(1987)}]{Riesterer1987}%
  \BibitemOpen
  \bibfield  {author} {\bibinfo {author} {\bibfnamefont {T.}~\bibnamefont
  {Riesterer}},\ }\href@noop {} {\bibfield  {journal} {\bibinfo  {journal}
  {Zeitschrift f{\"{u}}r Physik B Condensed Matter}\ }\textbf {\bibinfo
  {volume} {66}},\ \bibinfo {pages} {441} (\bibinfo {year} {1987})}\BibitemShut
  {NoStop}%
\bibitem [{\citenamefont {Scofield}(1973)}]{Scofield1973}%
  \BibitemOpen
  \bibfield  {author} {\bibinfo {author} {\bibfnamefont {J.~H.}\ \bibnamefont
  {Scofield}},\ }\href@noop {} {\emph {\bibinfo {title} {{Theoretical
  Photoionization Cross Sections from 1 to 1500 keV}}}},\ \bibinfo {type}
  {Tech. Rep.}\ (\bibinfo  {institution} {Lawrence Livermore Laboratory},\
  \bibinfo {year} {1973})\BibitemShut {NoStop}%
\bibitem [{\citenamefont {Fuggle}\ and\ \citenamefont
  {Alvarado}(1980)}]{Fuggle_1980}%
  \BibitemOpen
  \bibfield  {author} {\bibinfo {author} {\bibfnamefont {J.~C.}\ \bibnamefont
  {Fuggle}}\ and\ \bibinfo {author} {\bibfnamefont {S.~F.}\ \bibnamefont
  {Alvarado}},\ }\href@noop {} {\bibfield  {journal} {\bibinfo  {journal}
  {Physical Review A}\ }\textbf {\bibinfo {volume} {22}},\ \bibinfo {pages}
  {1615} (\bibinfo {year} {1980})}\BibitemShut {NoStop}%
\bibitem [{\citenamefont {Nyholm}\ \emph {et~al.}(1981)\citenamefont {Nyholm},
  \citenamefont {Martensson}, \citenamefont {Lebugle},\ and\ \citenamefont
  {Axelsson}}]{Nyholm_1981}%
  \BibitemOpen
  \bibfield  {author} {\bibinfo {author} {\bibfnamefont {R.}~\bibnamefont
  {Nyholm}}, \bibinfo {author} {\bibfnamefont {N.}~\bibnamefont {Martensson}},
  \bibinfo {author} {\bibfnamefont {A.}~\bibnamefont {Lebugle}}, \ and\
  \bibinfo {author} {\bibfnamefont {U.}~\bibnamefont {Axelsson}},\ }\href
  {\doibase 10.1088/0305-4608/11/8/025} {\bibfield  {journal} {\bibinfo
  {journal} {Journal of Physics F: Metal Physics}\ }\textbf {\bibinfo {volume}
  {11}},\ \bibinfo {pages} {1727} (\bibinfo {year} {1981})}\BibitemShut
  {NoStop}%
\bibitem [{\citenamefont {Berens}\ \emph {et~al.}(2020)\citenamefont {Berens},
  \citenamefont {Bichelmaier}, \citenamefont {Fernando}, \citenamefont
  {Thakur}, \citenamefont {Lee}, \citenamefont {Mascheck}, \citenamefont
  {Wiell}, \citenamefont {Eriksson}, \citenamefont {Kahk}, \citenamefont
  {Lischner}, \citenamefont {Mistry}, \citenamefont {Aichinger}, \citenamefont
  {Pobegen},\ and\ \citenamefont {Regoutz}}]{Berens_2020}%
  \BibitemOpen
  \bibfield  {author} {\bibinfo {author} {\bibfnamefont {J.}~\bibnamefont
  {Berens}}, \bibinfo {author} {\bibfnamefont {S.}~\bibnamefont {Bichelmaier}},
  \bibinfo {author} {\bibfnamefont {N.~K.}\ \bibnamefont {Fernando}}, \bibinfo
  {author} {\bibfnamefont {P.~K.}\ \bibnamefont {Thakur}}, \bibinfo {author}
  {\bibfnamefont {T.-L.}\ \bibnamefont {Lee}}, \bibinfo {author} {\bibfnamefont
  {M.}~\bibnamefont {Mascheck}}, \bibinfo {author} {\bibfnamefont
  {T.}~\bibnamefont {Wiell}}, \bibinfo {author} {\bibfnamefont {S.~K.}\
  \bibnamefont {Eriksson}}, \bibinfo {author} {\bibfnamefont {J.~M.}\
  \bibnamefont {Kahk}}, \bibinfo {author} {\bibfnamefont {J.}~\bibnamefont
  {Lischner}}, \bibinfo {author} {\bibfnamefont {M.~V.}\ \bibnamefont
  {Mistry}}, \bibinfo {author} {\bibfnamefont {T.}~\bibnamefont {Aichinger}},
  \bibinfo {author} {\bibfnamefont {G.}~\bibnamefont {Pobegen}}, \ and\
  \bibinfo {author} {\bibfnamefont {A.}~\bibnamefont {Regoutz}},\ }\href
  {\doibase 10.1088/2515-7655/ab8c5e} {\bibfield  {journal} {\bibinfo
  {journal} {Journal of Physics: Energy}\ }\textbf {\bibinfo {volume} {2}},\
  \bibinfo {pages} {035001} (\bibinfo {year} {2020})}\BibitemShut {NoStop}%
\bibitem [{\citenamefont {Jain}\ \emph {et~al.}(2013)\citenamefont {Jain},
  \citenamefont {Ong}, \citenamefont {Hautier}, \citenamefont {Chen},
  \citenamefont {Richards}, \citenamefont {Dacek}, \citenamefont {Cholia},
  \citenamefont {Gunter}, \citenamefont {Skinner}, \citenamefont {Ceder},\ and\
  \citenamefont {Persson}}]{Anubhav_2013}%
  \BibitemOpen
  \bibfield  {author} {\bibinfo {author} {\bibfnamefont {A.}~\bibnamefont
  {Jain}}, \bibinfo {author} {\bibfnamefont {S.~P.}\ \bibnamefont {Ong}},
  \bibinfo {author} {\bibfnamefont {G.}~\bibnamefont {Hautier}}, \bibinfo
  {author} {\bibfnamefont {W.}~\bibnamefont {Chen}}, \bibinfo {author}
  {\bibfnamefont {W.~D.}\ \bibnamefont {Richards}}, \bibinfo {author}
  {\bibfnamefont {S.}~\bibnamefont {Dacek}}, \bibinfo {author} {\bibfnamefont
  {S.}~\bibnamefont {Cholia}}, \bibinfo {author} {\bibfnamefont
  {D.}~\bibnamefont {Gunter}}, \bibinfo {author} {\bibfnamefont
  {D.}~\bibnamefont {Skinner}}, \bibinfo {author} {\bibfnamefont
  {G.}~\bibnamefont {Ceder}}, \ and\ \bibinfo {author} {\bibfnamefont {K.~A.}\
  \bibnamefont {Persson}},\ }\href {\doibase 10.1063/1.4812323} {\bibfield
  {journal} {\bibinfo  {journal} {APL Materials}\ }\textbf {\bibinfo {volume}
  {1}},\ \bibinfo {pages} {011002} (\bibinfo {year} {2013})}\BibitemShut
  {NoStop}%
\bibitem [{\citenamefont {Shinotsuka}\ \emph {et~al.}(2015)\citenamefont
  {Shinotsuka}, \citenamefont {Tanuma}, \citenamefont {Powell},\ and\
  \citenamefont {Penn}}]{Shinotsuka_2015}%
  \BibitemOpen
  \bibfield  {author} {\bibinfo {author} {\bibfnamefont {H.}~\bibnamefont
  {Shinotsuka}}, \bibinfo {author} {\bibfnamefont {S.}~\bibnamefont {Tanuma}},
  \bibinfo {author} {\bibfnamefont {C.~J.}\ \bibnamefont {Powell}}, \ and\
  \bibinfo {author} {\bibfnamefont {D.~R.}\ \bibnamefont {Penn}},\ }\href
  {\doibase https://doi.org/10.1002/sia.5789} {\bibfield  {journal} {\bibinfo
  {journal} {Surface and Interface Analysis}\ }\textbf {\bibinfo {volume}
  {47}},\ \bibinfo {pages} {871} (\bibinfo {year} {2015})}\BibitemShut
  {NoStop}%
\bibitem [{\citenamefont {Scanlon}\ \emph {et~al.}(2013)\citenamefont
  {Scanlon}, \citenamefont {Dunnill}, \citenamefont {Buckeridge}, \citenamefont
  {Shevlin}, \citenamefont {Logsdail}, \citenamefont {Woodley}, \citenamefont
  {Catlow}, \citenamefont {Powell}, \citenamefont {Palgrave}, \citenamefont
  {Parkin}, \citenamefont {Watson}, \citenamefont {Keal}, \citenamefont
  {Sherwood}, \citenamefont {Walsh},\ and\ \citenamefont
  {Sokol}}]{Scanlon2013}%
  \BibitemOpen
  \bibfield  {author} {\bibinfo {author} {\bibfnamefont {D.~O.}\ \bibnamefont
  {Scanlon}}, \bibinfo {author} {\bibfnamefont {C.~W.}\ \bibnamefont
  {Dunnill}}, \bibinfo {author} {\bibfnamefont {J.}~\bibnamefont {Buckeridge}},
  \bibinfo {author} {\bibfnamefont {S.~A.}\ \bibnamefont {Shevlin}}, \bibinfo
  {author} {\bibfnamefont {A.~J.}\ \bibnamefont {Logsdail}}, \bibinfo {author}
  {\bibfnamefont {S.~M.}\ \bibnamefont {Woodley}}, \bibinfo {author}
  {\bibfnamefont {C.~R.~A.}\ \bibnamefont {Catlow}}, \bibinfo {author}
  {\bibfnamefont {M.~J.}\ \bibnamefont {Powell}}, \bibinfo {author}
  {\bibfnamefont {R.~G.}\ \bibnamefont {Palgrave}}, \bibinfo {author}
  {\bibfnamefont {I.~P.}\ \bibnamefont {Parkin}}, \bibinfo {author}
  {\bibfnamefont {G.~W.}\ \bibnamefont {Watson}}, \bibinfo {author}
  {\bibfnamefont {T.~W.}\ \bibnamefont {Keal}}, \bibinfo {author}
  {\bibfnamefont {P.}~\bibnamefont {Sherwood}}, \bibinfo {author}
  {\bibfnamefont {A.}~\bibnamefont {Walsh}}, \ and\ \bibinfo {author}
  {\bibfnamefont {A.~A.}\ \bibnamefont {Sokol}},\ }\href {\doibase
  10.1038/nmat3697} {\bibfield  {journal} {\bibinfo  {journal} {Nature
  Materials}\ }\textbf {\bibinfo {volume} {12}},\ \bibinfo {pages} {798}
  (\bibinfo {year} {2013})}\BibitemShut {NoStop}%
\bibitem [{\citenamefont {Wang}\ \emph {et~al.}(2009)\citenamefont {Wang},
  \citenamefont {Badylevich}, \citenamefont {Afanas’ev}, \citenamefont
  {Stesmans}, \citenamefont {Adelmann}, \citenamefont {Van~Elshocht},
  \citenamefont {Kittl}, \citenamefont {Lukosius}, \citenamefont {Walczyk},\
  and\ \citenamefont {Wenger}}]{Wang_2009}%
  \BibitemOpen
  \bibfield  {author} {\bibinfo {author} {\bibfnamefont {W.~C.}\ \bibnamefont
  {Wang}}, \bibinfo {author} {\bibfnamefont {M.}~\bibnamefont {Badylevich}},
  \bibinfo {author} {\bibfnamefont {V.~V.}\ \bibnamefont {Afanas’ev}},
  \bibinfo {author} {\bibfnamefont {A.}~\bibnamefont {Stesmans}}, \bibinfo
  {author} {\bibfnamefont {C.}~\bibnamefont {Adelmann}}, \bibinfo {author}
  {\bibfnamefont {S.}~\bibnamefont {Van~Elshocht}}, \bibinfo {author}
  {\bibfnamefont {J.~A.}\ \bibnamefont {Kittl}}, \bibinfo {author}
  {\bibfnamefont {M.}~\bibnamefont {Lukosius}}, \bibinfo {author}
  {\bibfnamefont {C.}~\bibnamefont {Walczyk}}, \ and\ \bibinfo {author}
  {\bibfnamefont {C.}~\bibnamefont {Wenger}},\ }\href {\doibase
  10.1063/1.3236536} {\bibfield  {journal} {\bibinfo  {journal} {Applied
  Physics Letters}\ }\textbf {\bibinfo {volume} {95}},\ \bibinfo {pages}
  {132903} (\bibinfo {year} {2009})}\BibitemShut {NoStop}%
\bibitem [{\citenamefont {Chernikov}\ \emph {et~al.}(1987)\citenamefont
  {Chernikov}, \citenamefont {Savin}, \citenamefont {Fadeev}, \citenamefont
  {Landin},\ and\ \citenamefont {Izhvanov}}]{CHERNIKOV1987441}%
  \BibitemOpen
  \bibfield  {author} {\bibinfo {author} {\bibfnamefont {A.}~\bibnamefont
  {Chernikov}}, \bibinfo {author} {\bibfnamefont {V.}~\bibnamefont {Savin}},
  \bibinfo {author} {\bibfnamefont {V.}~\bibnamefont {Fadeev}}, \bibinfo
  {author} {\bibfnamefont {N.}~\bibnamefont {Landin}}, \ and\ \bibinfo {author}
  {\bibfnamefont {L.}~\bibnamefont {Izhvanov}},\ }\href {\doibase
  https://doi.org/10.1016/0022-5088(87)90139-1} {\bibfield  {journal} {\bibinfo
   {journal} {Journal of the Less Common Metals}\ }\textbf {\bibinfo {volume}
  {130}},\ \bibinfo {pages} {441} (\bibinfo {year} {1987})}\BibitemShut
  {NoStop}%
\bibitem [{\citenamefont {Dantzer}(1983)}]{DANTZER1983913}%
  \BibitemOpen
  \bibfield  {author} {\bibinfo {author} {\bibfnamefont {P.}~\bibnamefont
  {Dantzer}},\ }\href {\doibase https://doi.org/10.1016/0022-3697(83)90130-0}
  {\bibfield  {journal} {\bibinfo  {journal} {Journal of Physics and Chemistry
  of Solids}\ }\textbf {\bibinfo {volume} {44}},\ \bibinfo {pages} {913}
  (\bibinfo {year} {1983})}\BibitemShut {NoStop}%
\bibitem [{\citenamefont {wei Zhao}\ \emph {et~al.}(2008)\citenamefont {wei
  Zhao}, \citenamefont {Ding}, \citenamefont {feng Tian}, \citenamefont {juan
  Zhao},\ and\ \citenamefont {liang Hou}}]{JZhao_2008}%
  \BibitemOpen
  \bibfield  {author} {\bibinfo {author} {\bibfnamefont {J.}~\bibnamefont {wei
  Zhao}}, \bibinfo {author} {\bibfnamefont {H.}~\bibnamefont {Ding}}, \bibinfo
  {author} {\bibfnamefont {X.}~\bibnamefont {feng Tian}}, \bibinfo {author}
  {\bibfnamefont {W.}~\bibnamefont {juan Zhao}}, \ and\ \bibinfo {author}
  {\bibfnamefont {H.}~\bibnamefont {liang Hou}},\ }\href {\doibase
  10.1088/1674-0068/21/06/569-574} {\bibfield  {journal} {\bibinfo  {journal}
  {Chinese Journal of Chemical Physics}\ }\textbf {\bibinfo {volume} {21}},\
  \bibinfo {pages} {569} (\bibinfo {year} {2008})}\BibitemShut {NoStop}%
\bibitem [{\citenamefont {Lee}\ and\ \citenamefont
  {Duncan}(2018)}]{Duncan_2018}%
  \BibitemOpen
  \bibfield  {author} {\bibinfo {author} {\bibfnamefont {T.-L.}\ \bibnamefont
  {Lee}}\ and\ \bibinfo {author} {\bibfnamefont {D.~A.}\ \bibnamefont
  {Duncan}},\ }\href {\doibase 10.1080/08940886.2018.1483653} {\bibfield
  {journal} {\bibinfo  {journal} {Synchrotron Radiation News}\ }\textbf
  {\bibinfo {volume} {31}},\ \bibinfo {pages} {16} (\bibinfo {year}
  {2018})}\BibitemShut {NoStop}%
\bibitem [{\citenamefont {Griessen}\ and\ \citenamefont
  {Driessen}(1984)}]{Griessen_1984}%
  \BibitemOpen
  \bibfield  {author} {\bibinfo {author} {\bibfnamefont {R.}~\bibnamefont
  {Griessen}}\ and\ \bibinfo {author} {\bibfnamefont {A.}~\bibnamefont
  {Driessen}},\ }\href {\doibase 10.1103/PhysRevB.30.4372} {\bibfield
  {journal} {\bibinfo  {journal} {Physical Review B}\ }\textbf {\bibinfo
  {volume} {30}},\ \bibinfo {pages} {4372} (\bibinfo {year}
  {1984})}\BibitemShut {NoStop}%
\bibitem [{\citenamefont {Kalita}\ \emph {et~al.}(2010)\citenamefont {Kalita},
  \citenamefont {Sinogeikin}, \citenamefont {Lipinska-Kalita}, \citenamefont
  {Hartmann}, \citenamefont {Ke}, \citenamefont {Chen},\ and\ \citenamefont
  {Cornelius}}]{Kalita_2010}%
  \BibitemOpen
  \bibfield  {author} {\bibinfo {author} {\bibfnamefont {P.~E.}\ \bibnamefont
  {Kalita}}, \bibinfo {author} {\bibfnamefont {S.~V.}\ \bibnamefont
  {Sinogeikin}}, \bibinfo {author} {\bibfnamefont {K.}~\bibnamefont
  {Lipinska-Kalita}}, \bibinfo {author} {\bibfnamefont {T.}~\bibnamefont
  {Hartmann}}, \bibinfo {author} {\bibfnamefont {X.}~\bibnamefont {Ke}},
  \bibinfo {author} {\bibfnamefont {C.}~\bibnamefont {Chen}}, \ and\ \bibinfo
  {author} {\bibfnamefont {A.}~\bibnamefont {Cornelius}},\ }\href {\doibase
  10.1063/1.3455858} {\bibfield  {journal} {\bibinfo  {journal} {Journal of
  Applied Physics}\ }\textbf {\bibinfo {volume} {108}},\ \bibinfo {pages}
  {043511} (\bibinfo {year} {2010})}\BibitemShut {NoStop}%
\bibitem [{\citenamefont {Pebler}\ and\ \citenamefont
  {Wallace}(1962)}]{Pebler_1962}%
  \BibitemOpen
  \bibfield  {author} {\bibinfo {author} {\bibfnamefont {A.}~\bibnamefont
  {Pebler}}\ and\ \bibinfo {author} {\bibfnamefont {W.~E.}\ \bibnamefont
  {Wallace}},\ }\href {\doibase 10.1021/j100807a033} {\bibfield  {journal}
  {\bibinfo  {journal} {The Journal of Physical Chemistry}\ }\textbf {\bibinfo
  {volume} {66}},\ \bibinfo {pages} {148} (\bibinfo {year} {1962})}\BibitemShut
  {NoStop}%
\bibitem [{\citenamefont {Smr\v{c}ok}(1989)}]{Smrcok_1989}%
  \BibitemOpen
  \bibfield  {author} {\bibinfo {author} {\bibfnamefont {L.}~\bibnamefont
  {Smr\v{c}ok}},\ }\href {\doibase https://doi.org/10.1002/crat.2170240609}
  {\bibfield  {journal} {\bibinfo  {journal} {Crystal Research and Technology}\
  }\textbf {\bibinfo {volume} {24}},\ \bibinfo {pages} {607} (\bibinfo {year}
  {1989})}\BibitemShut {NoStop}%
\bibitem [{\citenamefont {Baur}\ and\ \citenamefont {Khan}(1971)}]{Baur_19715}%
  \BibitemOpen
  \bibfield  {author} {\bibinfo {author} {\bibfnamefont {W.~H.}\ \bibnamefont
  {Baur}}\ and\ \bibinfo {author} {\bibfnamefont {A.~A.}\ \bibnamefont
  {Khan}},\ }\href {\doibase 10.1107/S0567740871005466} {\bibfield  {journal}
  {\bibinfo  {journal} {Acta Crystallographica Section B}\ }\textbf {\bibinfo
  {volume} {27}},\ \bibinfo {pages} {2133} (\bibinfo {year}
  {1971})}\BibitemShut {NoStop}%
\bibitem [{\citenamefont {Miwa}\ and\ \citenamefont
  {Fukumoto}(2002)}]{Miwa_2002}%
  \BibitemOpen
  \bibfield  {author} {\bibinfo {author} {\bibfnamefont {K.}~\bibnamefont
  {Miwa}}\ and\ \bibinfo {author} {\bibfnamefont {A.}~\bibnamefont
  {Fukumoto}},\ }\href {\doibase 10.1103/PhysRevB.65.155114} {\bibfield
  {journal} {\bibinfo  {journal} {Physical Review B}\ }\textbf {\bibinfo
  {volume} {65}},\ \bibinfo {pages} {155114} (\bibinfo {year}
  {2002})}\BibitemShut {NoStop}%
\bibitem [{\citenamefont {Jackson}\ \emph {et~al.}(2018)\citenamefont
  {Jackson}, \citenamefont {Ganose}, \citenamefont {Regoutz}, \citenamefont
  {Egdell},\ and\ \citenamefont {Scanlon}}]{Jackson2018}%
  \BibitemOpen
  \bibfield  {author} {\bibinfo {author} {\bibfnamefont {A.~J.}\ \bibnamefont
  {Jackson}}, \bibinfo {author} {\bibfnamefont {A.~M.}\ \bibnamefont {Ganose}},
  \bibinfo {author} {\bibfnamefont {A.}~\bibnamefont {Regoutz}}, \bibinfo
  {author} {\bibfnamefont {R.~G.}\ \bibnamefont {Egdell}}, \ and\ \bibinfo
  {author} {\bibfnamefont {D.~O.}\ \bibnamefont {Scanlon}},\ }\href {\doibase
  10.21105/joss.00773} {\bibfield  {journal} {\bibinfo  {journal} {Journal of
  Open Source Software}\ }\textbf {\bibinfo {volume} {3}},\ \bibinfo {pages}
  {773} (\bibinfo {year} {2018})}\BibitemShut {NoStop}%
\end{thebibliography}%
\bibliographystyle{apsrev4-1}

\end{document}


\preprint{APS/123-QED}

\title[PRB]{Revealing the bonding nature and electronic structure of early transition metal dihydrides \newline Supplemental Material}

\author{Curran~Kalha}%
\affiliation{%
Department of Chemistry, University College London, 20 Gordon Street, London WC1H 0AJ, United Kingdom.
}%

\author{Laura~E.~Ratcliff}
\affiliation{ 
Centre for Computational Chemistry, School of Chemistry, University of Bristol, Bristol BS8 1TS, United Kingdom.
}

\author{Giorgio~Colombi}%
\affiliation{%
Materials for Energy Conversion and Storage, Department of Chemical Engineering, Delft University of Technology, NL-2629HZ Delft, The Netherlands.
}%
\author{Christoph~Schlueter}
\affiliation{%
Deutsches Elektronen-Synchrotron DESY, Notkestra{\ss}e 85, 22607 Hamburg, Germany.
}%

\author{Bernard~Dam}%
\affiliation{%
Materials for Energy Conversion and Storage, Department of Chemical Engineering, Delft University of Technology, NL-2629HZ Delft, The Netherlands.
}%

\author{Andrei~Gloskovskii}
\affiliation{%
Deutsches Elektronen-Synchrotron DESY, Notkestra{\ss}e 85, 22607 Hamburg, Germany.
}%

\author{Tien-Lin~Lee}%
\affiliation{%
Diamond Light Source Ltd., Diamond House, Harwell Science and Innovation Campus, Didcot, OX11 0DE, United Kingdom.
}%

\author{Pardeep~K.~Thakur}%
\affiliation{%
Diamond Light Source Ltd., Diamond House, Harwell Science and Innovation Campus, Didcot, OX11 0DE, United Kingdom.
}%

\author{Prajna~Bhatt}%
\affiliation{%
Department of Chemistry, University College London, 20 Gordon Street, London WC1H 0AJ, United Kingdom.
}%

\author{Yujiang~Zhu}%
\affiliation{%
Department of Chemistry, University College London, 20 Gordon Street, London WC1H 0AJ, United Kingdom.
}%

\author{J\"urg~Osterwalder}%
\affiliation{Physik-Institut, Universitat Z{\"u}rich, CH-8057 Z{\"u}rich, Switzerland.
}%

\author{Francesco~Offi}%
\affiliation{%
Dipartimento di Scienze, Università di Roma Tre, 00146 Rome, Italy.
}%

\author{Giancarlo~Panaccione}%
 \email{giancarlo <panaccione@iom.cnr.it}
\affiliation{%
Istituto Officina dei Materiali (IOM)-CNR, Laboratorio TASC, in Area Science Park, S.S.14, Km 163.5, I-34149 Trieste, Italy.
}%

\author{Anna~Regoutz}%
 \email{a.regoutz@ucl.ac.uk}
\affiliation{%
Department of Chemistry, University College London, 20 Gordon Street, London WC1H 0AJ, United Kingdom.
}%

\date{\today}
\maketitle

\newpage

 \tableofcontents

\cleardoublepage

\section{Survey spectra} \label{sec:survey}

\begin{figure*}[ht]
\centering
    \includegraphics[keepaspectratio, width = 12.9cm]{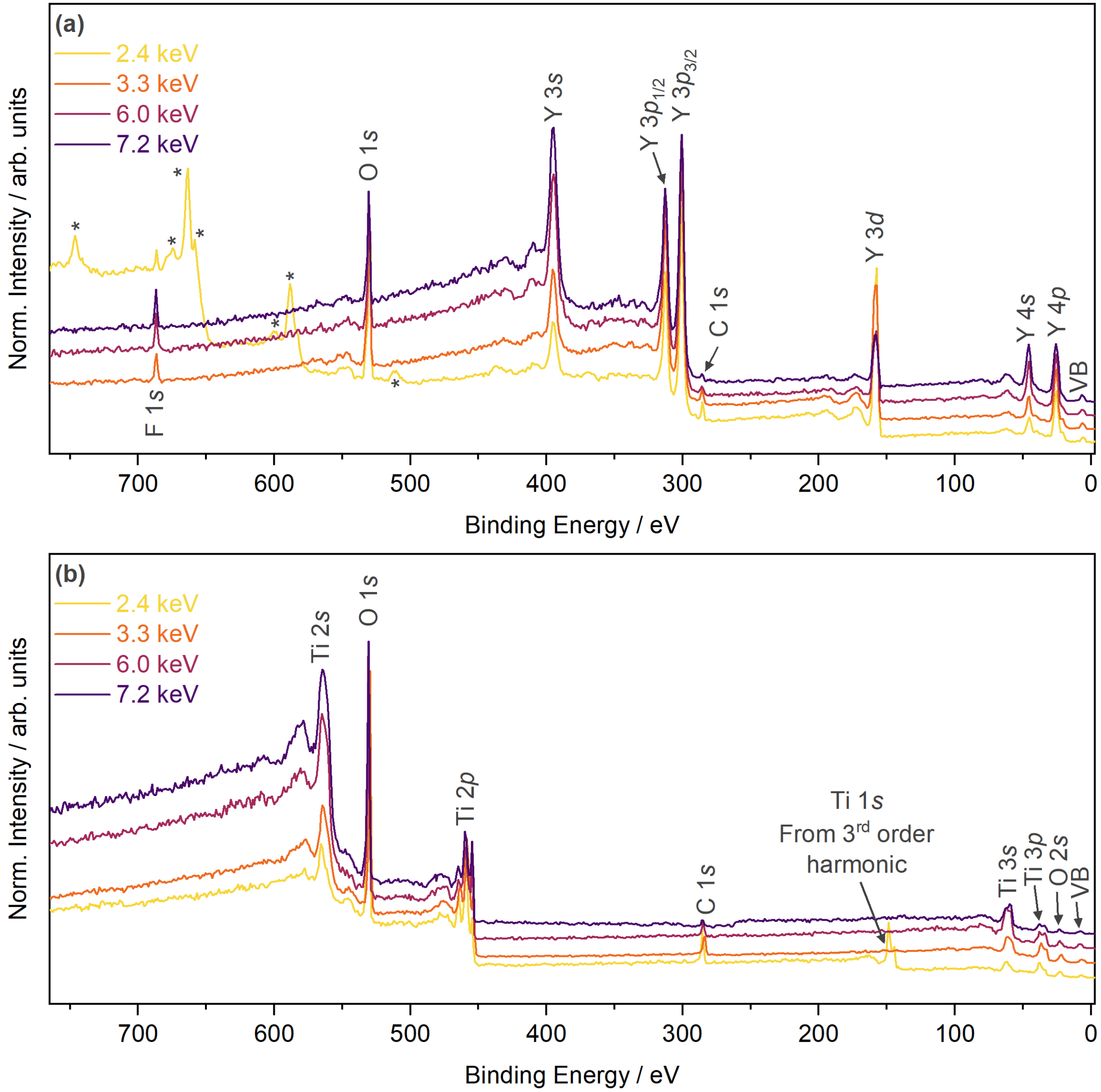}
    \caption{Survey spectra collected as a function of X-ray photon energy for (a) YH\textsubscript{2-$\delta$} and (b) TiH\textsubscript{2-$\delta$}. Spectra are normalised to the most intense peak, and the binding energy scale is calibrated to the Fermi edge of a polycrystalline gold foil reference.}
    \label{fig:survey}
\end{figure*}

Survey spectra for TiH\textsubscript{2-$\delta$} and YH\textsubscript{2-$\delta$} at the four excitation energies are shown in Fig.~\ref{fig:survey}. For the YH\textsubscript{2-$\delta$} sample only, the 2.4~keV spectrum was recorded at a photon energy of 2400.4~eV, approximately 10~eV lower than the core level and valence band spectra (2410.6~eV). For YH\textsubscript{2-$\delta$}, fluorine, oxygen and carbon signals are observed in addition to the expected yttrium signals. Fluorine originates from the synthesis procedure. In addition, Y Auger lines (labelled with an asterisk) are observed at 2.4~keV. For TiH\textsubscript{2-$\delta$}, oxygen and carbon signals are observed in addition to the expected Ti signals. The 2.4~keV energy uses the first harmonic of the Si(111) DCM, but reflections from the third harmonic allow for the Ti~1\textit{s} to also appear in the survey spectrum.

\cleardoublepage

\section{Additional core level spectra}

\begin{figure*}[ht]
\centering
    \includegraphics[keepaspectratio, width =\linewidth]{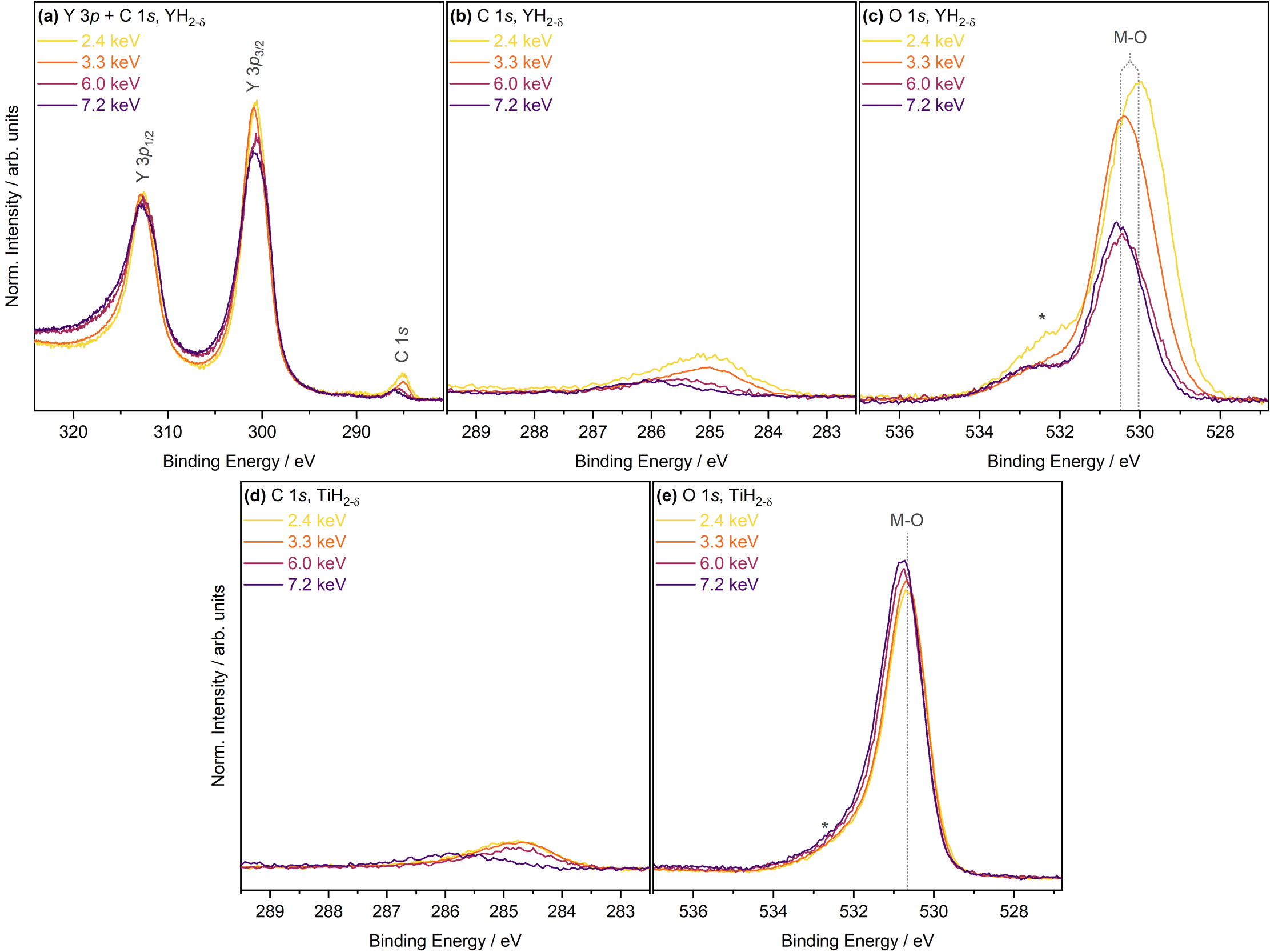}
    \caption{Additional core level spectra collected as a function of X-ray photon energy for the two samples, including (a) Y~3\textit{p} + C~1\textit{s}, (b) C~1\textit{s} and (c) O~\textit{s} spectra collected on the YH\textsubscript{2-$\delta$} sample and (d) C~1\textit{s} and (e) O~1\textit{s} core level spectra collected on the TiH\textsubscript{2-$\delta$}. All spectra are aligned to their intrinsic $E_F$. Core level spectra collected on sample YH\textsubscript{2-$\delta$} are normalised to the total Y~3\textit{p} spectral area (after the removal of a Shirley-type background), whereas the total Ti~2\textit{p} spectral area (after the removal of a Shirley-type background) was used to normalise the C~1\textit{s} and O~1\textit{s} spectra recorded on sample TiH\textsubscript{2-$\delta$}. The $y$-scale is arbitrary and the spectra were scaled until a similar signal-to-noise ratio was obtained to highlight the low intensity of the C signal. The Y~3\textit{p} spectral area was used to normalise the C~1\textit{s} and O~1\textit{s} signal intensities as the orbital has a similar photoionisation cross section decay to the C~1\textit{s} and O~1\textit{s} orbitals, whereas the Y~3\textit{d} cross section decay is not so similar, which would skew the intended effect of normalisation.}
    \label{fig:Other}
\end{figure*}

\cleardoublepage

\section{T\lowercase{i~1\textit{s}}} \label{sec:Ti1s}

In addition to the Ti~2\textit{p} core level, the Ti~1\textit{s} core level was explored for TiH\textsubscript{2-$\delta$} at 6.0 and 7.2~keV. HAXPES enables access to this deep core level, which does not include spin-orbit-splitting (SOS) and is, therefore, easier to interpret than the Ti~2\textit{p} spectrum. It confirms the chemical states observed in the Ti~2\textit{p}, namely both the titanium dihydride (Ti-H) state at a BE of 4964.8~eV, along with multiple titanium oxidation valence states (Ti-O). The main metal oxide state contributions are from the +4 (Ti(IV)-O) and +3 (Ti(III)-O) oxide states at BE positions of 4968.9 and 4967.3~eV, respectively (determined from the peak-fit analysis of the 7.2~keV spectrum). Additionally, a change in the satellite structure (between 4977-4985~eV) is observed between 6.0 and 7.2~keV. The main intensity satellite (S\textsubscript{2}) is observed at an approximate BE of 4981.9$\pm$0.5~eV and accompanied by a less intense satellite (S\textsubscript{1}) on the lower BE side at approximately 4978.9$\pm$0.5~eV.   \par

\begin{figure*}[ht!]
\centering
    \includegraphics[keepaspectratio, width =0.4\linewidth]{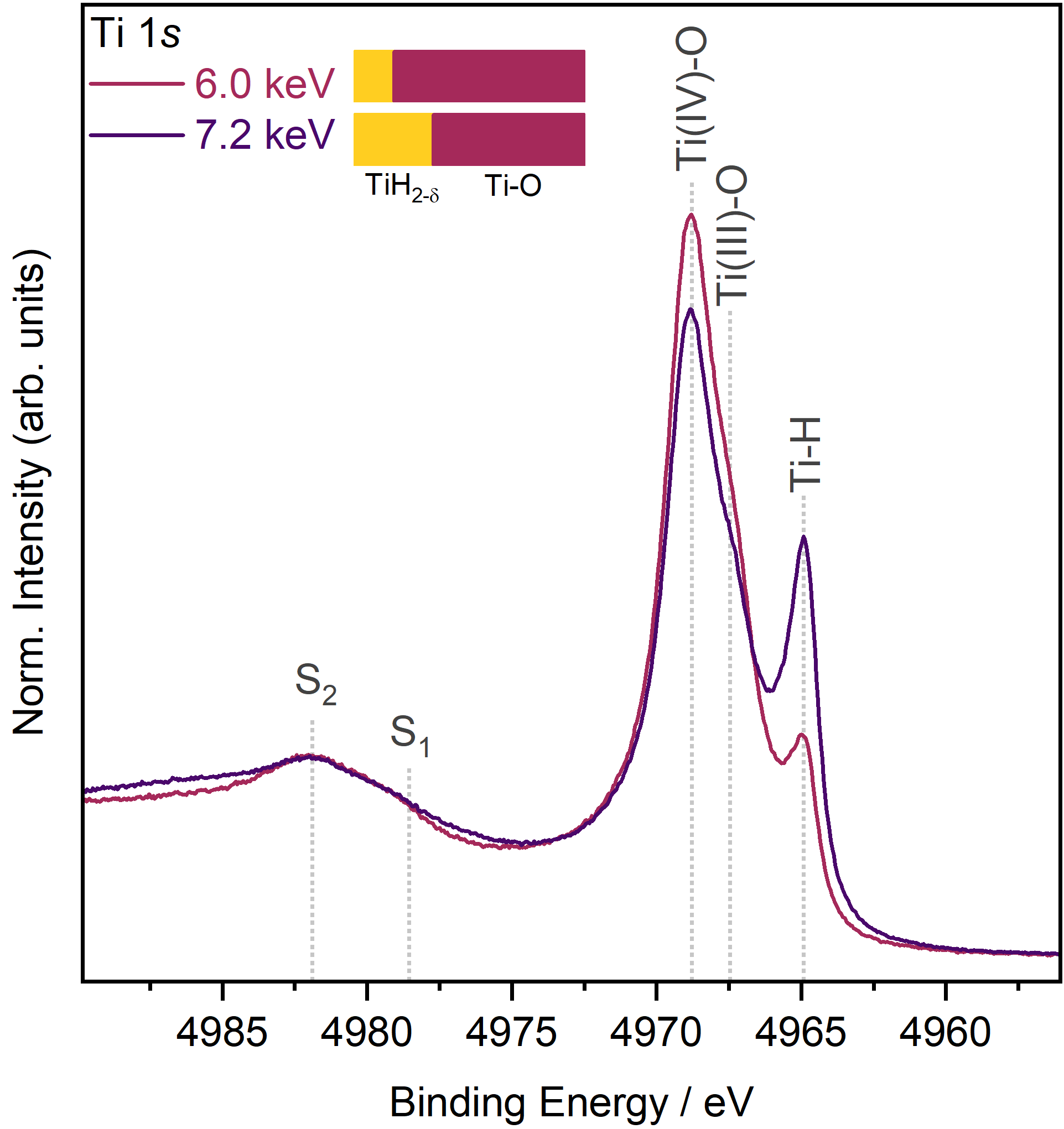}
    \caption{Ti~1\textit{s} recorded spectrum collected on TiH\textsubscript{2-$\delta$} using the 6.0 and 7.2~keV photon energies. Spectra are normalised to their respective areas. Through peak-fit analysis, the metal hydride contribution to the total spectral area was determined and is represented by bar charts adjacent to the legend. In the bar charts, Ti-O refers to all other chemical states besides the metal hydride state (i.e.\ non metal hydride states). }
    \label{fig:Ti1s}
\end{figure*}

Based on peak fit analysis of the Ti~1\textit{s} core state spectra displayed in Fig.~\ref{fig:Ti1s}, the hydride contribution to the 7.2~keV spectrum is found to be 34.5\% of the total signal. The significant difference in BE of the Ti~1\textit{s} and Ti~2\textit{p} core levels (over 4.5~keV) leads to a significant difference in kinetic energy (KE) and therefore probing depth, making the Ti~1\textit{s} core level more surface sensitive, which is why a smaller hydride contribution is found compared to that obtained with the Ti~2\textit{p} core level at the same photon energy.

\cleardoublepage

\section{Core level analysis}

Tab.~\ref{YH2_BE} lists selected reported binding energy (BE) values of the Y~3\textit{d}\textsubscript{5/2} peak for yttrium metal, hydride, hydroxide and oxide. Tab.~\ref{TiH2_BE} lists selected reported BE values of the Ti~2\textit{p}\textsubscript{3/2} peak for titanium metal, hydride and oxide. Fujimori~\textit{et al.} measured yttrium metal and compared the BE position shifts of yttrium hydride with increasing hydride content relative to the metal.~\cite{Fujimori_1984} Lamartine~\textit{et al.} also conducted a similar experiment but for titanium hydrides.~\cite{LAMARTINE1980537} Both show that with increasing H content, the BE position of the main core line peak shifts to a higher BE relative to the metal. Our values are in good agreement with the reported values, with discrepancies attributed to differences in the BE scale calibration and experimental resolution. 

\begin{table}[ht!]
     \caption{\label{YH2_BE}Reported BE positions of the Y~3\textit{d}\textsubscript{5/2} peak for yttrium, yttrium hydride, yttrium oxide, and yttrium hydroxide. The BE positions determined from peak-fit analysis of the Y~3\textit{d} core level spectra collected at 2.4 and 7.2~keV in this work are also included.}
     \begin{ruledtabular}
    \begin{tabular}{ccc}

Compound &  Y~3\textit{d}\textsubscript{5/2} BE / eV & Ref.      \\
   \hline

Y & 155.8 & \cite{Fujimori_1984} \\
Y & 155.6 & \cite{Hayoz_2000} \\
Y & 156.06$\pm$0.1 & \cite{Mongstad2014TheOxide} \\
YH\textsubscript{2} & 156.0 & \cite{Hayoz_2000} \\
YH\textsubscript{2} & 156.53$\pm$0.1 & \cite{Mongstad2014TheOxide} \\
YH\textsubscript{2.1} & 156.5 & \cite{Fujimori_1984} \\
YH\textsubscript{3} & 157.7 & \cite{Fujimori_1984} \\
YH\textsubscript{3} & 157.55$\pm$0.1 & \cite{Mongstad2014TheOxide} \\
&& \\
YH\textsubscript{2-$\delta$} & 156.3 & This work $@$ 2.4~keV \\
YH\textsubscript{2-$\delta$} & 156.4 & This work $@$ 7.2~keV \\
&& \\
Y\textsubscript{2}O\textsubscript{3} & 156.2 & \cite{Majumdar_1991} \\
Y\textsubscript{2}O\textsubscript{3} & 156.8 & \cite{UWAMINO198467} \\
Y\textsubscript{2}O\textsubscript{3} & 156.86 & \cite{Mitrovic_Veal_2014} \\
Y\textsubscript{2}O\textsubscript{3} & 157.2 & \cite{Ingo_1990} \\
Y\textsubscript{2}O\textsubscript{3} & 157.4 & \cite{Barreca_2001} \\
Y\textsubscript{2}O\textsubscript{3} & 158.5 & \cite{REICHL1986196} \\
Y hydroxide & 157.4 & \cite{Majumdar_1991} \\
&& \\
Y(III) oxide & 157.5 & This work $@$ 2.4~keV \\
Y(III) oxide & 157.9 & This work $@$ 7.2~keV \\
\end{tabular}
     \end{ruledtabular}

\end{table}

\begin{table}[ht!]
     \caption{\label{TiH2_BE}Reported BE positions of the Ti~2\textit{p}\textsubscript{3/2} peak for titanium, titanium hydride, and titanium oxide. The BE positions determined from peak-fit analysis of the Y~3\textit{d} core level spectra collected at 2.4 and 7.2~keV in this work are also included.}
     \begin{ruledtabular}
    \begin{tabular}{ccc}

Compound &  Ti~2\textit{p}\textsubscript{3/2} BE / eV & Ref.      \\
   \hline

Ti & 452.8 & \cite{LAMARTINE1980537} \\
Ti & 453.8 & \cite{BIWER1986207} \\
Ti & 454.0 & \cite{SLEIGH199641} \\
TiH\textsubscript{1.5} & 453.3 & \cite{LAMARTINE1980537} \\
TiH\textsubscript{1.8} & 453.4 & \cite{LAMARTINE1980537} \\
TiH\textsubscript{2} & 453.5 & \cite{Kaciulis_2018} \\
TiH\textsubscript{2} & 453.9 & \cite{Ren_2014} \\
TiH\textsubscript{2} & 453.9 & \cite{MA20092250} \\
&& \\
TiH\textsubscript{2-$\delta$} & 454.3 & This work $@$ 2.4~keV \\
TiH\textsubscript{2-$\delta$} & 454.3 & This work $@$ 7.2~keV \\
&& \\
TiO\textsubscript{2} & 458.5 & \cite{Gonzalez_1989} \\
TiO\textsubscript{2} & 459.3 & \cite{DIEBOLD200353} \\
&& \\
Ti(IV) oxide & 459.2 & This work $@$ 2.4~keV \\
Ti(IV) oxide & 459.3 & This work $@$ 7.2~keV \\

\end{tabular}
     \end{ruledtabular}

\end{table}

\cleardoublepage

\section{Peak-fitting Procedure} \label{sec:RPF}

\subsection{Y~3\textit{d}}

Few studies have been presented on the core level spectra of yttrium dihydride,~\cite{Fujimori_1984, Hayoz_2000} however, the consensus is that a systematic positive binding energy (BE) shift relative to the yttrium metal BE peak position occurs with increasing H content. Specifically for yttrium dihydride, most studies also report features on the high BE side of the Y~3\textit{d} doublet core lines, leading to a large BE tail. However, the strong overlapping metal oxide contribution in our spectra shown in Fig.~1(a) of the main manuscript means that identifying these features is difficult. Therefore, when peak-fitting, it is assumed that only one metal hydride contribution exists, and we omit a description of additional features.\par 

\begin{figure*}
\centering
    \includegraphics[keepaspectratio, width =0.68\linewidth]{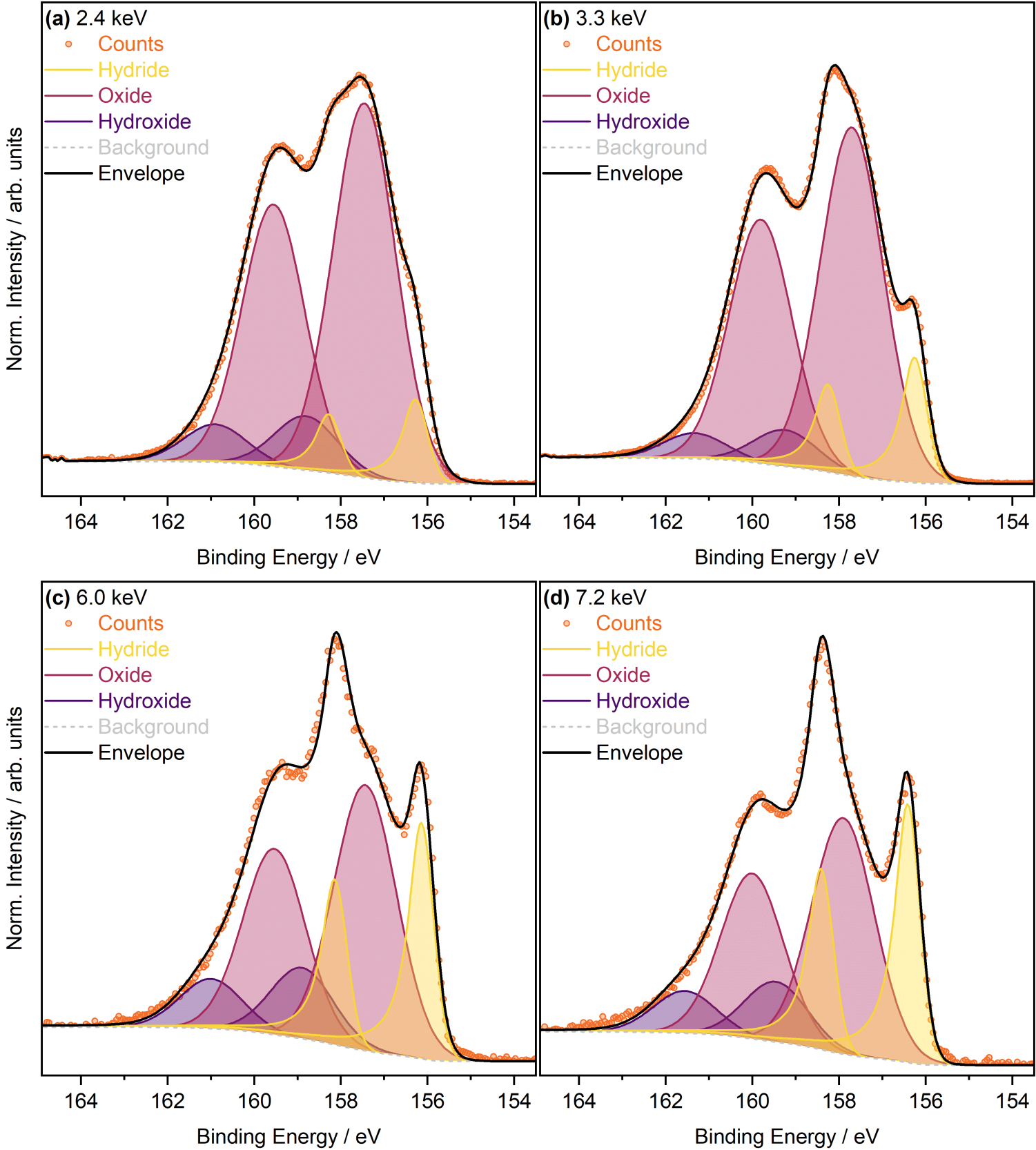}
    \caption{Peak-fit models of the Y~3\textit{d} core level spectra as a function of photon energy, including (a) 2.4, (b) 3.3, (c) 6.0, and (d) 7.2~keV.}
    \label{fig:Y3d_pf}
\end{figure*}

Peak-fit analysis of the Y~3\textit{d} core level spectrum was conducted using the CasaXPS software package and a Shirley-type background was used to remove the secondary background from the spectrum. The peak-fit models for each Y~3\textit{d} spectrum are displayed in Fig.~\ref{fig:Y3d_pf}. Three environments were expected - hydride, oxide and hydroxide. Metal hydride core level peaks are known to exhibit an asymmetric line shape owing to the coupling of the core hole with conduction band electrons. The degree of asymmetry is governed by the local density of states at the Fermi energy (N(E\textsubscript{F})), and compared to metals, metal hydrides should have a lower N(E\textsubscript{F}), and by extension, a smaller asymmetry parameter.~\cite{Riesterer1987} However, without a clean metal hydride reference spectrum, knowing the true line shape of the metal hydride peak is difficult. Therefore, an assumption was made to describe the hydride peaks with an asymmetric Doniach Sunjic (DS) line shape convoluted with a product mix of Gaussian and Lorentzian (DS(0.01, 10)GL(20)). The oxide and hydroxide environments should display a Voigt profile line shape and therefore were described with a product mix of Gaussian and Lorentzian (GL(20), i.e.\ 80~\% Gaussian, 20~\% Lorentzian). The doublet peaks of each environment were constrained to have the same line shape and full width at half maximum (FWHM). The oxide and hydroxide environments were constrained to have the same peak qualities. To account for the degeneracy of the \textit{d} orbital, the area ratio between Y~3\textit{d}\textsubscript{5/2} and Y~3\textit{d}\textsubscript{3/2} was set to 0.7 (determined by using the tabulated Scofield cross section values).~\cite{Scofield1973} This ratio was applied to all three environments. The spin-orbit splitting (SOS) for the hydride and oxide/hydroxide peaks was set to 2.1 and 2.0~eV, respectively. The value of the SOS for the hydride was taken from the work by Fujimori~\textit{et al.} who characterised YH\textsubscript{2.1} using XPS ($h\nu$ = Mg~K$\alpha$) after in-situ cleaning of the sample until the O~1\textit{s} to Y~3\textit{d} ratio was approximately 0.04.~\cite{Fujimori_1984} The SOS value of the oxide Y~3\textit{d} peaks was taken from the work by Majumdar~\textit{et al.} who studied standard cubic Y\textsubscript{2}O\textsubscript{3} with XPS ($h\nu$ = Al~K$\alpha$), and the hydroxide environment was assumed to have the same SOS as the oxide.~\cite{Majumdar_1991} Finally, the peak-fit analysis was conducted with these constraints to determine the FWHM of the hydride and oxide/hydroxide peaks. The 2.4 and 7.2~keV spectra give the best representation of the oxide and hydride environments, respectively. The FWHM of the oxide peak extracted from the 2.4~keV spectrum and the FWHM of the hydride peak extracted from the 7.2 keV spectrum was found to be 1.71 and 0.67~eV, respectively. These FWHMs were then applied to all spectra, and the peak-fit analysis was re-run adding this additional FWHM constraint. The ratio of hydride to non-hydride (i.e.\ oxide and hydroxide sum) was determined by comparing the raw Y~3\textit{d}\textsubscript{5/2} peak areas of each environment. No escape depth correction or any other relative atomic sensitivity factor was applied to the peak areas. The error associated with the quantification is assumed to be $\pm$0.2~at.\% because of the difficulty in peak-fit analysis. 

\subsection{Ti~1\textit{s}}

The Ti~1\textit{s} line is only accessible with the 6.0, and 7.2~keV photon energies as the BE of the core line is approximately 4966~eV. Due to the high BE (low kinetic energy) of the Ti~1\textit{s} core level, it offers a more surface-sensitive probing depth than the equivalent Ti~2\textit{p} core level measured at the same photon energy. For this reason, the Ti~1\textit{s} displays a greater oxide contribution to the total spectral line shape, however, it shows good agreement with the Ti~2\textit{p} in the sense that it too displays evidence of multiple valence state metal oxide environments and a lower BE metal-hydride environment. Here, we present the first reported Ti~1\textit{s} BE position of a TiH\textsubscript{2-$\delta$} environment at 4964.8~eV (determined from the 7.2~keV peak-fit). \par

\begin{figure*}
\centering
    \includegraphics[keepaspectratio, width =0.4\linewidth]{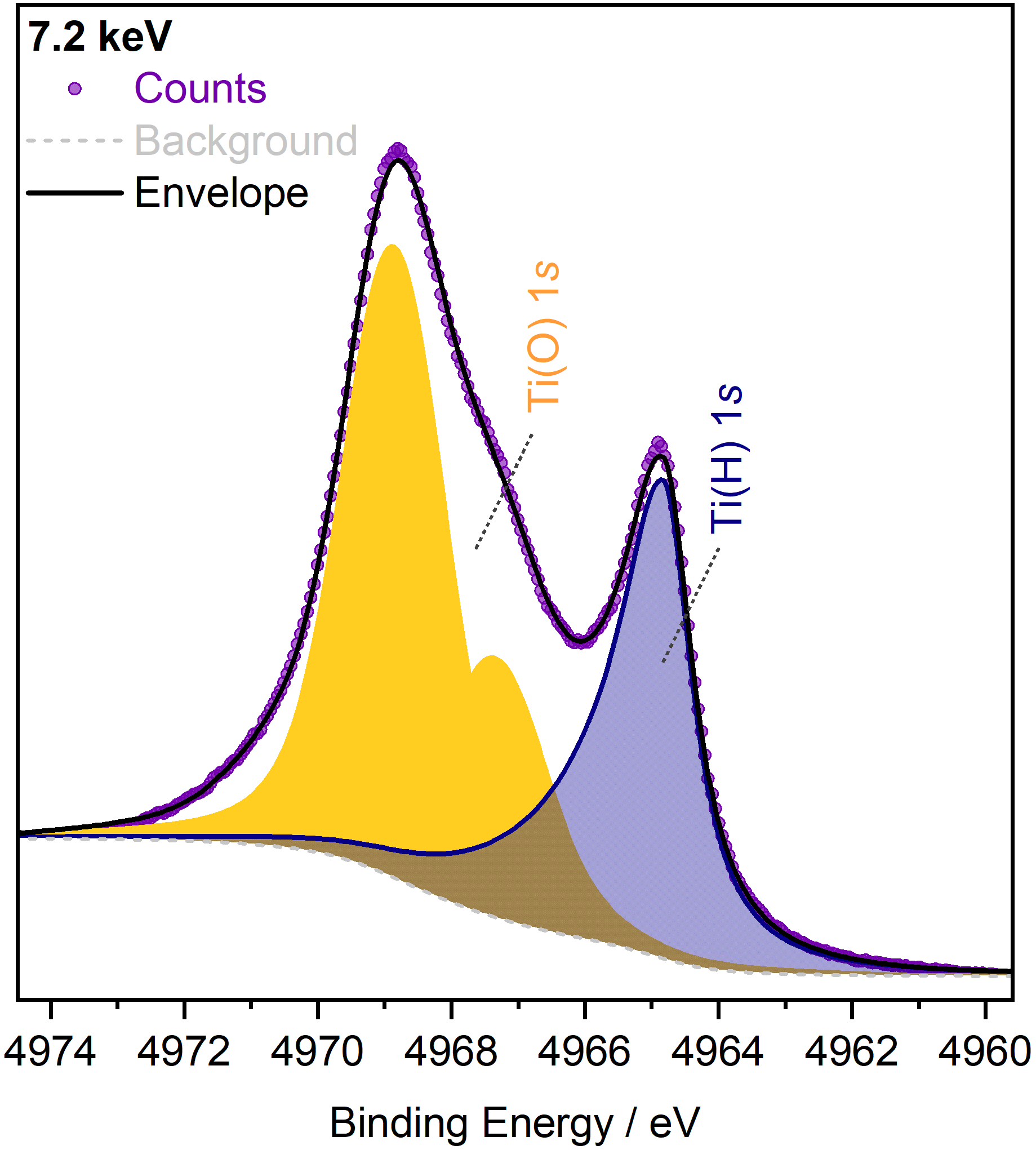}
    \caption{Representative peak-fit model of the Ti~1\textit{s} spectrum for the 7.2~keV collected spectrum. (H) refers to the hydride components, and (O) refers to the oxide components.}
    \label{fig:Ti1s_pf}
\end{figure*}

Ti~1\textit{s} was fitted using the Thermo Scientific Avantage software package v5.9925, and a Shirley-type background was applied to remove the secondary background. A representative peak-fit of the 7.2~keV spectrum is displayed in Fig.~\ref{fig:Ti1s_pf}. To date, the Ti~1\textit{s} core level spectrum has not been reported for titanium hydride. However, our assumptions on the expected environments are corroborated using the Ti~2\textit{p} core level spectrum. Similar to yttrium, contributions from a hydride, oxide and hydroxide environment were expected. Following a similar procedure to the Y~3\textit{d} fits, the hydride contribution was described with an asymmetric line shape, whereas the oxide/hydroxide peaks were described with a pseudo-Voigt profile line shape described by a convolution of a Gaussian and Lorentzian function. The only constraint implemented in the peak-fitting process is that the line shape and FWHM of the oxide/hydroxide environments must be the same. Four environments were needed to provide a suitable physically-meaningful fit: Ti-H, Ti(III) oxide, Ti(IV) oxide and titanium hydroxide. The resultant line shapes after the least-squares fitting process are as follows. For the Ti-H peak, the line shape was described with an 80.11~\% Lorentzian function and 19.89~\% Gaussian function convolution, FWHM of 1.05~eV, tail exponent = 0.0351 and tail mix = 24.98~\%. Whereas, for the oxide/hydroxide environments, a 32.88~\% Lorentzian function and 67.12~\% Gaussian function convolution, and an FWHM of 1.85~eV was found. The error associated with the quantification is assumed to be $\pm$0.2~at.\% because of the difficulty in peak-fit analysis. No escape depth correction or other relative atomic sensitivity factor was applied to the peak areas.

\subsection{Ti~2\textit{p}}

\begin{figure*}
\centering
    \includegraphics[keepaspectratio, width =0.4\linewidth]{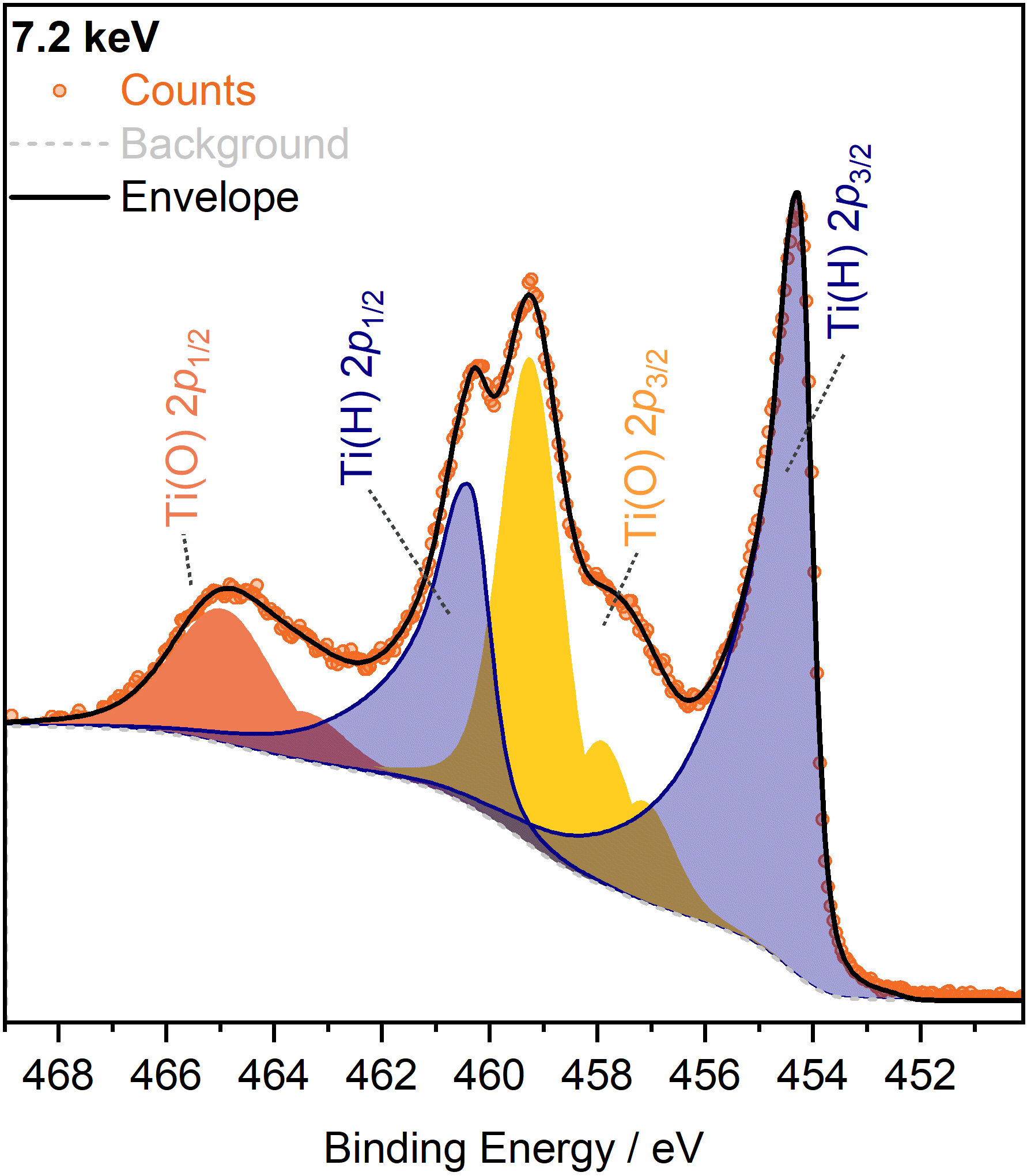}
    \caption{Representative peak-fit model of the Ti~2\textit{p} spectrum for the 7.2~keV collected spectrum. (H) refers to the hydride components, and (O) refers to the oxide components.}
    \label{fig:Ti2p_pf}
\end{figure*}

Ti~2\textit{p} was fitted using the Thermo Scientific Avantage software package v5.9925 and a Shirley-type background was applied to remove the secondary background. A representative peak-fit (of the 7.2~keV spectrum) is displayed in Fig.~\ref{fig:Ti2p_pf}. A similar approach used to fit the Ti~1\textit{s} was also applied for the Ti~2\textit{p} peak fitting procedure. The Ti~2\textit{p} displays a broadening of the lower spin state peak, owing to a Coster-Kronig decay transition,~\cite{Fuggle_1980, Nyholm_1981} which makes peak-fit analysis challenging. The SOS of the hydride doublet peak was constrained to be 6.1$\pm$0.3~eV, similar to titanium metal. The FWHM of the doublet peaks were set independently so that the 2\textit{p}\textsubscript{1/2} could be broader than the 2\textit{p}\textsubscript{3/2} to account for the Coster-Kronig decay, however, the asymmetric line shape of the pair of hydride peaks was constrained to be the same as each other. To ensure a physically meaningful fit, the doublet area ratio was constrained to 0.48 in accordance with the Scofield cross section tabulated data.~\cite{Scofield1973} The resultant line shape of the hydride peak taken from the 7.2~keV data set was then applied to the remaining datasets as the hydride peak was best resolved in this data-set compared to the others, owing to the higher photon energy and consequently deeper probing depth. For ease of analysis, the line shapes of the oxide and hydroxide environments were set to a convolution of 70\% Gaussian and 30\% Lorentzian. Four environments were required to be added, correlating well with the Ti~1\textit{s} peak-fitting result. The resultant peak-fit showed that the SOS of the hydride doublet was, on average, 6.1~eV with a standard deviation of 0.05~eV. The resultant FWHMs of the 2\textit{p}\textsubscript{3/2} and 2\textit{p}\textsubscript{1/2} hydride peaks were 0.6 and 0.9~eV, respectively, with the metal oxide 2\textit{p}\textsubscript{3/2} peaks having a broader FWHM of 1.4~eV. To determine the hydride contribution to the spectrum, the 2\textit{p}\textsubscript{3/2} peak area of the hydride was compared to the sum of the 2\textit{p}\textsubscript{3/2} peak areas of the remaining environments. No escape depth correction or other relative atomic sensitivity factor was applied to the peak areas. The error associated with the quantification is assumed to be $\pm$0.2~at.\% because of the difficulty in peak-fit analysis.

\cleardoublepage

\section{Air exposure tests}

\begin{figure}[ht!]
\centering
    \includegraphics[keepaspectratio, width=0.7\linewidth]{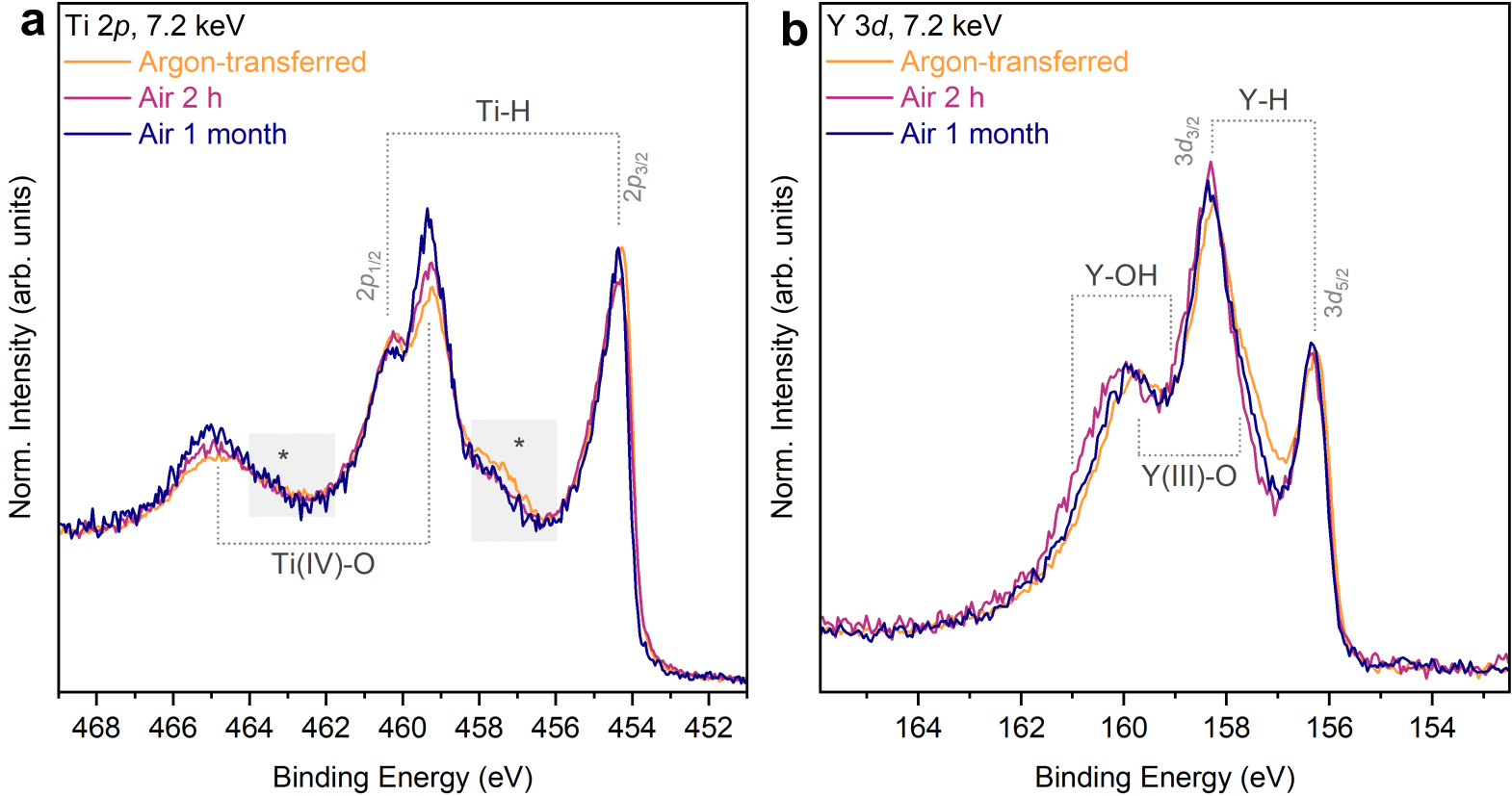}
    \caption{Core level spectra collected after exposing the samples to the ambient environment for 2~h and 1~month, and comparing them to the spectra collected without exposing the samples to air (labelled as Argon-transferred). (a) and (b) show the Ti~2\textit{p} and Y~3\textit{d} core levels collected for the TiH\textsubscript{2-$\delta$} and YH\textsubscript{2-$\delta$} samples, respectively. The spectra are normalised to their respective areas (after the removal of a Shirley-type background). Due to time constraints, the valence band spectra of the samples after exposure to air could not be included. The air-exposed core level spectra are aligned to the metal hydride peak of the Argon-transferred spectrum, with the latter already being aligned to the intrinsic $E_F$.}
    \label{fig:air}
\end{figure}

\cleardoublepage

\section{Depth distribution function}
To calculate the depth distribution function (DDF), the approach taken by Berens~\textit{et al.} was followed.~\cite{Berens_2020} This approach assumes that the samples are considered a bilayer, whereby the hydride is the bottom layer and is covered by an oxide overlayer. The required input parameters to the DDF are (a) the inelastic mean free path (IMFP) of the photoelectrons originating from their respective layers and travelling through their respective materials, (b) the number density of both layers, and (c) an estimate of the percentage of the signal originating from each layer. Several additional assumptions have been made for the calculation of the DDF:

\begin{itemize}
    \item The TPP-2M formula is suitable for calculating IMFPs when using hard X-ray photon energies.
    \item The hydrides are stoichiometric TiH\textsubscript{2} and YH\textsubscript{2}.
    \item The oxide overlayer is the highest valence state oxide only (i.e.\ TiO\textsubscript{2} and Y\textsubscript{2}O\textsubscript{3}), and all other non-hydride contributions to the spectral area are considered to be the highest valence state oxide.
    \item The photoelectrons are travelling along the surface normal.
\end{itemize}

The number density can be calculated by dividing the bulk density of the material by its atomic mass. Bulk density values of TiH\textsubscript{2} and YH\textsubscript{2} were taken from the Materials Project,~\cite{Anubhav_2013} whereas the bulk density for the oxides was taken from the QUASES software package, which implements the TPP-2M IMFP formula.~\cite{Shinotsuka_2015} This software package was also used to calculate the IMFPs of photoelectrons from each material. The densities are listed in Tab.~\ref{Densities}.

\begin{table}[h]
\caption{\label{Densities}Parameters used for the calculation of the DDF.}
\begin{ruledtabular}
\begin{tabular}{cccc}
Material & Bulk Density / g~cm\textsuperscript{-3} & Atomic Mass / g mol\textsuperscript{-1} & Number Density / mol~cm\textsuperscript{-3} \\
\hline
TiH\textsubscript{2}     & 3.83         & 49.88           & 0.077          \\
TiO\textsubscript{2}     & 4.26         & 79.90           & 0.053          \\
YH\textsubscript{2}      & 4.26         & 90.92           & 0.047          \\
Y\textsubscript{2}O\textsubscript{3}     & 5.01         & 225.81          & 0.022  \\
\end{tabular}
\end{ruledtabular}
\end{table}

Using the BE values of the Y~3\textit{d}\textsubscript{5/2} or Ti~2\textit{p}\textsubscript{3/2} hydride and main oxide peaks from the core level analysis, the kinetic energy (KE) of photoelectrons from each environment could be calculated. Using the KEs and inputting them into QUASES, the relativistic IMFP of the photoelectron in their respective material could be estimated. TiH\textsubscript{2} and YH\textsubscript{2} were not in the QUASES database and so were created as new materials. Within the material parameters of the QUASES database, the band gap of the existing QUASES database entries of TiO\textsubscript{2} and Y\textsubscript{2}O\textsubscript{3} was changed from their default of 0~eV to 3.03~\cite{Scanlon2013} and 5.60~eV~\cite{Wang_2009}, respectively. The values for this calculation are listed in Tab.~\ref{IMFP}. 

\begin{table}[h]
\caption{\label{IMFP}Input values required for calculating the IMFP and the resulting IMFP values.}
\begin{ruledtabular}
\begin{tabular}{cccccc}
Photon Energy / keV & Core Level & KE hydride~/~eV & KE oxide~/~eV & IMFP hydride / nm & IMFP oxide / nm \\
\hline
2.4     & Ti~2\textit{p}\textsubscript{3/2}         & 1956.3          & 1951.4      & 3.59 & 3.54   \\
2.4     & Y~3\textit{d}\textsubscript{5/2}         & 2254.3          & 2253.2   & 4.57  &  3.96   \\
\\
3.3     & Ti~2\textit{p}\textsubscript{3/2}          & 2814.8          & 2809.9    & 4.81 &  4.75   \\
3.3     & Y~3\textit{d}\textsubscript{5/2}           & 3112.8           & 3111.4      & 5.93 & 5.14    \\
\\
6.0     & Ti~1\textit{s}         & 1040.5         & 1036.5    & 2.19 &  2.17   \\
6.0     & Ti~2\textit{p}\textsubscript{3/2}         & 5551.0           & 5546.1    & 8.38 &  8.27   \\
6.0     & Y~3\textit{d}\textsubscript{5/2}          & 5849.2          & 5847.9   & 9.94 &  8.60   \\
\\
7.2     & Ti~1\textit{s}         & 2266.1          & 2262.1    & 4.03 &  3.99   \\
7.2    & Ti~2\textit{p}\textsubscript{3/2}         & 6776.6          & 6771.6    & 9.88 &  9.75   \\
7.2    & Y~3\textit{d}\textsubscript{5/2}         & 7074.6         & 7073.1    & 11.63 &  10.06   \\
\end{tabular}
\end{ruledtabular}
\end{table}

The final parameter required is the percentage of the signal originating from each layer. These were taken from the peak-fit analysis results and displayed in Tab.~\ref{ratio}. The hydride:non-hydride ratio refers to the hydride/(non-hydride + hydride) to non-hydride/(non-hydride + hydride) values.

\begin{table}[h]
\caption{\label{ratio}Non-hydride:hydride percentage contribution from peak-fit analysis (assuming that M-O is the area sum of all non-hydride contributions).}
\begin{ruledtabular}
\begin{tabular}{ccc}

Photon Energy / keV & Ti-O:Ti-H  & Y-O:Y-H   \\
\hline
2.4     &     70.8:29.2   & 90.5:9.5            \\
3.3     & 61.9:38.1          & 85.1:14.9         \\
6.0     & 49.1:50.9         & 72.6:27.4           \\
7.2    & 47.1:52.9        & 68.0:32.0          \\
\end{tabular}
\end{ruledtabular}
\end{table}

\begin{figure}[ht!]
\centering
    \includegraphics[keepaspectratio, width=0.4\linewidth]{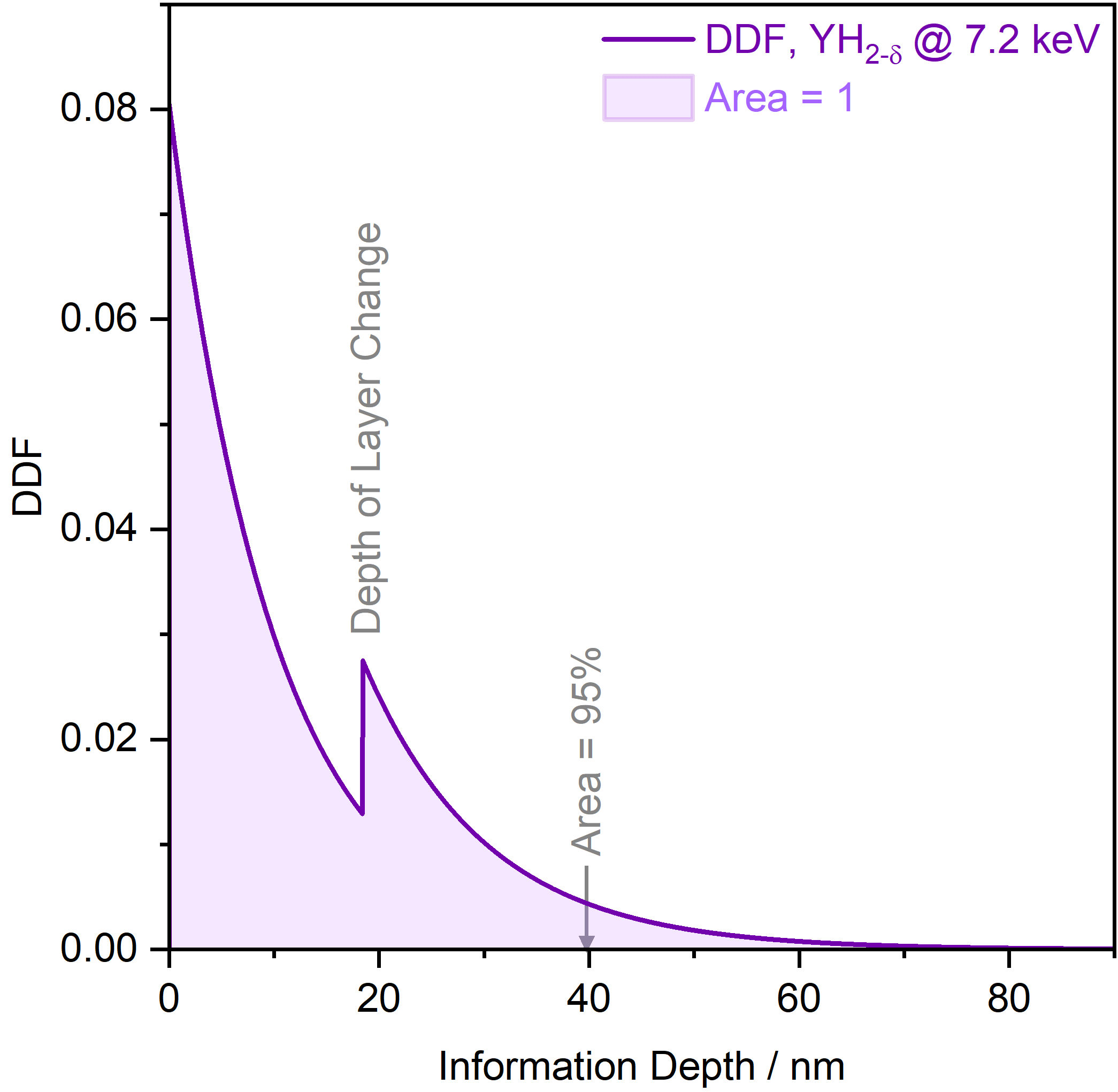}
    \caption{Determination of the information depth in the YH\textsubscript{2-$\delta$} sample when measured at a 7.2~keV photon energy. The area under the DDF curve equates to one and finding the depth at which the area is 95\% of the total area (i.e. 0.95), marks the estimate probing depth or information depth of the measurement.}
    \label{fig:DDF_95}
\end{figure}

Using the input values listed in Tables~\ref{Densities} and \ref{IMFP} above and a goal-seeking tool to set the non-hydride:hydride ratio to the values listed in Tab.~\ref{ratio}, the oxide overlayer thickness could be estimated. From the resultant DDFs, the estimated probing depth or information depth at each X-ray photon energy was calculated by integrating over 95\% of the area as shown in Fig.~\ref{fig:DDF_95}.

\cleardoublepage

\section{Complete valence band spectra}

\begin{figure*}[h]
\centering
    \includegraphics[keepaspectratio, width = 0.7\linewidth]{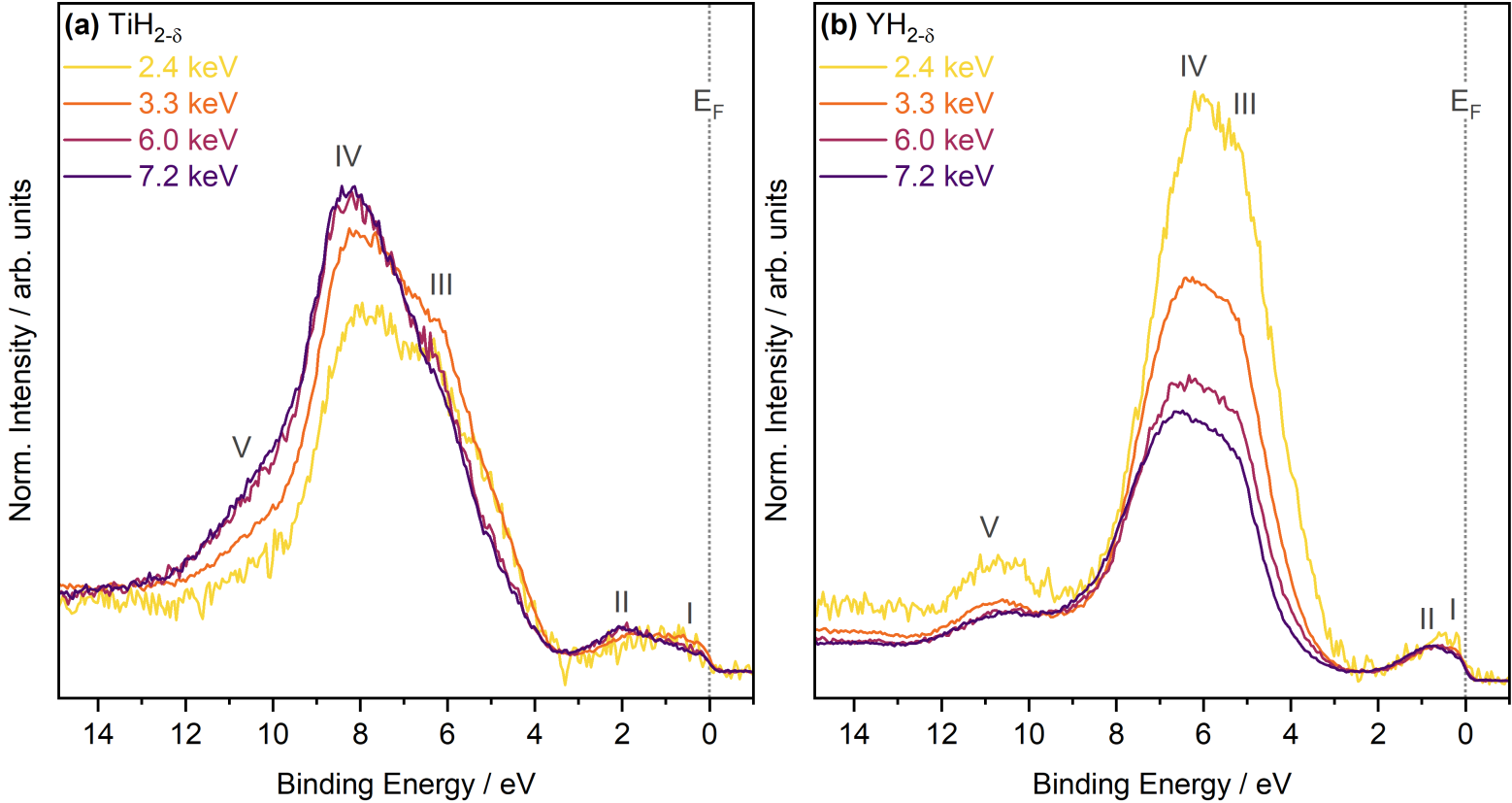}
    \caption{Valence band spectra collected as a function of photon energy for (a) TiH\textsubscript{2-$\delta$} and (b) YH\textsubscript{2-$\delta$}. Spectra are normalised to the area of the features adjacent to the $E_F$ between 0-4 and 0-2~eV (determined after the removal of a Shirley-type background), for TiH\textsubscript{2-$\delta$} and YH\textsubscript{2-$\delta$}, respectively. (a) and (b) are plotted on different $y$-scales. }
    \label{fig:All_VBs}
\end{figure*}

\cleardoublepage

\section{Shallow core levels} \label{sec:SCL}

\begin{figure*}[h]
\centering
    \includegraphics[keepaspectratio, width = 0.7\linewidth]{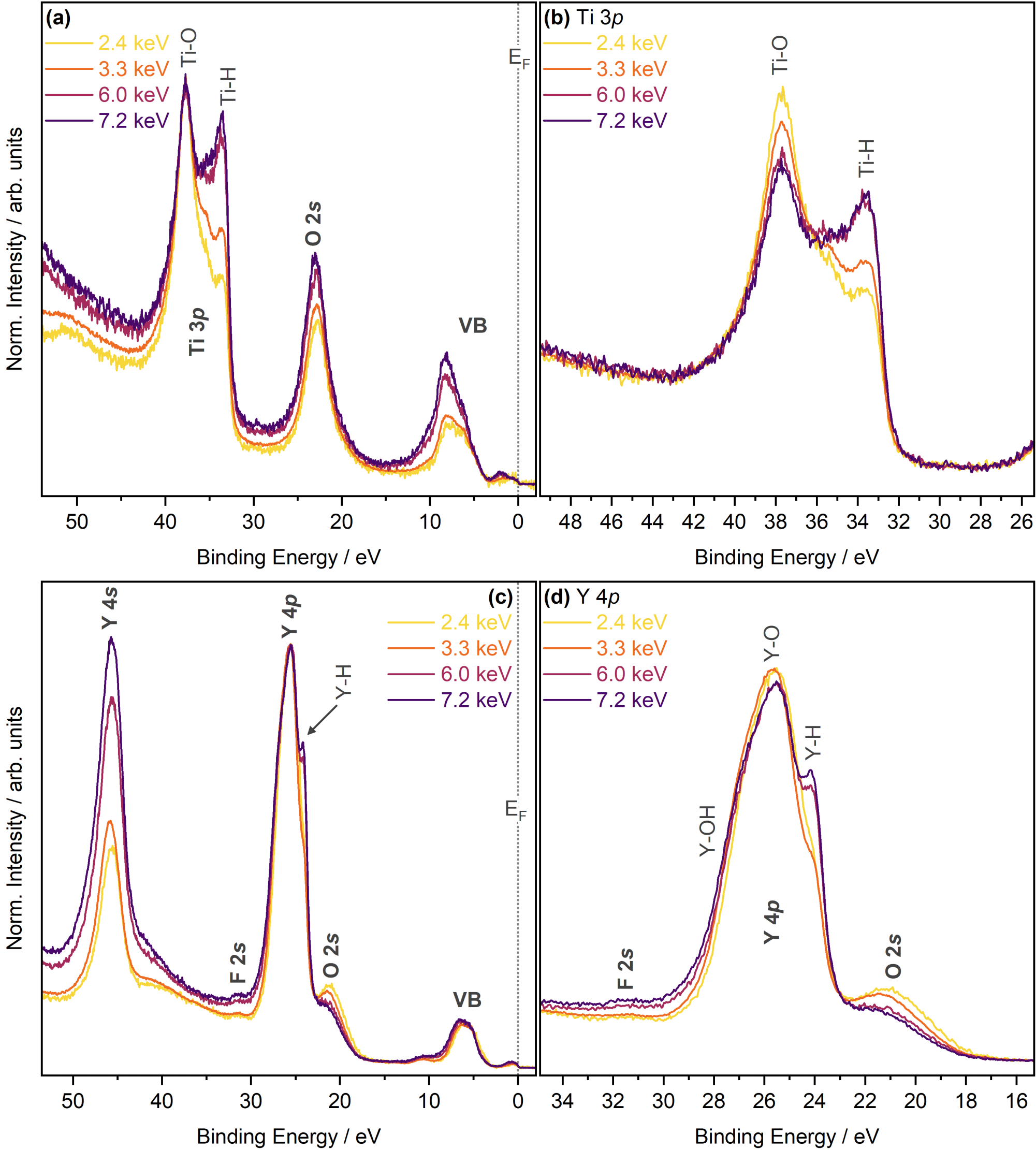}
    \caption{Shallow core level and valence band spectra of (a) TiH\textsubscript{2-$\delta$} and (c) YH\textsubscript{2-$\delta$}, including magnified views of the (b) Ti~3\textit{p} and (d) Y~4\textit{p} shallow core levels. (a) and (c) are normalised to the maximum intensity of the Ti~3\textit{p} and Y~4\textit{p}, respectively, whereas (b) and (d) are normalised to their respective areas.}
    \label{fig:SCL}
\end{figure*}

\cleardoublepage

\section{3.3 \lowercase{ke}V weighted PDOS} \label{sec:3.3_Weighted}

\begin{figure*}[h]
\centering
    \includegraphics[keepaspectratio, width = 0.875\linewidth]{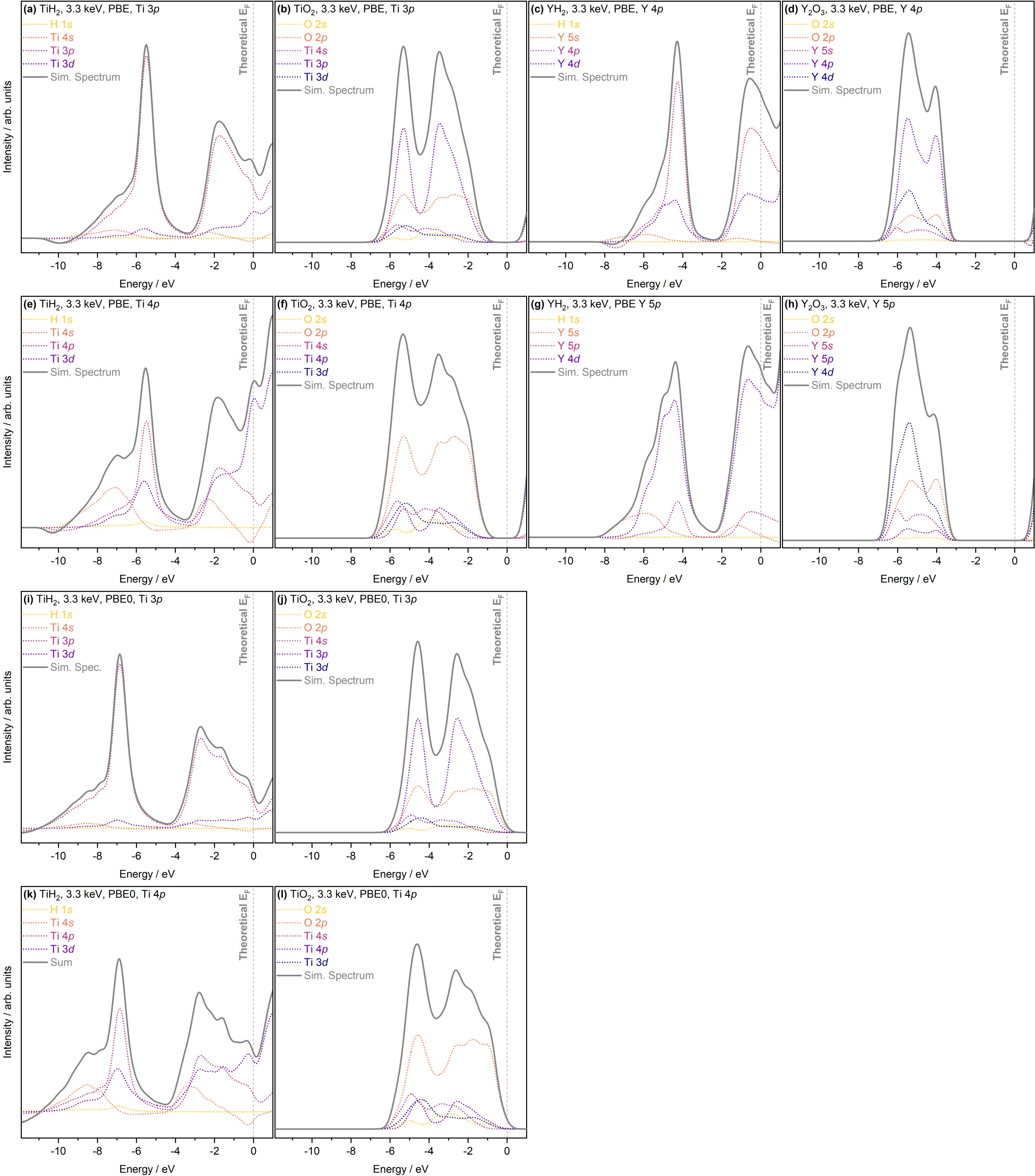}
    \caption{3.3~keV photoionisation cross section weighted PDOS calculated with PBE and PBE0. The top two rows (a-h) are calculated using PBE, with the bottom two rows (i-l) calculated using PBE0. Counting from the top, rows 1 and 3 were weighted according to the tabulated Scofield cross section database, and rows 2 and 4 were weighted using estimated cross sections for unoccupied orbitals. Going from (L-R) TiH\textsubscript{2}, TiO\textsubscript{2}, YH\textsubscript{2} and Y\textsubscript{2}O\textsubscript{3} calculated PDOS. Sim. Spectrum (simulated spectrum) corresponds to the sum of the weighted PDOS contributions. All spectra are aligned to the theoretically determined Fermi energy. The cross sections used to weight the orbitals are one electron corrected. Spectra are broadened with a 240 meV Gaussian smearing.}
    \label{fig:3.3_PDOS}
\end{figure*}
\cleardoublepage

\section{7.2 \lowercase{ke}V weighted PDOS} \label{sec:7.2_Weighted}

\begin{figure*}[h]
\centering
    \includegraphics[keepaspectratio, width = 0.875\linewidth]{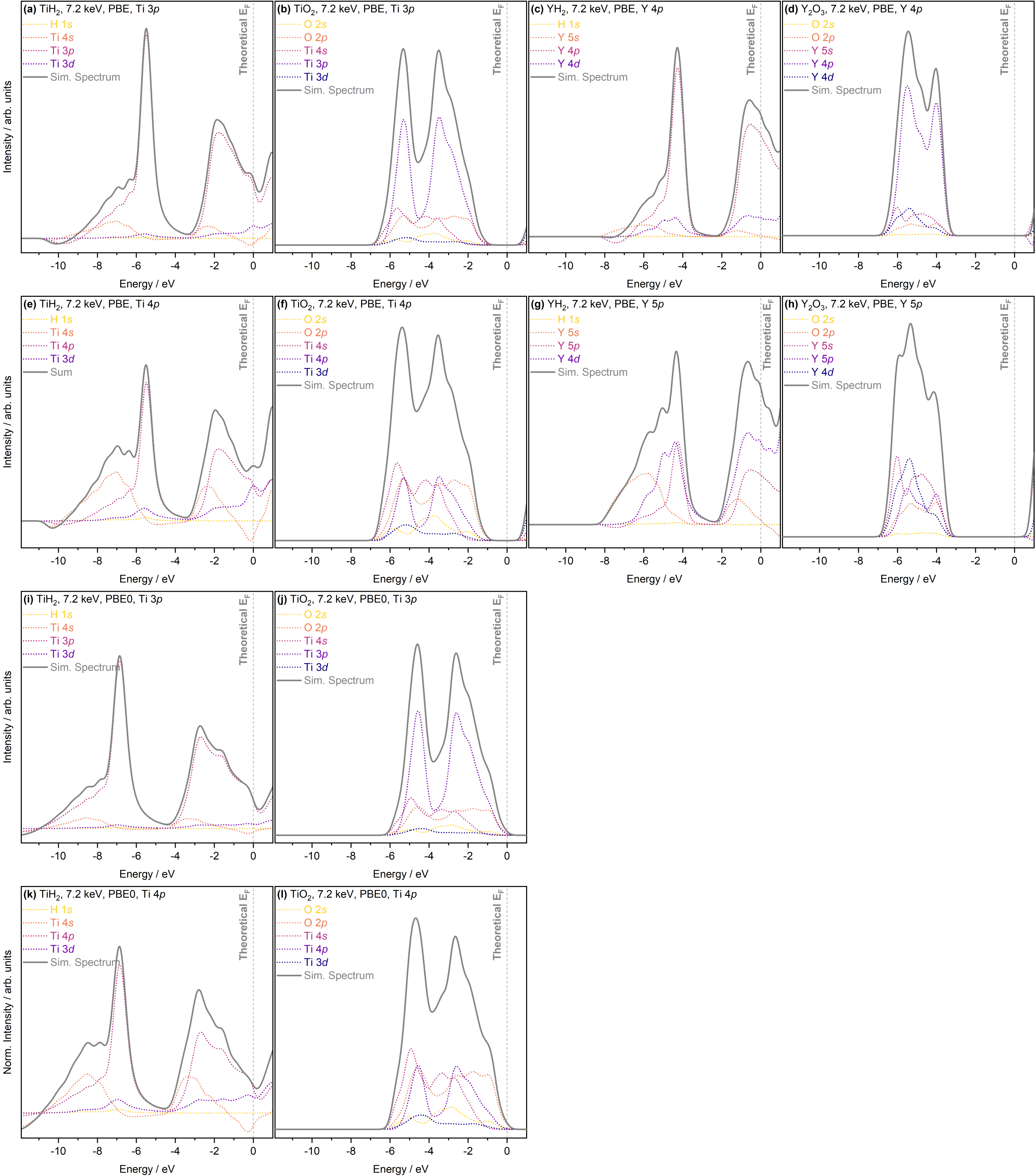}
    \caption{7.2~keV photoionisation cross section weighted PDOS calculated with PBE and PBE0. The top two rows (a-h) are calculated using PBE, with the bottom two rows (i-l)calculated using PBE0. Counting from the top, rows 1 and 3 were weighted according to the tabulated Scofield cross section database, and rows 2 and 4 were weighted using estimated cross sections for unoccupied orbitals. Going from (L-R) TiH\textsubscript{2}, TiO\textsubscript{2}, YH\textsubscript{2} and Y\textsubscript{2}O\textsubscript{3} calculated PDOS. Sim. Spectrum (simulated spectrum) corresponds to the sum of the weighted PDOS contributions. All spectra are aligned to the theoretically determined Fermi energy. The cross sections used to weight the orbitals are one electron corrected. Spectra are broadened with a 200 meV Gaussian smearing.}
    \label{fig:7.2_PDOS}
\end{figure*}
\cleardoublepage

\section{Unweighted PDOS} \label{sec:Unweighted}

\begin{figure*}[h]
\centering
    \includegraphics[keepaspectratio, width = 0.875\linewidth]{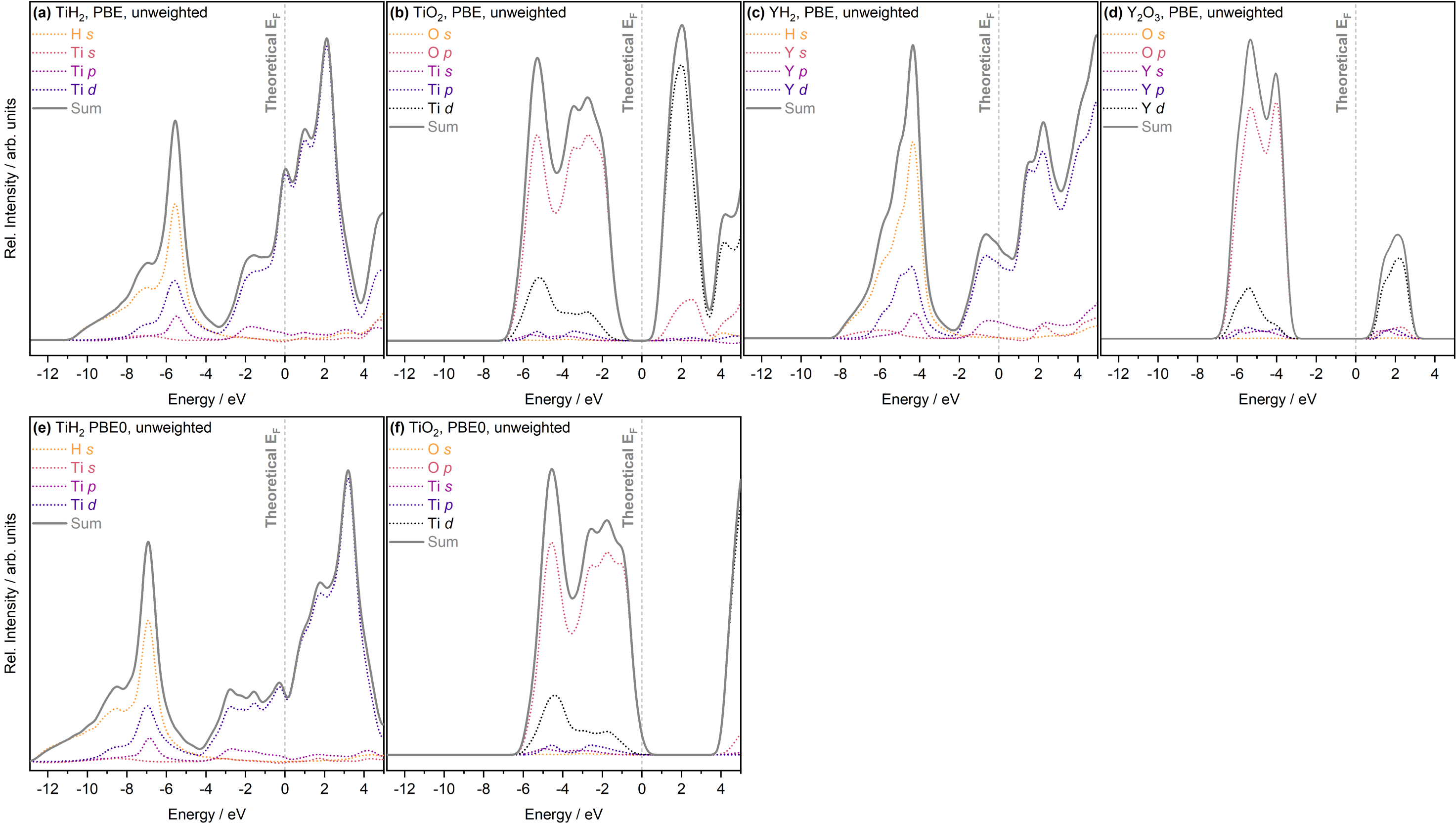}
    \caption{Unweighted PDOS spectra calculated using (a)-(d) PBE and (e)-(f) PBE0 for (a) and (e) TiH\textsubscript{2}, (b) and (f) TiO\textsubscript{2}, (c) YH\textsubscript{2}, and (d)  Y\textsubscript{2}O\textsubscript{3}. Spectra are aligned to the theoretical $E_F$ and ``Sum'' refers to the sum of all contributions to the PDOS. Spectra are broadened with a 240 meV Gaussian smearing.}
    \label{fig:Unweight_PDOS}
\end{figure*}

\cleardoublepage

\section{Tabulated charges, bond populations and bond lengths extracted from DFT calculations} \label{sec:tab_charges}

\begin{table*}[h]
\caption{\label{Charges}Summary of calculated charges, bond populations and bond lengths. For systems where inequivalent metal-O bonds exist, the bond populations and corresponding bond lengths are given in the same order. Therefore, while the bond lengths are not directly of interest in this work, they are useful for interpreting the bond populations. Calculations with both PBE and PBEO functionals were performed for the metal dihydride and oxide Y and Ti systems. The effective valence charge was taken as the difference between the formal ionic charge and the Mulliken/Hirshfeld/Bader charge on the anion species in the crystal.}
\begin{ruledtabular}
\begin{tabular}{lr|cccc|cc}

\multicolumn{1}{l}{}                &        & \multicolumn{4}{c|}{\textbf{PBE}}                               & \multicolumn{2}{c}{\textbf{PBE0}} \\
\hline
\multicolumn{1}{l}{}                &        & \textbf{TiH\textsubscript{2}}  & \textbf{TiO\textsubscript{2} }      & \textbf{YH\textsubscript{2}}   & \textbf{Y\textsubscript{2}O\textsubscript{3}  }                   & \textbf{TiH\textsubscript{2}}     & \textbf{TiO\textsubscript{2} }         \\
\hline

\textbf{Mulliken   Charges }                 & Ti/Y   & 0.66  & 1.34       & 0.76  & 1.20                     & 0.77     & 1.54          \\
\textbf{}                           & H      & -0.33 & -          & -0.38 & -                        & -0.39    & -             \\
\textbf{}                           & O      & -     & -0.67      & -     & -0.80                    & -        & -0.77         \\
\textbf{}                           & Effective Valence      & 1.34     & 2.66      & 1.24     & 3.60   & 1.22        & 2.46         \\

\hline
\textbf{Hirshfeld   Charges }                 & Ti/Y   & 0.22  & 0.57       & 0.29  & 0.56/0.57                     & -     & -          \\
\textbf{}                           & H      & -0.11 & -          & -0.14 & -                        & -    & -             \\
\textbf{}                           & O      & -     & -0.29      & -     & -0.38                    & -        & -         \\
\textbf{}                           & Effective Valence      & 1.78     & 3.43      & 1.71     & 4.88   & -        & -        \\

\hline
\textbf{Bader   Charges }                 & Ti/Y   & 1.26  & 2.25       & 1.60  & 2.13/2.14                     & -     & -          \\
\textbf{}                           & H      & -0.63 & -          & -0.80 & -                        & -    & -             \\
\textbf{}                           & O      & -     & -1.13      & -     & -1.42                    & -        & -        \\
\textbf{}                           & Effective Valence      & 0.74     & 1.74      & 0.40     & 1.74   & -        & -         \\

\hline
\textbf{Bond Population}                     & Ti/Y-H & 0.08  & -          & 0.12  & -                        & 0.06     & -             \\
\textbf{}                           & Ti/Y-O & -     & 0.38, 0.24 & -     & 0.33, 0.31, 0.30,   0.23 & -        & 0.38, 0.12    \\
\hline
\multicolumn{1}{l}{\textbf{Bond Length / {\AA}}} & Ti/Y-H & 1.91  & -          & 2.25  & -                        & 1.91     &               \\
\multicolumn{1}{l}{\textbf{}}       & Ti/Y-O & -     & 1.96, 2.01 & -     & 2.26,   2.28, 2.29, 2.34 & -        & 1.96, 2.01   \\

\end{tabular}
\end{ruledtabular}

\end{table*}

\cleardoublepage

\section{DFT enthalpy of formation calculations} \label{sec:DFT_calc_FE}

The additional calculations for H\textsubscript{2} and bulk Ti and Y were performed using PBE, following the same procedure as the calculations from the manuscript, and employing the same cut-off energy and type of pseudopotentials. A 25~{\AA} supercell size was used to run the calculations for H\textsubscript{2}. Both Ti and Y metals were calculated using an HCP crystal structure and a 2-atom cell. For Ti, a 5$\times$5$\times$5 \textit{k}-point grid was selected, whereas 6$\times$6$\times$6 was selected for Y, both of which are lower than the 10$\times$10$\times$10 \textit{k}-point grid used to generate the PDOS.

\begin{table}[h]
\caption{\label{dft_calc}Energies of the hydrogen, metal, and metal dihydride systems calculated with DFT.}
\begin{ruledtabular}
\begin{tabular}{cccc}
\textbf{System} & \textbf{Energy, $E$ / eV} & \textbf{Number of Ti/Y atoms} & \textbf{Number of H atoms} \\
\hline
H\textsubscript{2} & -31.7395 & 0 & 2  \\
&&&\\
TiH\textsubscript{2} cubic & -6469.4131 & 4 & 8  \\
TiH\textsubscript{2} tetragonal & -6469.4212 & 4 & 8  \\
Ti & -3168.2677 & 2 & 0  \\
&&&\\
YH\textsubscript{2} cubic & -4487.1987 & 4 & 8  \\
Y & -2175.7426 & 2 & 0  \\

\end{tabular}
\end{ruledtabular}
\end{table}

Using the values listed in Tab.~\ref{dft_calc} and accounting for the number of atoms, the enthalpy of formation for titanium dihydride is calculated as follows:

\begin{align}
    & Ti + 2H \rightleftharpoons TiH_2 \\
    & {\Delta}H_f = E_{products} - E_{reactants} \\
    & {\Delta}H_f = -6469.4131 - [(-31.7395 \times \frac{8}{2}) + (-3168.2677 \times \frac{4}{2})] \\
   & {\Delta}H_f = -5.92~\text{eV} \\
    & {\Delta}H_f = -5.92~\text{eV} \times 1.602\times10^{-22}~\text{kJ} \times 6.022\times10^{23}~\text{mol\textsuperscript{-1}} = -571.12~\text{kJ/mol} \\
   & {\Delta}H_f = -571.12~\text{kJ/mol} \div \text{8~H~atoms} = \textbf{-71.39~kJ/mol H }
\end{align}

Using the same approach for yttrium dihydride gives a value of -105.58~kJ/mol H. Both are in good agreement with the values determined using the Griessen model ($\Delta H_{f, YH_{2-\delta}}$ = -~111~kJ/mol~H, $\Delta H_{f, TiH_{2-\delta}}$ = -~76~kJ/mol~H) and those extracted from the literature wherein thermodynamical methods were used to determine the enthalpy of formation ($\Delta H_{f, YH_2}$ = -112.25~kJ/mol~H (between 873-1073~K),~\cite{CHERNIKOV1987441} $\Delta H_{f, TiH_2}$ = -68.47~kJ/mol~H (at 737~K)~\cite{DANTZER1983913, JZhao_2008}).

\cleardoublepage

\section{Comparison between the metal and dihydride PDOS} \label{sec:PDOS_met_dih}

\begin{figure*}[h]
\centering
    \includegraphics[keepaspectratio, width = 0.6\linewidth]{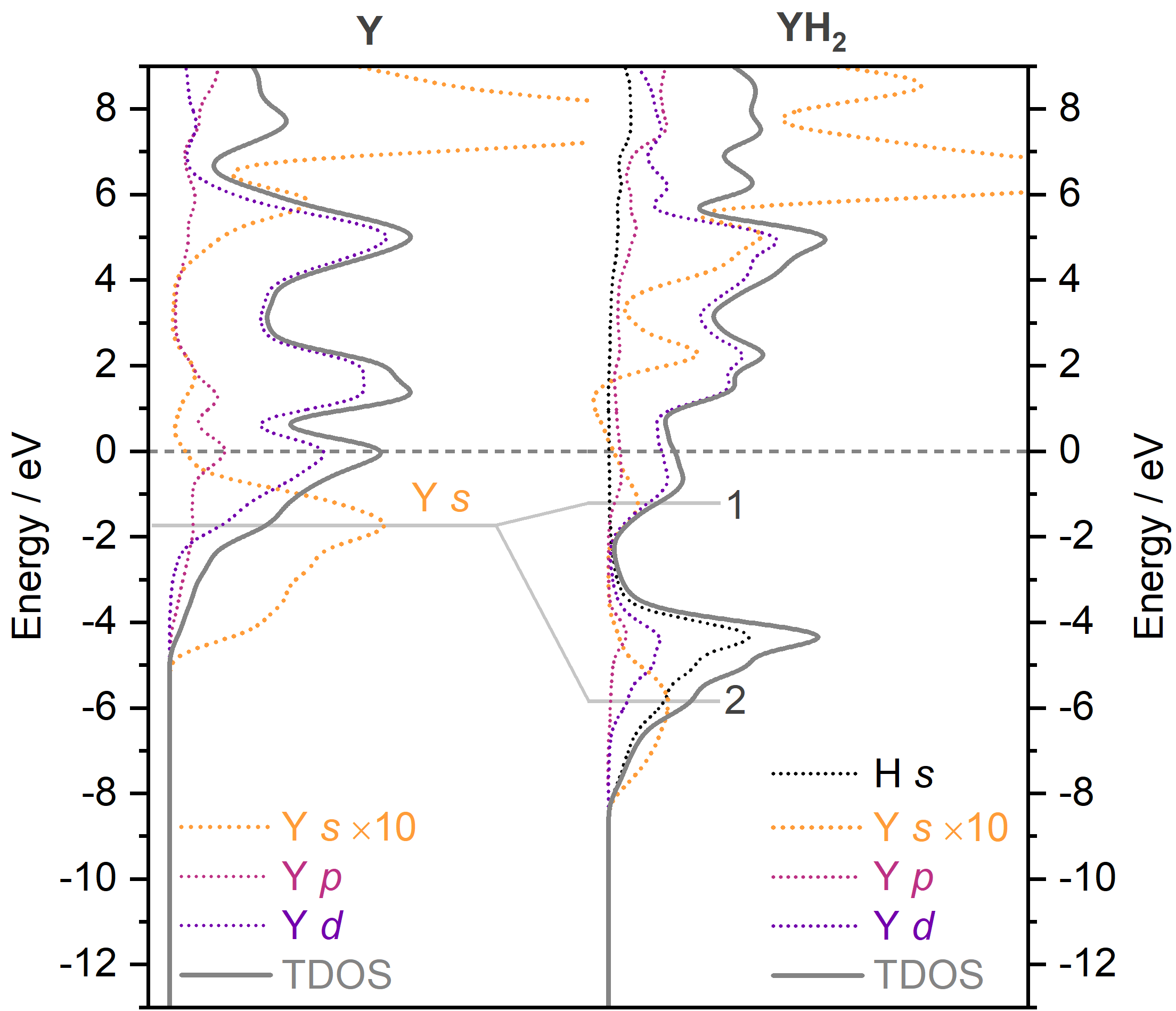}
    \caption{Comparison between the projected density of states (PDOS) of Y metal (left) and YH\textsubscript{2} (right). The PDOS of both have not been weighted with photoionisation cross sections (i.e.\ unweighted), but they have been aligned to the theoretical $E_F$, normalised to the maximum intensity of the occupied states, and applied with the same level of Gaussian broadening (240 meV). To aid with the identification of the \textit{s} band peak positions, the \textit{s} states for both PDOS were magnified by a factor of ten, but the total density of states (TDOS) does not account for this magnification. Grey solid guidelines are shown to highlight the lowering of the main intensity \textit{s} band (2), and the formation of a small \textit{s} band (1) pulled below $E_F$, upon hydriding.}
    \label{fig:PDOS_met_dih_Y}
\end{figure*}

\begin{figure*}[h]
\centering
    \includegraphics[keepaspectratio, width = 0.6\linewidth]{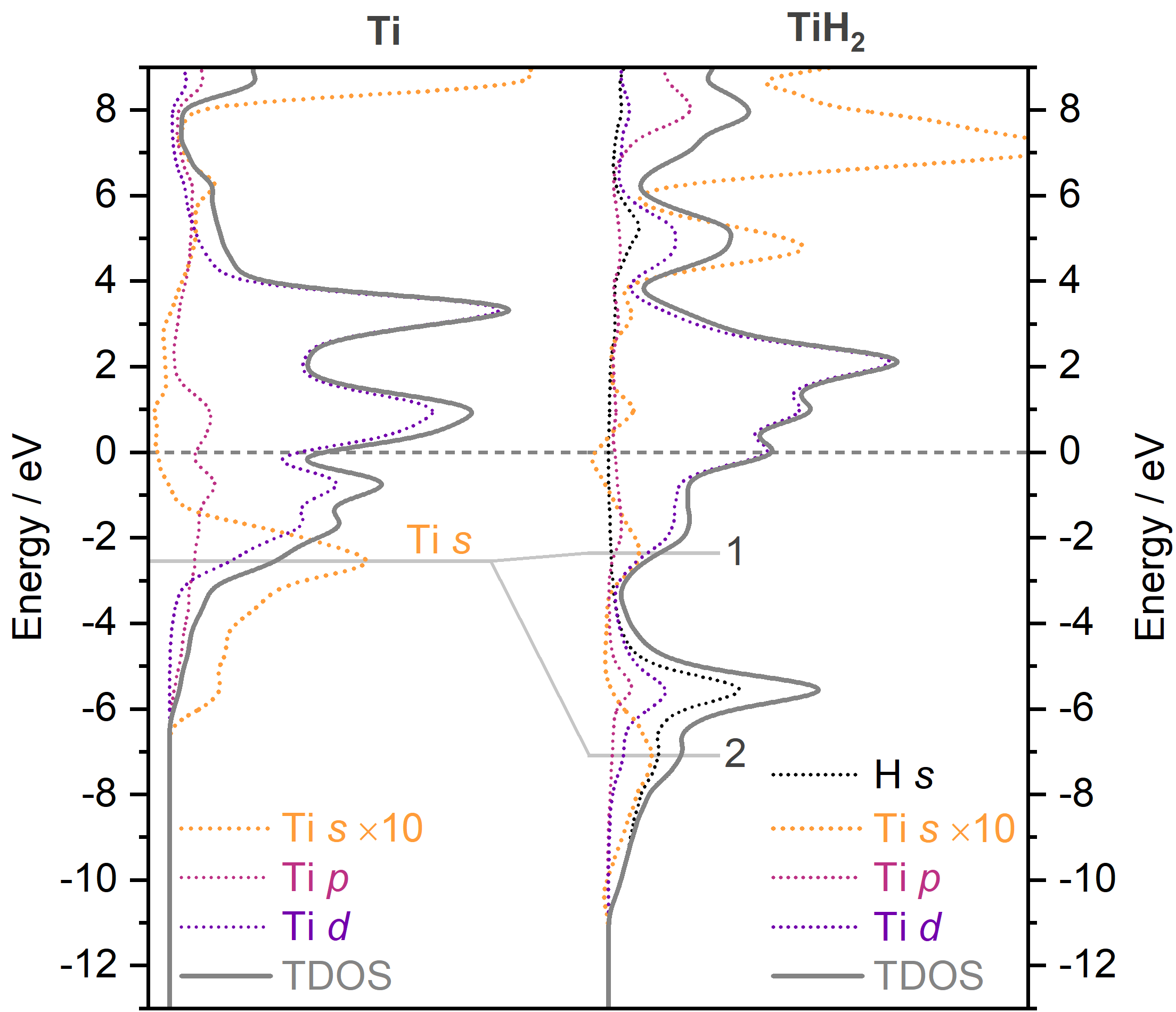}
    \caption{Comparison between the projected density of states (PDOS) of Ti metal (left) and TiH\textsubscript{2} (right). The PDOS have not been weighted with photoionisation cross sections (i.e.\ unweighted), but they have been aligned to the theoretical $E_F$, normalised to the maximum intensity of the occupied states, and applied with the same level of Gaussian broadening (240 meV). To aid with the identification of the \textit{s} band peak positions, the \textit{s} states for both PDOS were magnified by a factor of ten, but the total density of states (TDOS) does not account for this magnification. Grey solid guidelines are shown to highlight the lowering of the main intensity \textit{s} band (2), and the formation of a small \textit{s} band (1) pulled below $E_F$, upon hydriding.}
    \label{fig:PDOS_met_dih_Ti}
\end{figure*}

\cleardoublepage

\section{Ti and Y metal valence band spectra collected with HAXPES} \label{sec:Metals}

\begin{figure*}[h]
\centering
    \includegraphics[keepaspectratio, width = 0.6\linewidth]{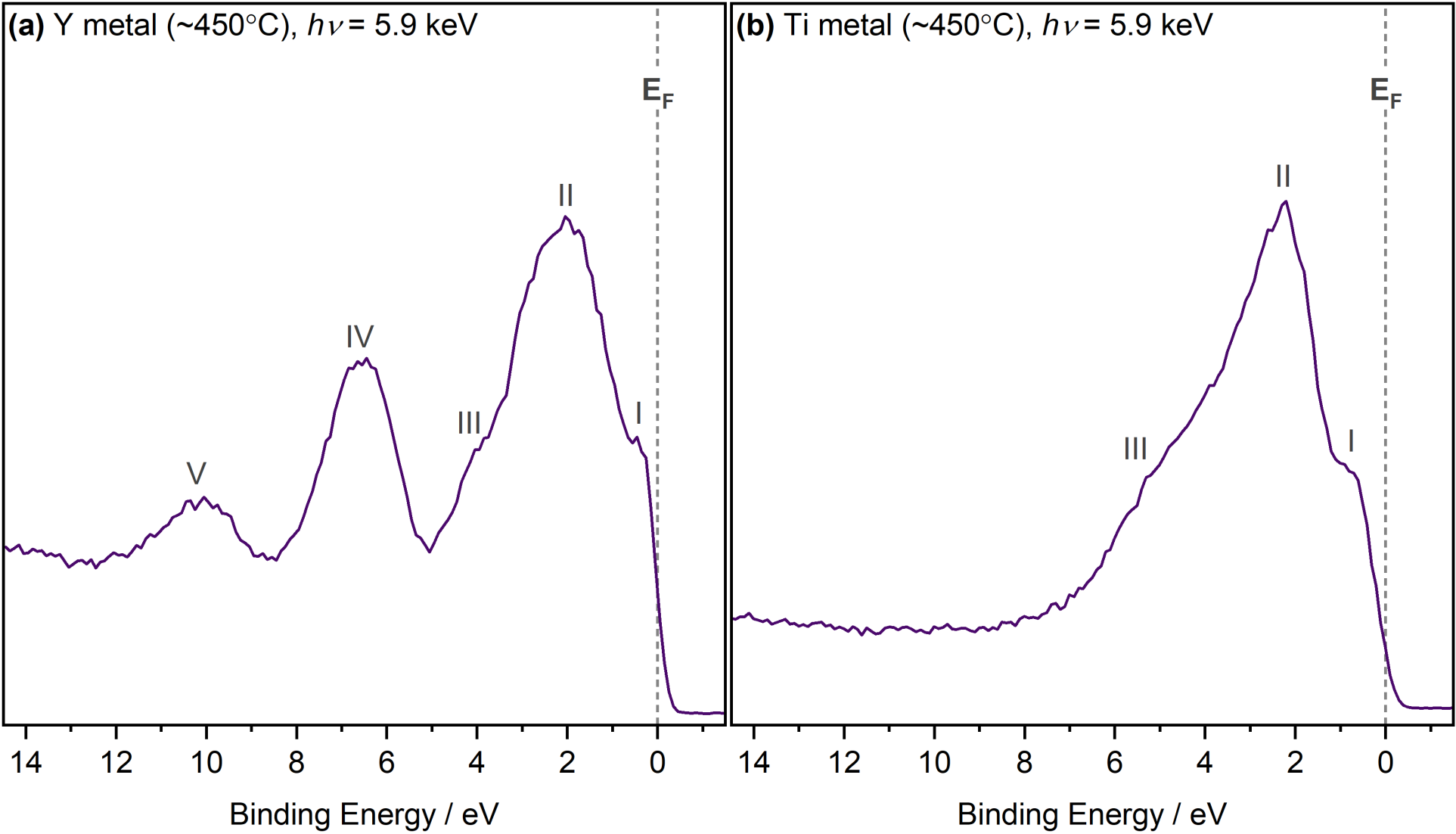}
    \caption{Valence band spectra of (a) Y metal and (b) Ti metal, collected with HAXPES ($h\nu$ = 5.9~keV) at beamline I09 (Diamond Light Source, U.K.). The samples (which were polycrystalline high-purity metal foils) were actively heated to approximately 450$\degree$C so that a clean surface was maintained during the collection of spectra. The main features of the VBs are annotated with Roman numerals. The spectra are aligned to their intrinsic $E_F$.}
    \label{fig:Metals}
\end{figure*}

Fig.~\ref{fig:Metals} displays the valence band spectra collected with HAXPES on Ti and Y metal foils. The HAXPES measurements were conducted at beamline I09 of the Diamond Light Source (UK).~\cite{Duncan_2018} High-purity (99+\% metal basis) foils were acquired from Goodfellow Cambridge Ltd. (UK) for the measurements. A photon energy, $h\nu$ of 5.9266~keV (5.9~keV) was selected using a Si(111) double crystal monochromator and S(004) channel-cut monochromator, achieving a room temperature total energy resolution of 320~meV. The end station is equipped with a high-voltage hemispherical VG Scienta EW4000 electron analyser, providing a wide $\pm$28{\textdegree} acceptance angle and operates under a base pressure of 3$\times$10\textsuperscript{-10}~mbar. Both metals are notorious getterers of oxygen and so the main difficulty of these measurements was to obtain and then maintain a clean metal surface free from oxide contributions. To achieve this, both metals were first acid etched ex-situ to destabilise the native oxide layer, and then run through cycles of in-situ argon ion sputtering and heating cycles. The samples were measured separately and during the collection of spectra, the samples were heated to an approximate temperature of 450$\degree$C to limit the recombination of residual gases adsorbing to their surfaces. Oxygen and carbon signals were minimised to nearly zero when measuring Ti, however, for Y, the signals although minimised as much as possible still remained.

\cleardoublepage

\section{Tabulated enthalpy of formation values} \label{sec:Delta_H_EF}

\begin{table*}[ht!]
\caption{\label{enthalpy_formation}Enthalpy of formation ($\Delta{H_f}$) values calculated using the various methods. For all rows where $\Delta{E}$ has been calculated, the corresponding $\Delta{H_f}$ has been estimated using the empirical formula by Griessen and Driessen (eqn.~(2) in the manuscript).}
\begin{ruledtabular}
\begin{tabular}{cccc}
\textbf{Hydride} & \textbf{Method} & \textbf{$\Delta{E}$ / eV} & \textbf{$\Delta{H_f}$ / kJ/mol H} \\
\hline
& Extracted from Fig.~4 in Ref.~\cite{Griessen_1984} & 2.41 & -63.6 \\
& Optimised $E_s$ value taken from Tab.~II in Ref.~\cite{Griessen_1984} & 2.55 & -59.5 \\

& $E_s$ taken as the Ti metal \textit{s} band position from PDOS & 2.54 & -59.8 \\
& $E_s$ taken as the pulled \textit{s} band position near $E_F$ from the TiH\textsubscript{2} PDOS & 2.34 & -65.7 \\
\textbf{TiH\textsubscript{2}} & $E_s$ taken as the TiH\textsubscript{2} lowest-lying H-induced \textit{s} band position from PDOS & 7.08 & +74.7 \\

& $E_s$ taken as feature II in TiH\textsubscript{2-$\delta$} HAXPES VB spectrum & 2.0 & -75.8 \\
& $E_s$ taken as feature II in Ti metal HAXPES VB spectrum & 2.2 & -69.8 \\

& DFT & - & -71.39 \\
& Literature (Microcalorimetry measurements)~\cite{DANTZER1983913} & - & -68.47 \\
& Literature (Extracted from Fig.~4 in Ref.~\cite{Griessen_1984}) & - & -67.6 \\
\\

& Extracted from Fig.~4 in Ref.~\cite{Griessen_1984} & 1.47 & -91.5 \\
& Optimised $E_s$ value taken from Tab.~II in Ref.~\cite{Griessen_1984} & 0.85 & -109.8 \\

& $E_s$ taken as the Y metal \textit{s} band position from PDOS & 1.72 & -84.1 \\
& $E_s$ taken as the pulled \textit{s} band position near $E_F$ from the YH\textsubscript{2} PDOS & 1.19 & -99.8 \\
\textbf{YH\textsubscript{2}}  & $E_s$ taken as the YH\textsubscript{2} lowest-lying H-induced \textit{s} band position from PDOS & 5.88 & +39.2 \\

& $E_s$ taken as feature II in YH\textsubscript{2-$\delta$} HAXPES VB spectrum & 0.8 & -111.3 \\
& $E_s$ taken as feature II in Y metal HAXPES VB spectrum & 2.0 & -75.8 \\
& DFT & - & -105.58 \\
& Literature (Dissociation pressure measurements)~\cite{CHERNIKOV1987441} & - & -112.25 \\
& Literature (Extracted from Fig.~4 in Ref.~\cite{Griessen_1984}) & - & -137.7 \\

\end{tabular}
\end{ruledtabular}
\end{table*}

\cleardoublepage

\section{Resolution} \label{sec:Res_Au}

\begin{figure*}[ht]
\centering
    \includegraphics[keepaspectratio, width = 0.65\linewidth]{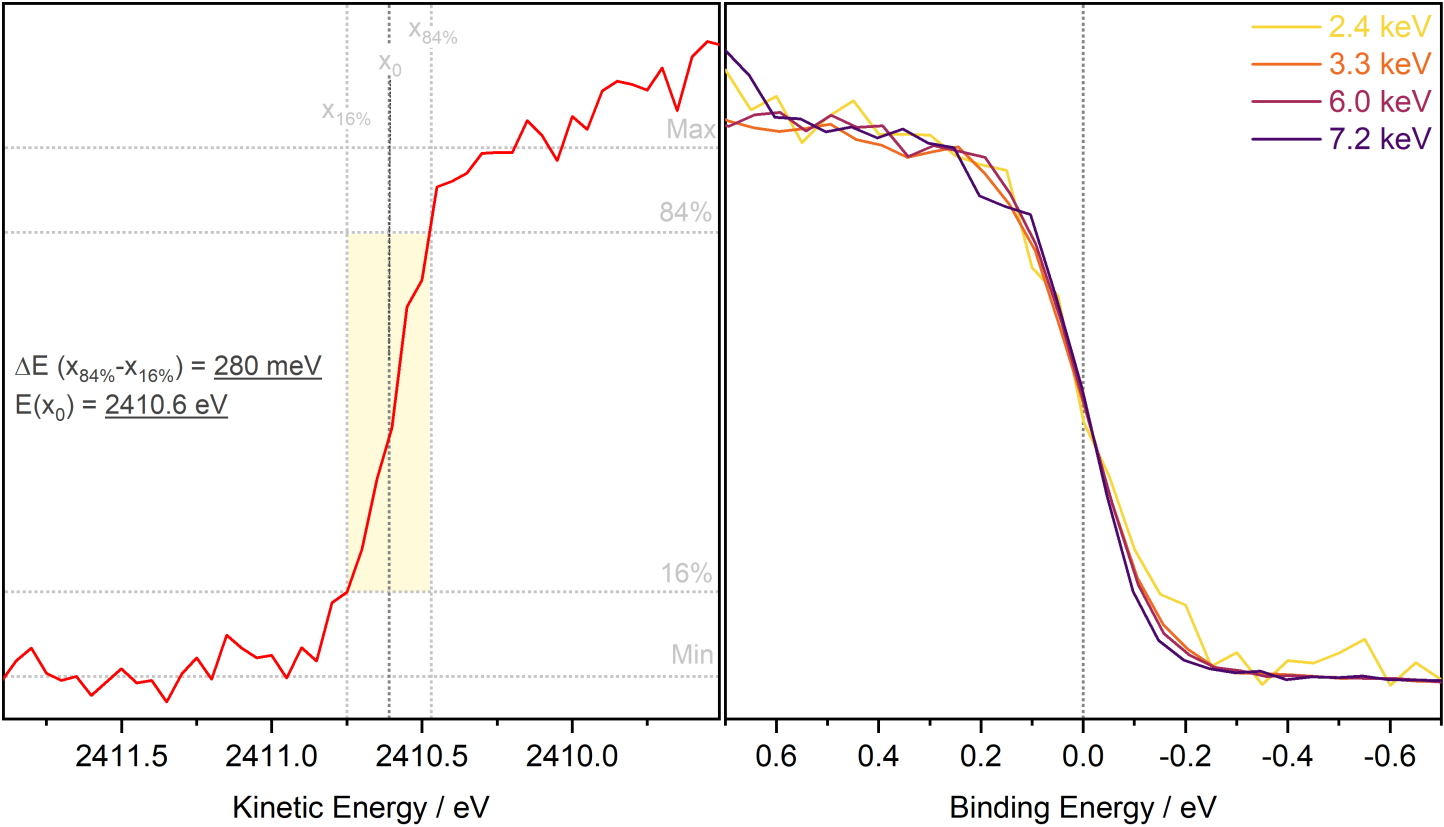}
    \caption{Fermi edge energy resolution measurements and analysis, including the measured Fermi edge width of a polycrystalline gold foil determined using the 16/84\% method for the 2410.6~eV photon energy and plotted on the raw kinetic energy scale (left), and the Au Fermi edges collected as a function of photon energy, normalised to their maximum height and plotted on the corrected binding energy scale (right).}
    \label{fig:Au_Res}
\end{figure*}

\cleardoublepage

\section{Influence of cubic versus tetragonal crystal structure on the PDOS of titanium dihydride} 

\begin{figure*}[h]
\centering
    \includegraphics[keepaspectratio, width = \linewidth]{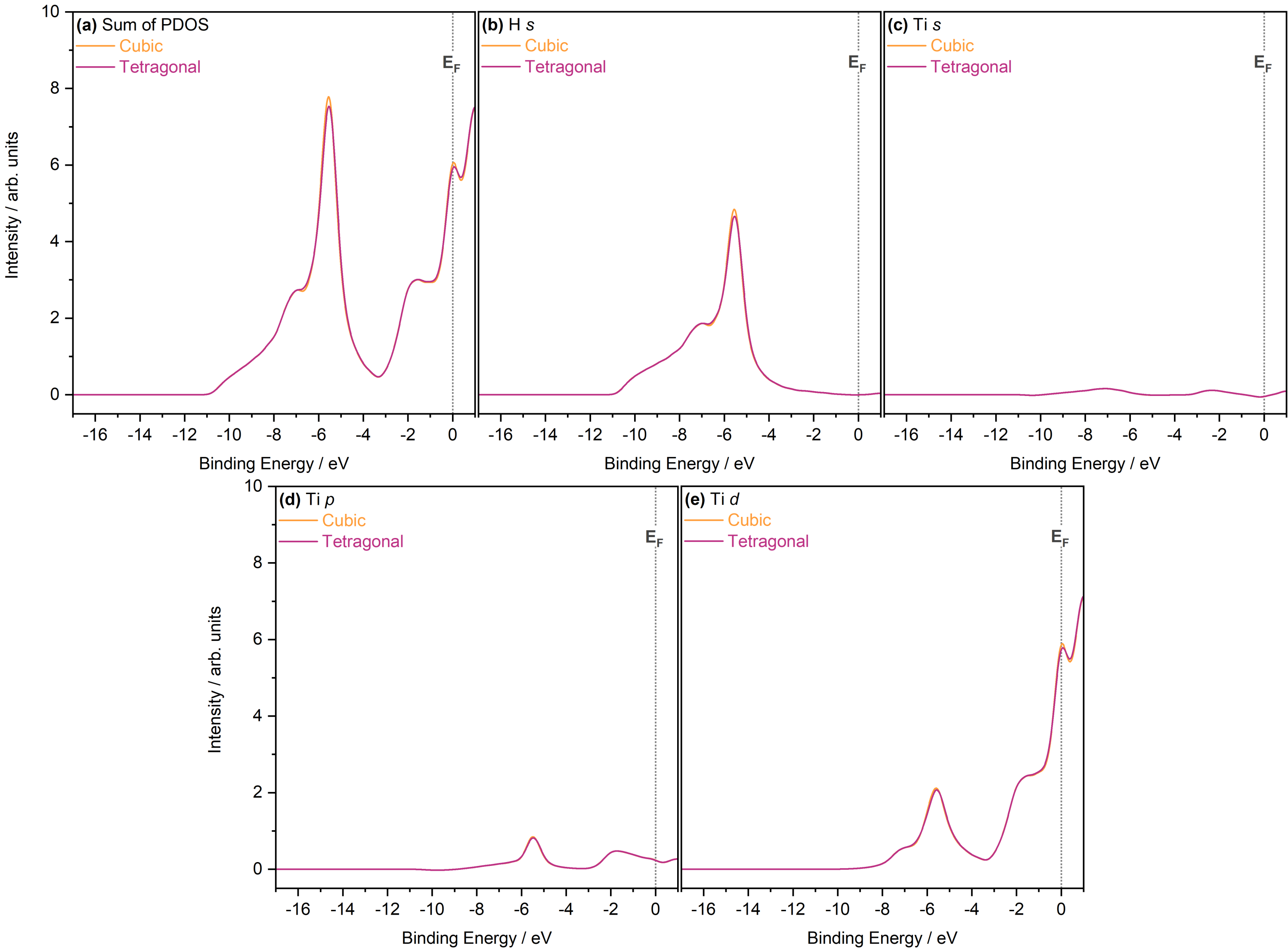}
    \caption{Comparison of the PDOS calculated using the PBE functional for TiH\textsubscript{2} when either the cubic or tetragonal crystal structure is used, including (a) the sum of the PDOS, (b) H~\textit{s}, (c) Ti~\textit{s}, (d) Ti~\textit{p} and (e) Ti~\textit{d}. The spectra are plotted as they were calculated and on the same \textit{y}-axis. The energy \textit{x}-axis is aligned to the theoretically calculated $E_F$.}
    \label{fig:Tetra}
\end{figure*}

\cleardoublepage
\section{Lattice parameters of relaxed structures} \label{sec:Relaxed}

\begin{table}[ht]
\caption{\label{Lattice parameters}Lattice parameters of relaxed structures for PBE calculations}
\begin{ruledtabular}
\begin{tabular}{cccccccc}
\textbf{}  & \textbf{}  & \textbf{$a$} (\AA)     & \textbf{$b$} (\AA)     & \textbf{$c$} (\AA)    & \textbf{$\alpha$ ($^{\circ}$)} & \textbf{$\beta$ ($^{\circ}$)}  & \textbf{$\gamma$ ($^{\circ}$)} \\
\hline
      & exp. & 4.45  & 4.45  & 4.45  & 90    & 90    & 90    \\

\textbf{TiH\textsubscript{2} Cubic}            & theory     & 4.42  & 4.42  & 4.42  & 90    & 90    & 90    \\

           & $\Delta$  (\%)      & -0.61 & -0.61 & -0.61 & 0     & 0     & 0     \\

      & exp. & 4.53  & 4.53  & 4.28  & 90    & 90    & 120    \\

\textbf{TiH\textsubscript{2} Tetragonal}            & theory     & 4.47  & 4.47  & 4.34  & 90    & 90    & 120    \\

           & $\Delta$ (\%)      & -1.3 & -1.3 & 1.4 &    0  &   0   &   0   \\

       & exp. & 4.59  & 4.59  & 2.96  & 90    & 90    & 90    \\
    
\textbf{TiO\textsubscript{2} }          & theory     & 4.65  & 4.65  & 2.97  & 90    & 90    & 90    \\
           & $\Delta$  (\%)       & 1.25  & 1.25  & 0.23  & 0     & 0     & 0     \\

      & exp. & 5.16  & 5.16  & 13.61 & 90    & 90    & 120   \\
\textbf{Ti\textsubscript{2}O\textsubscript{3}}           & theory     & 5.11  & 5.11  & 14.03 & 90    & 90    & 120   \\
           & $\Delta$  (\%)      & -0.99 & -0.99 & 3.1   & 0     & 0     & 0     \\

        & exp. & 5.21  & 5.21  & 5.21  & 90    & 90    & 90    \\
\textbf{YH\textsubscript{2} Cubic}           & theory     & 5.19  & 5.19  & 5.19  & 90    & 90    & 90    \\
           & $\Delta$  (\%)       & -0.29 & -0.29 & -0.29 & 0     & 0     & 0     \\

       & exp. & 10.6  & 10.6  & 10.6  & 90    & 90    & 90    \\
\textbf{Y\textsubscript{2}O\textsubscript{3}}           & theory     & 10.64 & 10.64 & 10.64 & 90    & 90    & 90    \\
           & $\Delta$  (\%)       & 0.33  & 0.33  & 0.33  & 0     & 0     & 0    
\end{tabular}
\end{ruledtabular}
\end{table}

The experimental lattice parameters for TiH\textsubscript{2} cubic, YH\textsubscript{2} cubic, Y\textsubscript{2}O\textsubscript{3} and TiO\textsubscript{2} were taken from the Inorganic Crystal Structure Database (ICSD). The collection codes used were 169601 (TiH\textsubscript{2} cubic),~\cite{Kalita_2010} 638537 (YH\textsubscript{2} cubic),~\cite{Pebler_1962} 66242 (Y\textsubscript{2}O\textsubscript{3}),~\cite{Smrcok_1989} and 9161 (TiO\textsubscript{2}, rutile)~\cite{Baur_19715}. The experimental lattice parameters for TiH\textsubscript{2} tetragonal was taken from Ref.~\cite{Miwa_2002}.

\cleardoublepage

\section{Valence band photoionisation cross sections} \label{sec:VB_cross_sections}

\begin{table}[h!]
     \caption{One electron photoionisation cross section values as a function of photon energy, taken from the Scofield cross section tabulated data~\cite{Scofield1973} using the Galore software package.~\cite{Jackson2018} The units of the values are in barns/electron. The Ti~4\textit{p}/Y~5\textit{p} cross sections were estimated by dividing the one-electron Ti~4\textit{s}/Y~5\textit{s} cross sections by a factor of 2.}
     \label{tab:Xsection_VB}

    \begin{tabular}{cccccccc}
    \hline \hline
$h\nu$ / eV & H~1\textit{s} & Ti~4\textit{s} & Ti~4\textit{p} & Ti~3\textit{d} & Y~5\textit{s} & Y~5\textit{p} & Y~4\textit{d}  \\
   \hline
3269.1 & 2.091e-01 & 4.454e+01 & 2.227e+01 & 3.891e+00 & 4.670e+01 & 2.335e+01 & 3.106e+01 \\
7231.0 & 1.389e-02 & 6.652e+00 & 3.326e+00 & 1.270e-01 & 8.963e+00 & 4.482e+00 & 1.606e+00 \\
    \hline \hline
    \end{tabular}

\end{table}

\cleardoublepage

\bibliography{references_SI.bib}
\bibliographystyle{apsrev4-1}